\documentclass[sigconf]{acmart}
\AtBeginDocument{%
  }

\copyrightyear{2025}
\acmYear{2025}
\setcopyright{cc}
\setcctype{by-nc}
\acmConference[UIST '25]{The 38th Annual ACM Symposium on User Interface Software and Technology}{September 28-October 1, 2025}{Busan, Republic of Korea}
\acmBooktitle{The 38th Annual ACM Symposium on User Interface Software and Technology (UIST '25), September 28-October 1, 2025, Busan, Republic of Korea}\acmDOI{10.1145/3746059.3747677}
\acmISBN{979-8-4007-2037-6/2025/09}

\usepackage{booktabs, multirow} 
\usepackage{soul}
\usepackage{xcolor,colortbl} 
\usepackage{changepage} 

\usepackage{listings}
\usepackage{tikz}

\usepackage{booktabs}
\usepackage{graphicx}
\usepackage{multirow}
\usepackage{tabularx}
\usetikzlibrary{shapes,arrows,positioning}
\usepackage{caption}
\usepackage{subcaption}

\usepackage{color}
\usepackage{tabularray}
\definecolor{Silver}{rgb}{0.752,0.752,0.752}
\usepackage{float}

\newcommand{\sysname}{\textsc{BloomIntent}}

\begin{document}






\title{\sysname{}: Automating Search Evaluation with LLM-Generated Fine-Grained User Intents}


\author{Yoonseo Choi}
\email{yoonseo.choi@kaist.ac.kr}
\orcid{0000-0001-6808-2848}
\affiliation{%
  \institution{School of Computing, KAIST}
  \city{Daejeon}
  \country{Republic of Korea}
}

\author{Eunhye Kim}
\email{gracekim027@snu.ac.kr}
\orcid{0009-0004-1460-8532}
\affiliation{%
  \institution{Seoul National University}
  \city{Seoul}
  \country{Republic of Korea}
}

\author{Hyunwoo Kim}
\email{khw0726@kaist.ac.kr}
\orcid{0000-0002-4923-6650}
\affiliation{%
  \institution{School of Computing, KAIST}
  \city{Daejeon}
  \country{Republic of Korea}
}

\author{Donghyun Park}
\email{donghyun.park@navercorp.com}
\orcid{0009-0004-8917-5137}
\affiliation{%
  \institution{NAVER Corporation}
  \city{Seongnam-si}
  \country{Republic of Korea}
}

\author{Honggu Lee}
\email{honggu.lee@navercorp.com}
\orcid{0009-0004-4044-7189}
\affiliation{%
  \institution{NAVER Corporation}
  \city{Seongnam-si}
  \country{Republic of Korea}
}

\author{Jin Young Kim}
\email{jin.y.kim@navercorp.com}
\orcid{0009-0003-2678-3197}
\affiliation{%
  \institution{NAVER Corporation}
  \city{Bellevue, WA}
  \country{USA}
}

\author{Juho Kim}
\email{juhokim@kaist.ac.kr}
\orcid{0000-0001-6348-4127}
\affiliation{%
  \institution{School of Computing, KAIST}
  \city{Daejeon}
  \country{Republic of Korea}
}


\renewcommand{\shortauthors}{Choi et al.}

\begin{abstract}

If 100 people issue the same search query, they may have 100 different goals.
While existing work on user-centric AI evaluation highlights the importance of aligning systems with fine-grained user intents, current search evaluation methods struggle to represent and assess this diversity.
We introduce \sysname{}, a user-centric search evaluation method that uses user intents as the evaluation unit. \sysname{} first generates a set of plausible, fine-grained search intents grounded on taxonomies of user attributes and information-seeking intent types. Then, \sysname{} provides an automated evaluation of search results against each intent powered by large language models.
To support practical analysis, \sysname{} clusters semantically similar intents and summarizes evaluation outcomes in a structured interface.
With three technical evaluations, we showed that \sysname{} generated fine-grained, evaluable, and realistic intents and produced scalable assessments of intent-level satisfaction that achieved 72\% agreement with expert evaluators. 
In a case study (N=4), we showed that \sysname{} supported search specialists in identifying intents for ambiguous queries, uncovering underserved user needs, and discovering actionable insights for improving search experiences.
By shifting from query-level to intent-level evaluation, \sysname{} reimagines how search systems can be assessed---not only for performance but for their ability to serve a multitude of user goals.

\end{abstract}



\begin{teaserfigure}
    \includegraphics[width=1.00\textwidth]{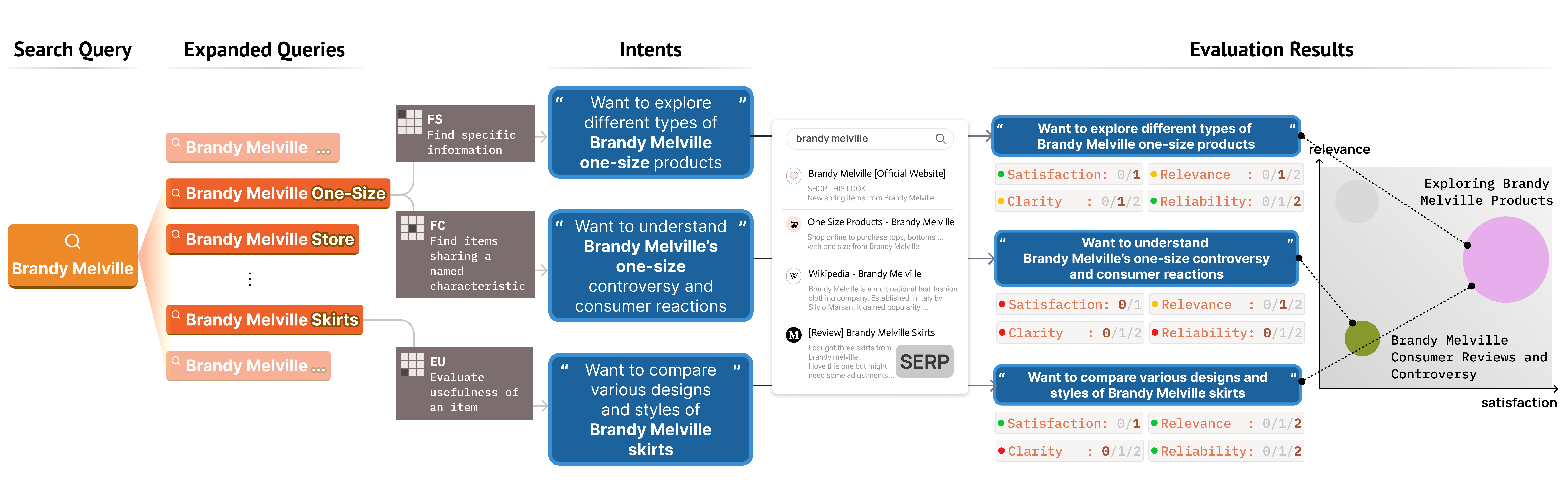}
    \centering
    \caption{A conceptual overview of \sysname{}, showing how a single search query is expanded into various queries, transformed into specific intents with intent types, evaluated against search engine results page (SERP) across different metrics, and clustered to reveal diverse user satisfaction patterns.}
    \Description{
    Teaser image. A conceptual diagram of the BloomIntent. On the left, a single search query "Brandy Melville" is shown expanding into multiple follow-up queries (e.g., “Brandy Melville One-Size”, “Brandy Melville Store”, “Brandy Melville Skirts”). Each expanded query is linked to a specific user intent, categorized by type: FS (Find specific information), FC (Find items sharing a named characteristic), and EU (Evaluate usefulness of an item). These intents are then evaluated against search engine results (SERP) on four metrics—Satisfaction, Relevance, Clarity, and Reliability. On the right, a 2D cluster visualization plots the evaluation results, illustrating how intents group are positioned with different satisfaction and relevance/clarity/reliability levels.
    }
    \label{fig:teaser}
\end{teaserfigure}


\maketitle

\section{Introduction}

The way people search the web shapes the information they encounter and how they make sense of the world.
Yet, the queries users enter are often underspecified and open-ended, leaving room for multiple interpretations depending on individual goals, preferences, and contexts~\cite{dou2007large}. 
For example, the query \textbf{\texttt{Hawaii honeymoon}} might imply various intents, such as exploring hotel options, checking available local activities, or looking for reviews from newlywed couples. 
Recognizing such intent diversity---the presence of multiple plausible user goals within a single query---is essential for understanding user search goals and evaluating how well systems meet users’ actual needs. 

Recent work on user-centric evaluation of AI systems highlights the importance of understanding fine-grained user goals and defining user-driven evaluation criteria to assess user satisfaction~\cite{kim2024evallm, joshi2025coprompter, szymanski2024comparing, sung2025verila, shankar2024spade}.
However, current methods for search evaluation and intent diversification have limitations in effectively measuring how well the search results satisfy each intent from a wide spectrum of possible user intents for a given query. 
Traditionally, search engine performance has been evaluated based on ranking metrics on benchmark datasets~\cite{voorhees2002Philosophy} or using behavioral signals from users~\cite{joachims2002optimizing, hassan2013beyond, dou2007large, kim2014modeling}.
While another thread of research seeks to incorporate multiple intents to improve ranking performance~\cite{macavaney2021intent5, zhou2024cognitive, li2023agent4rankingsemanticrobustranking, Radlinski2010Inferring}, their intent representations---often in the form of expanded queries or embeddings---lack the granularity needed to capture the fine-grained and contextual user goals, making it difficult to evaluate how well the search results satisfy specific user needs.
For example, even when an intent is represented via an expanded query such as \texttt{Hawaii honeymoon accommodations} derived from \texttt{Hawaii honeymoon}, the underlying user goals may vary--- such as comparing hotel options, looking for others' reviews, or seeking budget-friendly packages. This highlights the need for more explicit, intent-level representations to better capture diverse user needs and enable user-aligned evaluation.

In this work, we aim to enable diverse and scalable user-centric evaluation of search quality by using user intents as the unit of analysis.
We introduce \sysname{}, a multi-intent evaluation method that simulates a wide spectrum of plausible user intents for a given query, and automatically evaluates how well each is fulfilled by a search engine result page (SERP).
\sysname{} implements a two-stage pipeline: (1) intent generation and (2) intent contextualization.
To generate a diverse set of realistic, sentence-level intents, \sysname{} leverages large language models (LLMs) guided by a structured intent generation process, informed by query context, user attributes, and grounded in an information-seeking taxonomy drawn from IR literature~\cite{mitsui16extracting}.
For instance, the search query \texttt{``Hawaii Honeymoon''} can be expanded into specific sentence-level intents such as \texttt{``Check promotions for Hawaii honeymoons''} or \texttt{``Explore popular shopping destinations and recommended items for a Hawaii honeymoon.''} 
For contextualizing the generated user intents in terms of search result evaluations, \sysname{} first provides an LLM-powered automated evaluation of the SERP against each intent in terms of satisfaction, relevance, clarity, and reliability~\cite{liu2018satisfaction}.
Then, to provide high-level diagnostic insights, \sysname{} semantically clusters related intents and summarizes evaluation results through a web-based user interface, across individual intents, clusters of intents, and queries.

We validated the quality of \sysname{}-generated intents and evaluation results through three technical evaluations: (1) assessing the alignment between \sysname{}-generated follow-up queries and ground-truth follow-up queries from real-world search user logs; (2) comparing intent generation quality to a state-of-the-art baseline~\cite{jia-etal-2024-mill} in terms of evaluability and realism; and (3) evaluating the agreement between our automated evaluation and expert human judgments. 
We found that even the top 5\% of \sysname{}'s expanded queries most distant from the ground truth still captured valid search intents, with 84\% rated as relevant and 86\% as plausible.
The intent quality comparison showed that \sysname{} generated more realistic and evaluable intents with greater semantic diversity compared to the baseline.
Our automated SERP evaluation produced reliable results comparable to human evaluation, achieving 72.1\% agreement with expert raters on satisfaction.

To assess the practical utility of \sysname{}, we conducted a case study with four professional search specialists (e.g., product managers). Our participants valued that \sysname{} helped them uncover previously overlooked user needs, identify intent-specific issues often obscured by aggregate scores, and better understand SERP quality in context. They highlighted its value in inspecting ambiguous or low-performing queries, enabling early-stage diagnosis, and identifying actionable design signals (e.g., when a SERP might benefit from richer comparison formats or visual elements).

Overall, we propose intent-based evaluation as a scalable and user-centric method for diagnostic understanding of search quality considering user needs.
Our contributions are as follows:
\begin{itemize}
    \item \sysname{}, a novel LLM-powered evaluation method that generates and contextualizes diverse, fine-grained user intents to enable automatic, intent-level evaluation of search results.
    \item Empirical findings from three technical evaluations and a case study with search experts: 
    \begin{itemize}
        \item Technical evaluations demonstrating that \sysname{} generates diverse and realistic search intents, and provides reliable intent-level evaluation of SERPs.
        \item A case study showing how intent-level evaluation enables search specialists to uncover underserved user needs and gain a user-centric understanding of search performance.
    \end{itemize}
\end{itemize}

\section{Related Work}
We review existing work on: 1) taxonomizing and analyzing search intents, 2) evaluating search results through traditional information retrieval (IR) methods and recent LLM approaches, and 3) user-centric evaluation of AI systems.

\subsection{Search Intent Representation and Modeling}

Understanding user intent is central to the modern search domain. It not only helps retrieve relevant results, but also plays a critical role in evaluating whether the system effectively supports the user's information-seeking goals~\cite{Chapelle2011IntentbasedDOA}. However, queries alone often fail to capture these goals.  
Early research focused on categorizing queries based on high-level intent types---such as informational, navigational, transactional queries~\cite{broder2002taxonomy, rose2004understanding}. More recent taxonomies have evolved to capture more fine-grained search tasks~\cite{mitsui16extracting, cambazoglu2021intent, shah2023using}.

While these approaches focus on classifying or inferring a dominant intent per query, recent studies have highlighted the need to model the intent diversity within individual queries---capturing the presence of multiple user goals behind a single query string. 
Traditional IR research inferred user intents using external resources such as behavioral logs~\cite{Li2008learning, li2010learning, sadikov2010clustering, Radlinski2010Inferring}, knowledge bases~\cite{Jiang2017Generating, Bouchoucha2013Diversified}, or topics from retrieved documents~\cite{yu2022towards, hashemi2021learning}.

With the generative capabilities of LLMs, several lines of work has explored generating diverse query variants~\cite{macavaney2021intent5, li2023agent4rankingsemanticrobustranking, Alaofi2023Can, zhou2024cognitive, jia-etal-2024-mill} capturing multiple facets of user intent, and incorporating such diversity into query-document matching for improved retrieval~\cite{li2024multi, bai2024intent, jia-etal-2024-mill, wang-etal-2023-query2doc, Sakai2021Retrieval}.
However, many of these methods represent intents using short phrases or keyword-based proxies, which offer limited insights into the users' specific intent and task with the search. 

In this work, we explore how to generate user intents that are more comprehensible and fine-grained to guide understanding of different user goals and design the search experience. 
Our approach goes beyond keyword-level proxies by generating sentence-level, behaviorally-grounded intent statements that are fine-grained and evaluable.


\subsection{Search Evaluation: From Traditional Approaches to LLM-Based Frameworks}

Traditional approaches for online search evaluation have focused on evaluating the relevance of search results using a variety of user behavior signals such as dwell time, click ranks, query reformulations, or combinations of them~\cite{vaughan2004new, hassan2013beyond, hofmann2016Online}. By comparing these behavioral signals with user feedback, such metrics were shown to be closely related to user satisfaction in general~\cite{fox2005Evaluating, liu2018satisfaction, huffman2007well, mao2016does}. However, these approaches often struggle to capture the full spectrum of user intents behind a query. Therefore, the need for flexible evaluation methods that model diverse user perspectives has been a long-standing challenge in information retrieval research~\cite{ali2011overview, patel2024aime, anand2024understanding}. 

Recent work demonstrates the growing use of LLMs as evaluators across tasks like text generation and information retrieval~\cite{li2024llms}, with studies confirming that their evaluations align well with expert judgments~\cite{chiang2023can}. G-Eval~\cite{liu2023g} showed GPT-4 with chain-of-thought reasoning could assess outputs closely aligned with human judgments, while GPTScore~\cite{fu2023gptscore} introduced multi-faceted text evaluations using pre-trained models.
Specifically, for IR evaluation, LLMs have enabled improved user profiling, context modeling, and interaction decoding~\cite{zhang2024agentic, xiong2024search, ai2023information, malaviya2024contextualizedevaluationstakingguesswork}. Research has shown that LLMs can predict searcher preferences at levels comparable to human annotators~\cite{thomas2024large} and provide nuanced, context-aware evaluations for document ranking~\cite{niu2024judgerank}. Explorations of large-scale search relevance automation further demonstrate LLMs' potential as scalable alternatives to human annotators~\cite{rahmani2024llmjudge}.

Despite these advances, most prior work focuses on LLMs as scoring engines---replacing human assessors or automating labels for training and evaluation pipelines. While effective for aggregate relevance judgments, these approaches often overlook which specific user goals are unmet, especially in cases of ambiguous or underspecified queries. 
In contrast, we treat LLMs not merely as evaluators, but as enablers of intent-aware, fine-grained diagnosis of search performance. Rather than optimizing for agreement with human scores, our method leverages LLMs to uncover and assess a range of plausible user intents behind a query. This approach supports deeper diagnostic insight and user-centered analysis within the evaluation process.

\subsection{Customizable Evaluations of AI systems}


Recent studies in human-centered AI emphasize that evaluating AI systems should go beyond static, benchmark-based metrics and instead support user-defined, context-aware criteria. Specifically, several studies have focused on surfacing model weaknesses that may be hidden beneath aggregate performance scores. Zeno~\cite{Cabrera2023ZenoAIA}, Errudite~\cite{wu2019errudite}, and Scattershot~\cite{wu2023scattershot} support fine-grained inspection across input subgroups or examples, helping users detect systematic failures often overlooked by overall accuracy or relevance metrics. Similarly, Divisi~\cite{sivaraman2025divisi} introduces an exploratory subgroup evaluation framework that automatically proposes interpretable slices of data for deeper inspection, surfacing non-obvious behavior patterns. Angler~\cite{robertson2023angler} tackles the challenge of prioritization, offering interfaces that help users decide which subgroup failures are more critical or urgent to address. 

To this end, interactive tools have been developed that empower users to define, revise, and operationalize their own evaluation standards, leading to more tailored and interpretable assessments. Tools such as ChainForge~\cite{arawjo2024chainforge}, EvalLM~\cite{kim2024evallm}, EvalGen~\cite{shankar2024validates}, MetricMate~\cite{gebreegziabher2025metricmate}, and CoPrompter~\cite{joshi2025coprompter} exemplify this trend by supporting criteria-driven evaluation in LLM-based systems. For instance, EvalLM allows users to iteratively create and refine evaluation criteria for prompt engineering tasks~\cite{kim2024evallm}.

Our work aligns with and extends these directions by introducing a search evaluation method grounded in explicit user intents. In line with user-defined evaluation approaches, our method surfaces evaluation criteria through intent statements that capture diverse and nuanced user goals. This intent-level structure enables both detailed analysis and potential prioritization, supporting tasks such as discovering underserved informational needs or identifying high-impact failure points across queries.

\section{Challenges of Multi-Intent Evaluation}

Recent work on user-centric evaluation of AI systems in HCI emphasizes the importance of understanding fine-grained user intents for assessment. 
However, there exist gaps between the context of user-centric AI systems evaluation and the search evaluations, making it challenging to directly apply the existing approaches.
Bridging the gap requires determining how the intent should be elicited, represented, and utilized for search evaluation. 
Here, we introduce several practical challenges of implementing large-scale intent-based evaluation in the domain of search and propose our approaches for resolving them. 

\paragraph{Challenge 1: Eliciting and Representing User Intents}
In HCI, tools like ChainForge~\cite{arawjo2024chainforge}, EvalLM~\cite{kim2024evallm}, and MetricMate~\cite{gebreegziabher2025metricmate} have empowered users to define evaluation criteria aligned with their own goals, supporting more user-centered evaluation pipelines.
However, the context of search systems introduces distinct challenges for user-defined evaluation. 
Unlike creative or open-ended tasks, where users can directly articulate their preferences or criteria, search tasks often begin with underspecified queries that only implicitly reflect user goals. 
Current human evaluation practices (e.g., Google's evaluator guidelines~\cite{Google_2025}) rely heavily on the evaluator's common sense or imagination to infer the potential search goals, which could be distant from the actual user goals. To address this, we propose combining LLM-driven generation and structured representation of user intents, designed to meet four key requirements: \textit{explicit and clear goal expression}, \textit{broad coverage}, \textit{realistic reflection of actual user behaviors}, and \textit{capability to uncover unexpected but valid needs}.

\paragraph{Challenge 2: Scaling Evaluation Across Intents}
While existing approaches of user-centric AI evaluation focus on the end user evaluating the AI outcome to their own needs, the context of search evaluation leads to a different setting, as separate search specialists evaluate the satisfaction of the end-users at scale. 
Therefore, if multiple user intents surface, evaluating them individually introduces the challenge of scalability. 
Building upon prior work using LLMs as automated judges~\cite{niu2024judgerank} and scalable alternatives to human evaluations~\cite{rahmani2024llmjudge}, we explore repositioning existing LLM-powered text evaluation approaches for enabling more diagnostic and user-aligned evaluation per search intent. 
In our work, we explore how LLM-based automated evaluation might help support fine-grained intent assessments without requiring costly human annotation for each case. 
Instead of optimizing only for agreement with human scores, our framework leverages LLMs to uncover and assess a range of plausible user intents behind a query.

\paragraph{Challenge 3: Making Multi-Intent Evaluation Results Understandable and Actionable}
Prior work in human-centered AI has shown that as systems grow in complexity, raw performance scores often fail to reveal where and why models underperform.
Similarly, with dozens of intents per query, search specialists should be able to identify insights from the raw satisfaction scores, such as meaningful patterns or critical failures.
Inspired by existing HCI systems like Zeno~\cite{Cabrera2023ZenoAIA}, Errudite~\cite{wu2019errudite}, and Divisi~\cite{sivaraman2025divisi}, which support fine-grained subgroup exploration and prioritization, we expect that multi-intent evaluation would also benefit from having structured interfaces and workflow support. To be usable in practice, our framework incorporates a visual interface that clusters related intents, surfaces cross-intent evaluation patterns, and enables exploration from query-level overviews to intent-specific diagnostics. 


\section{\sysname{}}

\begin{figure}
\centering
\includegraphics[width=0.98\linewidth]{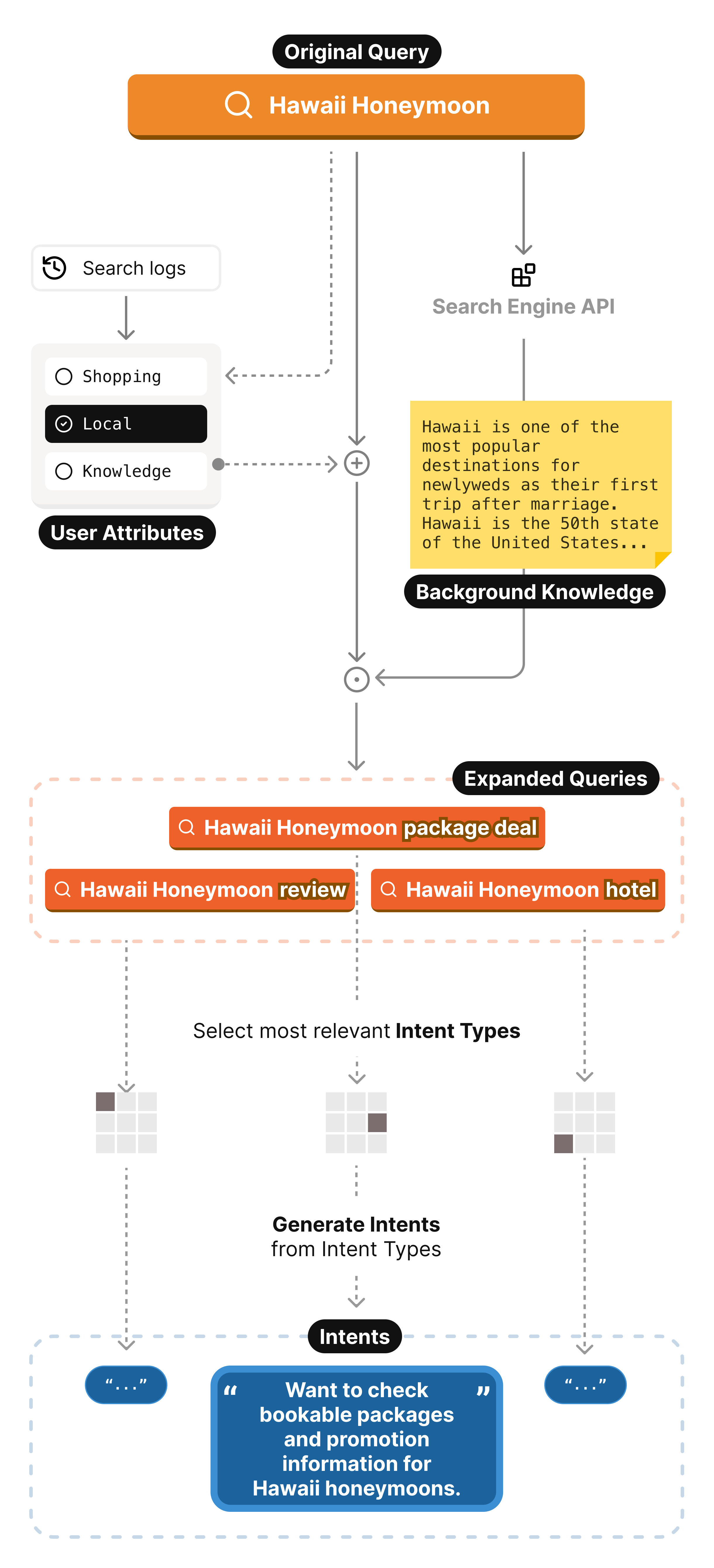}
\caption{Intent Generation Pipeline overview, illustrating the process from original query to final intent construction. The pipeline first generates expanded queries representing the search tasks by incorporating the user attributes per query category and the background context retrieved from external search engines. For each expanded query, the pipeline selects relevant intent types and constructs explicit intent statements that represent diverse user goals.}
\Description{
A diagram illustrating the Intent Generation Pipeline for the example query "Hawaii Honeymoon." The process begins with the original query, which is enriched with domain user attributes (e.g., Shopping, Location, Knowledge) derived from search logs, and background knowledge retrieved via search engine APIs. This combined information is used to generate expanded queries such as "Hawaii Honeymoon package deal," "Hawaii Honeymoon review," and "Hawaii Honeymoon hotel." These are then mapped to the most relevant intent types, which guide the generation of specific user intents, such as "Want to check bookable packages and promotion information for Hawaii honeymoons." The figure demonstrates how contextual attributes and query semantics contribute to fine-grained intent construction.
}
\label{fig:pipeline}
\end{figure}

We propose a two-stage framework, \sysname{}, for a fine-grained, scalable user intent-based evaluation of search results, consisting of: (a) an \emph{intent generation pipeline}, which produces a scalable yet comprehensive set of plausible intents for a given search query (Figure~\ref{fig:pipeline}), and (b) an \emph{intent contextualization pipeline}, which systematically assesses the SERP quality for each intent and constructs structured results that search specialists can act upon.

\subsection{Intent Generation Pipeline: Identifying Plausible and Diverse Intents at Scale}
\label{sec:pipeline1}

The intent generation pipeline is designed to simulate realistic user intents by leveraging actual search logs to generate topic-specific user attributes and query-specific background knowledge. 
It consists of two main components: (1) expanded query generation by incorporating user attributes and contextual knowledge to reflect diverse perspectives and realistic scenarios, and (2) final intent construction through the selection of best-matching intent types to further specify the search task.

\subsubsection{Expanded Query Generation} 
\label{sec:pipeline1-expandedquery}

As a first step toward bringing original queries to a form that better reflects the user's intents, we generate expanded queries simulating how users naturally reformulate or refine their initial inputs. 
To achieve this, we incorporate two types of information: (1) user attributes, which help us consider diverse search preferences and user contexts that shape users' information needs (e.g., price sensitivity may lead to comparing different price options, novice users would likely need more introductory information); and (2) query background knowledge, which ensures that the generated expansions are grounded in up-to-date and contextually accurate information. The combination of these components allows for expanded queries that remain contextually relevant while addressing the varied information needs of different users with the same original query.

\paragraph{User Attribute Generation}
\label{sec:pipeline1-userattribute}

Users with different goals and backgrounds may issue the same query but expect different outcomes. For example, a user searching for \texttt{``Hawaii Honeymoon''} may prefer authentic experiences from travel blogs, while another user may seek resort packages or price comparisons based on their budget constraints or planning preferences. These different user attributes lead to divergent information needs even for identical queries.

To incorporate this diversity, we leverage search logs of actual users. Search logs provide rich behavioral signals---including search queries, dwell times, and clicked URLs---which collectively capture how users explore information online~\cite{6514421}. We leverage these signals to infer user attributes and combine them to simulate realistic user profiles.

To construct the attribute taxonomy, we randomly sampled 100 search sessions from each query category (Shopping, Local, Knowledge) that occurred from November 21st to November 27th, 2024, from users who had issued over 100 queries during that period. Then, we prompted OpenAI's \texttt{GPT-4o-2024-11-20} model to generate user attributes from the logs.
To ensure that all user attributes are captured, we repeated the generation three times.
We then aggregated and synthesized overlapping behavioral traits through affinity diagramming to derive a comprehensive attribute set. 

The final user attribute set includes six to seven dimensions such as \textit{price sensitivity} (Shopping category; e.g., Discount seeker, Quality-focused, Value hunter), \textit{content format preference} (Location category; e.g., map-centric, list/directory, textual details), \textit{task complexity} (Knowledge category; e.g., single-step, multi-step, comparative, problem-solving, etc).
The prompt for generating user attributes from search logs and the final list of complete user attributes for the three query categories is provided in Appendix~\ref{appendix:user_attribute_generation}.

\paragraph{Query Background Knowledge Retrieval}
\label{sec:pipeline1-background}

To ensure intent generation with up-to-date query-specific information, we augment LLM's general world knowledge with targeted background retrieval. We collect background knowledge by retrieving search results from Google and Naver\footnote{https://www.naver.com}, the two most popular search engines in South Korea. We use Google Custom Search API\footnote{https://developers.google.com/custom-search/v1/overview} to collect the top 10 results from Google, and Naver Search API\footnote{https://developers.naver.com/docs/serviceapi/search/blog/blog.md} to collect the top 5 results across multiple categories (web, blogs, cafes, shopping, and maps). We then summarize the retrieved results into a 200-word paragraph using the \texttt{GPT-4o-2024-11-20} model, offering a concise and up-to-date contextual snapshot for the query. The prompt for generating query background is provided in Appendix~\ref{appendix:query_background_knowledge_retrieval}.

\paragraph{Expanded Query Formulation}
\label{sec:eq-formulation}

Each expanded query is generated by combining (a) the original query, (b) its summarized background context, and (c) a combination of user attributes.
To ensure that user attributes reflect real users' search needs as ingredients for our approach, we prompt the \texttt{GPT-4o-2024-11-20} model to create up to 10 distinct and realistic user profiles per query. Each profile consists of multiple attributes that are likely to co-occur in real user behavior to reflect a unique search persona (e.g., an exploratory user with low familiarity and high information need; a decisive expert looking for quick comparisons). The prompt explicitly instructs the model to generate \textit{meaningful, plausible, and mutually exclusive} profiles tailored to the query's context with a brief rationale describing how the generated profile would shape the search intent.

We prompt the \texttt{GPT-4o-2024-11-20} model to generate expanded queries using no more than two additional words, reflecting the natural refinement patterns observed from real-world search logs. 
For each original query, we generate 100 expanded queries: 50 guided by user attribute profiles and 50 without, to capture both user-attribute-based expansions and general intent variations. 
We then apply an additional LLM prompting step to remove duplicates using the same model. We empirically chose the number 100 as a balance point between encompassing the diversity of user tasks, especially for ambiguous and open-ended queries, and the practical usability of the evaluation result. This number can be adjusted depending on different evaluation goals. 
The full prompt used for expanded query generation is included in Appendix~\ref{appendix:expanded_query_prompt}.


\subsubsection{Intent Type Selection \& Construction}
\label{sec:pipeline1-intenttype}

We convert each expanded query into an explicit intent statement through a two-step process: (1) selecting one or more appropriate intent types from a predefined taxonomy, and (2) generating the intent statement grounded in the selected types.

We adopt an existing taxonomy of search intents~\cite{mitsui16extracting}, which captures common search goals. These intent types serve as intermediate semantic structures guiding the generation of clear, goal-oriented intent statements. Of the original 20 intent types, we excluded nine types that did not align with our evaluation goal based on our empirical observations. 
The final set of the 11 intent types and the selection prompt are included in Appendix~\ref{appendix:intent_types}. 

For each expanded query, we prompt the \texttt{GPT-4o-2024-11-20} model to select up to three of the most relevant intent types, along with a brief justification. Once intent types are selected, we prompt OpenAI's \texttt{o3-mini-2025-01-31} model to generate a single, explicit natural language intent statement that clearly articulates the user's underlying goal, constrained to a maximum of 15 words. For all intent outputs, we apply a filtering prompt to remove vague, redundant, or implausible outputs. The prompt for generating intents is included in Appendix~\ref{appendix:intent_gen_prompt}. 

\subsection{Intent Contextualization Pipeline: Structuring Scalable Intents for Search Results Evaluation}
\label{sec:pipeline2}

The intent contextualization pipeline aims to enable the discovery of actionable insights to improve search results using the generated search intents. 
This stage has two key goals: (1) translating user-centered intents into consistent SERP evaluation results, and (2) aggregating results into semantically coherent intent clusters for high-level analysis and actionable insight discovery.

\subsubsection{Automated SERP Evaluation with Search Intents} 
\label{sec:pipeline2-autoeval}
\sysname{} first contextualizes the search results in terms of each search intent by evaluating whether the given SERP satisfies it.

\sysname{} first parses the SERP given in HTML format to extract the title and the preview text of each result snippet, while skipping image-only and video-only results.
The pipeline evaluates the SERP quality using four dimensions, adapted from widely used search evaluation frameworks~\cite{liu2018satisfaction}:
\begin{itemize}
    \item \textbf{Satisfaction}: Assesses whether the search results fully fulfill the user’s intent by providing complete and useful information.
    \item \textbf{Relevance}: Measures topical alignment between the content and the expressed intent, independent of presentation or completeness.
    \item \textbf{Clarity}: Evaluates how well the content is organized and conveyed, including readability and appropriate use of language or structure.
    \item \textbf{Reliability}: Judges the trustworthiness and credibility of the information based on source authority and evidence quality.
\end{itemize}

We use a binary score for satisfaction (1 = satisfied, 0 = unsatisfied) and a three-point scale (0–2) for the other metrics:

\begin{itemize}
    \item \textbf{Score 2}: Fully meets the evaluation criteria.
    \item \textbf{Score 1}: Partially meets the criteria.
    \item \textbf{Score 0}: Fails to meet the core criteria.
\end{itemize}

Each evaluation is conducted with the \texttt{GPT-4o-2024-11-20} model, with a structured prompt consisting of a query, user intent, retrieved SERP content of the search query being evaluated, and the metric definition to produce a numeric score and a free-form text explanation behind the scoring.
These explanations not only justify the judgment but also introduce a form of reasoning akin to chain-of-thought prompting, which we found helpful in improving consistency. The full prompt templates for each metric are available in Appendix~\ref{appendix:evaluation_prompts}.

\begin{figure*}[t]
\centering
\includegraphics[width=\linewidth]{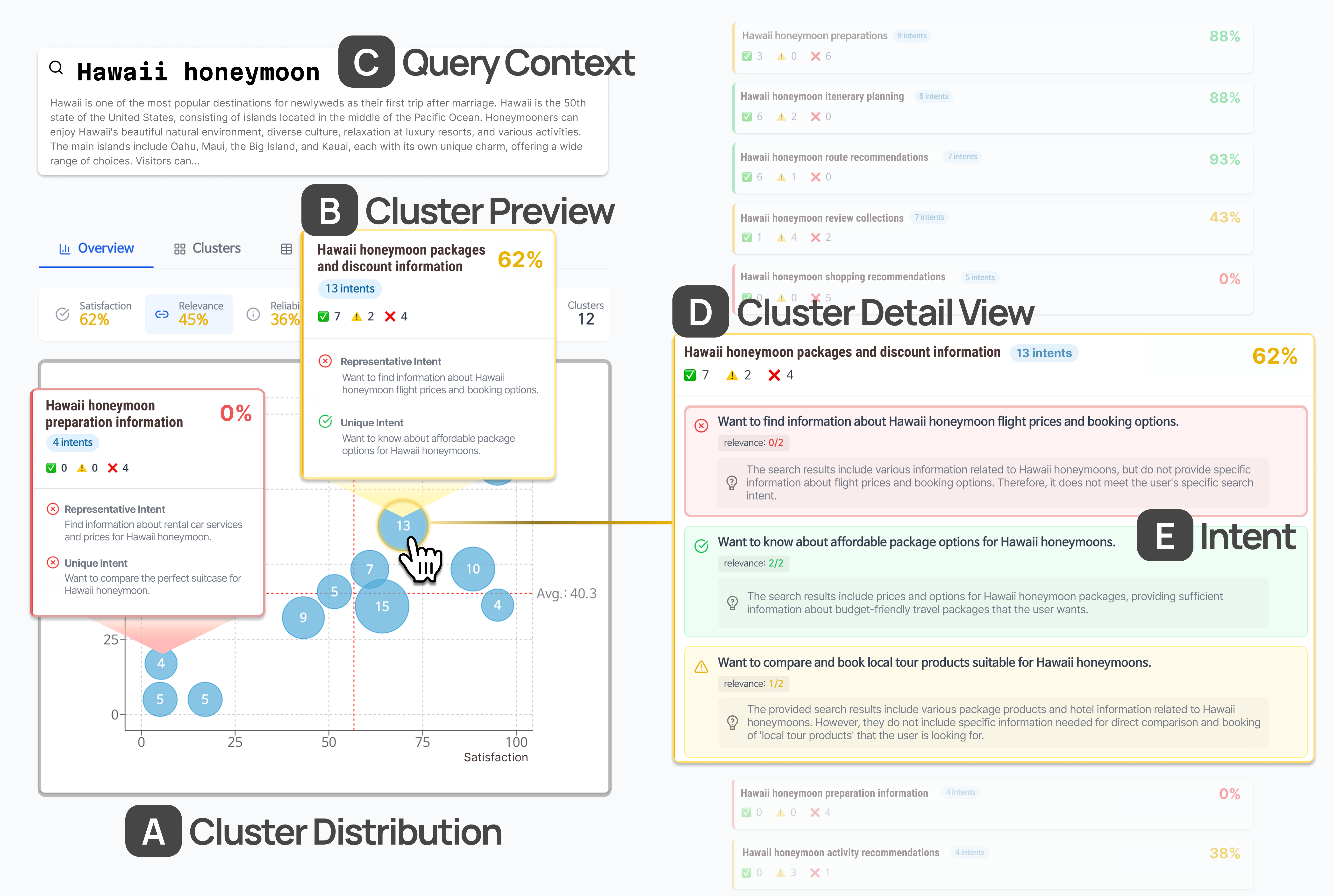}
\caption{The system interface of \sysname{} used in our case study. Given a query, the \emph{Query Context} (C) provides background information to help users better understand the search scenario and user expectations.
In \emph{Cluster Distribution} (A), users can investigate intent clusters with varying levels of evaluation metrics. Hovering over a cluster activates the \emph{Cluster Preview} (B), which summarizes representative and unique intents within that cluster along with key evaluation signals. Clicking on a cluster opens the \emph{Cluster Detail View} (D), where users can examine all underlying intents and their evaluation outcomes. Each \emph{Intent Card} (E) presents a specific user intent, the evaluation result of the current search result page, and natural language explanations, supporting user-centered diagnosis and interpretation.
}
\Description{
A screenshot of the BloomIntent user interface for analyzing search intent evaluation. (A) A scatterplot shows clusters of intents plotted by satisfaction (x-axis) and a selected metric (y-axis), with numeric labels indicating cluster sizes. (B) A highlighted cluster labeled "Hawaii honeymoon packages and discount information" displays a 62\% satisfaction score and 13 intents, along with representative and unique intents summarized. (C) The top section shows the original query context for "Hawaii honeymoon" with a snippet of background information. (D) A detailed cluster view lists all intents within the selected cluster, with associated relevance scores and evaluation notes. (E) Individual intents are shown with interpretations explaining whether the retrieved search results meet user needs. The interface enables exploration of evaluation statistics and satisfaction patterns across different clusters and intents.
}
\label{fig:system}
\end{figure*}

\subsubsection{Semantic Clustering for User-friendly Interpretation}
\label{sec:pipeline2-clustering}
While intent-level evaluation provides fine-grained insights, the comprehensive nature of this approach can generate numerous intents per query. To optimize both analytical depth and practical utility for search specialists who need to efficiently review data and make timely decisions, we implement semantic clustering of similar intents. This approach streamlines the evaluation process by combining intent evaluation outcomes at the cluster level.

We apply agglomerative clustering to group intents with semantically similar goals. We first created embeddings for each intent using OpenAI's \texttt{text-embedding-3-small} model. Then, we use the elbow method to determine the optimal number of clusters. We added a balancing mechanism to the cluster structure by splitting overly broad clusters and merging those that are too sparse.

To make each cluster interpretable, we provide three different representations. First, we identify the \textbf{centroid intent}, which is the closest intent from the mean cluster embedding, as the most representative intent of the cluster. We also identify the \textbf{outlier intent}, which is the most distant intent from the cluster center, as a boundary example. Finally, we use the \texttt{GPT-4o-2024-11-20} model to concisely capture all intents within the cluster to create a \textbf{descriptive cluster name}.

These clusters with their summaries help search specialists reason about the system performance at the level of intent categories (e.g., ``comparison-focused intents'') rather than isolated cases. For example, a cluster on ``resort comparisons'' for the query ``Hawaii honeymoon'' might benefit from tabular SERP formats, while a cluster of ``Hawaii honeymoon planning guides'' might call for structured itineraries or visual content featuring popular destinations. Similarly, a cluster on ``Hawaii honeymoon reviews'' could be better served with user testimonials and authentic experiences.

In this way, semantic clustering functions as a bridge between detailed intent-level analysis and practical evaluation workflows by reducing complexity while preserving information on user intents.

\subsection{Intent Exploration Interface}

To support search quality evaluators in understanding and acting on our multi-intent evaluation results, we developed an interactive interface that visualizes intents, clusters, and evaluation results at multiple levels of granularity. While our pipeline effectively generates diverse interpretations of user intents behind the query, using these results in real-world evaluation workflows remains challenging due to the volume of data and the complexity of the relationships between intents. Our custom interface addresses these challenges by organizing evaluation data into structured views that facilitate pattern recognition, cross-metric comparison, and identification of priority improvement areas. The interface enables search specialists to navigate from high-level query performance to specific intent clusters and individual evaluation details as described below.

\subsubsection{Query List View}
The \textit{Query List View} serves as an entry point to our system, providing a high-level overview of all evaluated queries. For each query, we display its topic category, the number of generated intents, the number of intent clusters, and aggregate scores across our four evaluation metrics. 
This view enables evaluators to quickly identify queries with potential issues---such as low satisfaction---and to compare performance across different query categories.
Evaluators can also enter new queries using the search bar to explore generated intents and their associated evaluation scores in real-time.

\subsubsection{Query View}
The Query View presents three complementary components that help evaluators interpret evaluation results of a specific query: \textit{Query Context panel}, \textit{Search Engine Preview}, and \textit{Cluster Overview}.

The central element is the \textit{Cluster Distribution} (Figure~\ref{fig:system}A), which plots intent clusters on a customizable two-dimensional scatter plot. 
Clusters are positioned on the x-axis by their average satisfaction (our primary metric), while the y-axis can be customized to reflect one of the other metrics---relevance, clarity, or reliability. The size of each plotted circle corresponds to the number of intents in that cluster. This visualization enables evaluators to quickly identify patterns such as clusters that perform well on relevance but poorly on clarity, suggesting specific improvement opportunities. For example, when evaluating the query ``best running shoes,'' evaluators might discover that a cluster of price-comparison intents shows high relevance but low satisfaction---indicating a need for clearer or more complete price information in the SERP.
When hovering on each cluster in the Cluster Distribution, the system presents a Cluster Preview (Figure~\ref{fig:system}B) showing the score summary and the representative intents within the cluster. 

\textit{Query Context panel} (Figure~\ref{fig:system}C) displays background information about the query, including potential ambiguities and common information, helping evaluators understand the query's scope and complexity. 
Adjacent to this, the \textit{Search Engine Preview} shows the actual search results that were evaluated, providing the necessary context for understanding the scores. 

\subsubsection{Cluster Detail View}
The \textit{Cluster Detail View} (Figure~\ref{fig:system}D) provides fine-grained access to individual intents and their evaluations. Intents are organized by their clusters, with each cluster displaying its representative (centroid) and outlier intents. For each intent, we display the user attributes that influenced its generation (e.g., ``price-conscious shopper''), along with the evaluation scores and the LLM's reasoning for each metric (Figure~\ref{fig:system}E).

Evaluators can filter the view by metric and sort by score to focus on specific aspects of search quality. Importantly, the interface allows evaluators to activate or deactivate specific intents or entire clusters, providing a mechanism to exclude edge cases or irrelevant intents from the aggregate evaluation. This feature acknowledges that automatically generated intents may occasionally miss the mark and gives evaluators greater control over which user perspectives are included in the overall quality judgment.

\subsubsection{Implementation.}
The interface was implemented as a web-based dashboard using React, and the backend was built using Python with FastAPI\footnote{https://fastapi.tiangolo.com/} using a Firestore Database\footnote{https://firebase.google.com/docs/firestore}. Evaluation results were stored with query-intent-cluster linkage for responsive filtering and aggregation.


\section{Technical Evaluation}

We describe a series of technical evaluations designed to assess the effectiveness and reliability of our framework across three dimensions: (1) the quality of generated expanded queries (Section~\ref{sec:pipeline1-expandedquery}), (2) the quality of generated intents (Section~\ref{sec:pipeline1-intenttype}), and (3) the alignment between our automated evaluation pipeline results (Section~\ref{sec:pipeline2-autoeval}) and expert human judgments. Each evaluation was grounded in both computational methods and qualitative assessments by human evaluators to provide a comprehensive understanding of how our system aligns with real-world user behavior and expert human evaluation standards.

Through these evaluations, we address two research questions:
\begin{itemize}
    \item \textbf{RQ1: Can our intent generation pipeline generate \emph{evaluative intents} that provide realistic, concrete and novel user perspectives?} 
    \item \textbf{RQ2: To what extent can our automated evaluation pipeline replicate expert human judgments of result quality across diverse user intents?} 
\end{itemize}

\paragraph{Evaluation Dataset Construction}
\label{sec:evaluation-dataset}
To evaluate our system's ability to generate realistic, diverse, and intent-rich query representations, we constructed a dataset of search queries that vary along two proxy measures of intent diversity: (1) click diversity---how diverse users' behaviors are in response to the query, and (2) reformulation diversity---how diverse the reformulated queries when initial results are unsatisfactory. A good evaluation dataset should include queries that are both straightforward (i.e., with a dominant or well-understood intent) and ambiguous or open-ended (i.e., likely to represent many possible user goals), in order to test the limits of our multi-intent generation and evaluation pipeline.

We began with a set of short-head queries---frequently searched, brief (1-3 words) queries that are often under-specified yet can exhibit substantial variability in user intent~\cite{dou2007large}. 
From the set of a year-long search logs of a commercial search engine (Dec 1st, 2023 - Nov 30th, 2024), we randomly selected 655 queries in three topic categories: Shopping, Location, and Knowledge. 
We chose a 1-year period to consider temporal and seasonal variance in search queries and behaviors.
After removing malformed expressions (e.g., incomplete phrases, misspellings, or unintended combinations of words), we retained a cleaned set of 623 queries (Shopping: 126, Location: 259, Knowledge: 238).

To quantify the click diversity and reformulation diversity, we computed two key metrics:
\begin{itemize} 
    \item \textbf{Click entropy}~\cite{duan2012click}, which captures the diversity of documents clicked by different users for the same query, indicating multiple possible interpretations.
    \item \textbf{Negative reformulation count}, defined as the number of distinct follow-up queries that were likely issued due to dissatisfaction with the initial results. We defined negative reformulations as follow-up queries that (a) included the original query as a strict substring, (b) were not auto-suggested by the system, (c) occurred within a short time window from the original query, and (d) meaningfully specified the search intent. To distinguish meaningful intent shifts from trivial edits, we applied a scoring function based on the Jaccard similarity of word sets between the original and reformulated query and the temporal gap between them.
\end{itemize}

To capture a representative range of search situations, we classified queries into four categories, based on the median of the two metrics: (1) high entropy / high reformulation, (2) high entropy / low reformulation, (3) low entropy / high reformulation, and (4) low entropy / low reformulation. These quadrants help us ensure that our evaluation covers both straightforward and ambiguous queries, with varied user satisfaction patterns.
From each quadrant, we randomly sampled 15 queries per topic category, resulting in 60 queries per category and 180 queries total.

\subsection{Evaluating the Quality of Generated Expanded Queries} 
\label{techeval-1}

To evaluate our expanded query generation pipeline, we assessed how well the generated queries reflect real-world user reformulation behavior. In particular, we focused on comparison with follow-up search queries---a natural indicator of unmet information needs~\cite{hassan2013beyond, chen2021towards}---as they signal implicit intents unfulfilled by search with the original query. By comparing \sysname{}-generated expanded queries against these real-world follow-ups, we aim to measure (a) \textbf{alignment}—how well the generated queries semantically match with real-world reformulations—and (b) \textbf{validity}—whether the generated queries are plausible, relevant, and useful.

For all 180 queries sampled in our dataset construction process, our pipeline generated up to 100 expanded queries per query. After filtering (Section~\ref{sec:eq-formulation}), we had 9,873 expanded queries in total. 
While this ground truth was limited mainly to unsatisfied intents, we argue that these cases are particularly important, as they reveal unmet information needs~\cite{chen2021towards, rha2016reformulations}. 



\subsubsection{Measuring Alignment via Similarity Scores}
\sysname{}-generated follow-up queries aimed to replicate real-world query reformulations. We measured semantic alignment against ground-truth follow-ups from real-world search logs across 180 queries (7,427 ground-truth follow-ups, \textmu = 41.2 per query).
Since direct lexical comparison was impractical due to varying word choices and reformulation strategies, we used BERTScore~\cite{bert-score}, a semantic similarity metric based on contextual embeddings. Specifically, we employed the KoBigBird model~\cite{park2021kobigbird}, one of the state-of-the-art Korean sentence embedding models.

For each pipeline-generated follow-up query, we computed similarity against all ground-truth follow-ups, and used the highest BERTScore as the \emph{similarity score}, capturing alignment with at least one real user reformulation.

\subsubsection{Evaluating Validity via Human Judgment}
We supplemented computational metrics with human evaluation, particularly focusing on queries that showed low similarity scores, to determine whether these \sysname{}-generated queries still represented valid and meaningful user intentions despite being semantically distant from real-world follow-up queries.
This assumption stems from \sysname{}'s core goal of exploring diverse and long-tail user intents beyond historical reformulations, potentially surfacing underserved needs not frequently observed in query logs but still highly relevant.
We selected the bottom 5\% of generated follow-up queries based on similarity scores ($\leq$ 0.743), resulting in 491 follow-up queries across three query categories (Shopping: 24, Location: 19, Knowledge: 28 original queries).

Five external annotators independently evaluated each query on the following three binary (0/1) criteria, which were aggregated by majority voting:
\begin{itemize}
    \item \textbf{Semantic Relevance}~\cite{rahimi2021explaining}: Is the generated expanded query meaningfully connected to the original query, such that a user would likely make this refinement?
    \item \textbf{Plausibility}: Would searching for this expanded query help a user accomplish a realistic task or goal? 
    \item \textbf{Novelty}: Does the query reflect a non-obvious, less typical angle that adds diversity to likely reformulations?
\end{itemize}

We measured semantic relevance and plausibility as core criteria to assess the validity of the expanded queries prior to intent generation. In addition, we included novelty to test whether the query contributed to achieving diversity~\cite{Clarke2008}. 

\subsection{Result 1: The Quality of Expanded Queries Generated} 

To evaluate the overall coverage, we computed similarity scores of all 9,873 generated expanded queries against ground truth follow-up queries derived from 180 original queries. The scores ranged from 0.663 to 1.0, with the average of \textbf{$0.83$}, a standard deviation of $0.067$ (Appendix~\ref{appendix:techeval1}).  

As it is challenging to define a clear threshold of similarity with similarity scores, we present some examples of different similarity score levels:
\begin{itemize}
    \item \textbf{High similarity (0.936)}: \textit{Health insurance enrollment}\\ (\sysname{}) vs. \textit{Health insurance enrollment procedure} (Ground Truth) — a near paraphrase.  
    \item \textbf{Low similarity (0.742)}: \textit{Beams limited edition} (\sysname{}) vs. \textit{Beams Japan} (Ground Truth) — low lexical overlap but still topically relevant and potentially novel.
\end{itemize}

Human evaluation of the bottom 5\% queries (N=491) showed 84\% (N=412) were semantically relevant and 86\% (N=423) were plausible search tasks. Notably, 12\% of plausible queries (50 queries) were also rated as novel, such as luxurious chiffon dress (from chiffon dress), Sapporo ryokan seasonal discount (from Sapporo ryokan), or name ranking visualization (from name ranking). Such cases show that \sysname{}-generated follow-up queries were able to surface uncommon user intents, which are valuable to understand and improve atypical or emerging user needs.

\subsection{Evaluating the Quality of Generated Intents} 
\label{techeval-2}

To assess whether our pipeline could generate evaluable, realistic, and diverse intents, we conducted both computational and human evaluations against an LLM-based baseline. We first computationally measured the diversity of our generated intents with semantic and lexical similarity measures. Adding to this, we collected human evaluations for each metric. 

\subsubsection{Intent Sampling and Comparison Setup.}
Among 180 sampled queries (Section~\ref{techeval-1}), we randomly selected 60 queries (20 per category). Our pipeline generated up to 100 expanded queries and up to three intents per expanded query. We randomly sampled one intent per expanded query to ensure coverage across semantically different reformulations and avoid over-representation of similar intents.

To enable a fair comparison with the baseline, we used the same original queries to generate an equal number of intent statements using the baseline method. This resulted in two intent sets---one from our pipeline and one from the baseline.

\subsubsection{Baseline}
We adapted the intent generation component from MILL~\cite{jia-etal-2024-mill}, a recent approach for LLM-based query expansion serving as a strong zero-shot baseline.
To make outputs directly comparable, we modified the baseline's prompt to produce natural language intent statements in declarative form: 
\begin{quotation}
\texttt{What intents should be searched to answer the following query?: \{query\}. Please express 10 intents in Korean, in declarative sentence form. I will generate the intent and write passages to answer these generated questions.}
\end{quotation}
All intents were generated using the \texttt{o3-mini-2025-01-31} model, to make a fair comparison with our pipeline.

\subsubsection{Comparing Semantic Diversity with Computational Evaluation}
We computationally measured the intent diversity with two metrics: semantic similarity and lexical diversity. For semantic similarity, we encoded each intent with OpenAI's \texttt{text-embedding-3-small} model and computed the average pairwise cosine similarity. A lower average similarity indicates higher internal diversity. For lexical diversity, we computed n-gram diversity---the proportion of unique n-grams in the set~\cite{li-etal-2023-contrastive}. We first tokenized each intent into morphemes using the Kiwi tokenizer~\cite{Lee2024Kiwi} and removed the stopwords based on the part-of-speech tags. Then, we used the NLTK library to produce n-grams (1 $\leq$ n $\leq$ 4).
For each metric, we compared the value between the two sets and determined the winner per query.

\subsubsection{Measuring Evaluativeness, Realism, Novelty with Human Evaluation}
To assess intent quality, we conducted a human evaluation on 33 randomly sampled queries (11 per query category) out of 60 queries used for computational analysis. For each query, we randomly sampled 20 intents from each intent set per query. 
Each intent was independently evaluated by three human evaluators using three binary (0/1) criteria:

\begin{itemize}
    \item \textbf{Evaluativeness:} Is the intent specific enough that a user or evaluator could judge whether it was fulfilled by a search result?
    \item \textbf{Realism:} Does the intent resemble a natural and plausible goal that a real user might have when issuing the query?
    \item \textbf{Novelty:} If the intent is realistic, does it reflect a fresh, unexpected, but valid perspective not typically covered in common search behavior?
\end{itemize}

The evaluators saw the original query and the relevant query context used by \sysname{} (Section~\ref{sec:pipeline1-background}) along with the generated intents. We asked them to assess novelty only if they considered the intent as realistic. 
We aggregated the final labels by majority voting.
We then counted how many intents per query passed each criterion and used the Wilcoxon signed-rank test~\cite{wilcoxon1970critical} to compare our system with the baseline.

\subsection{Result 2: The Quality of Generated Intents}

\begin{table*}[ht]
\caption{Comparison of the quality of intents generated by the baseline and BloomIntent in terms of evaluability, realism, and novelty across different query categories. Values indicate the mean number of intents (out of 20) meeting each criterion, with standard deviations in parentheses. We report the p-values and Wilcoxon signed-rank test (W) results; ** indicates p < 0.01.}
\Description{
Table comparing the baseline method and BloomIntent across three evaluation metrics—evaluability, realism, and novelty—for shopping, local, and knowledge query categories. Each cell shows the mean number of intents (out of 20) meeting the criterion, with standard deviations in parentheses. For evaluability, BloomIntent outperformed baseline in shopping (19.45 vs. 14.64), local (18.91 vs. 13.45), and matched baseline in knowledge (both 20). For realism, BloomIntent scored higher than baseline across all categories, notably in shopping (19.91 vs. 9.82). For novelty, BloomIntent exceeded baseline in shopping (5.00 vs. 1.55), was similar in local (0.36 vs. 0.55), and lower in knowledge (0.36 vs. 4.09). P-values from Wilcoxon signed-rank tests indicate significant differences (p < 0.01) for most comparisons except novelty in local and all queries. Overall, BloomIntent showed higher evaluability and realism but mixed results for novelty.
}
\label{tab:techeval2-results}
\small
\begin{tabular}{@{}l|lll|lll|lll@{}}
\toprule
          & \multicolumn{3}{l|}{\textbf{Evaluability}}           & \multicolumn{3}{l|}{\textbf{Realism}}               & \multicolumn{3}{l}{\textbf{Novelty}}              \\ \midrule
Category  & Baseline     & BloomIntent  & $p$ / $W$         & Baseline     & BloomIntent  & $p$ / $W$        & Baseline    & BloomIntent & $p$ / $W$        \\ \midrule
Shopping & 14.64 (4.99) & 19.45 (1.04) & 0.0078 (**) / 1.0 & 9.82 (3.54) & 19.91 (0.30) & 0.00097 (**) / 0.0 & 1.55 (1.21) & 5.00 (2.61) & 0.0020 (**) / 1.0 \\
Local     & 13.45 (5.07) & 18.91 (1.51) & 0.00098 (**) / 0.0 & 11.91 (6.09) & 17.27 (3.17) & 0.0020 (**) / 0.0 & 0.55 (0.69) & 0.36 (0.50) & 0.6875 / 7.0 \\
Knowledge & 20.00 (0.00) & 20.00 (0.00) & 1.0 / 0.0     & 16.82 (3.63) & 19.55 (1.21) & 0.0020 (**)/ 0.0 & 4.09 (3.18) & 0.36 (0.50) & 0.0020 (**) / 0.0 \\ \midrule
All       & 16.03 (4.94) & 19.45 (1.13) & 0.00011 (**) / 2.0 & 12.85 (5.42) & 18.91 (2.27) & 1.1e-6 (**) / 0.0 & 2.06 (2.49) & 1.91 (2.73) & 0.98 / 188.0 \\ \bottomrule
\end{tabular}
\end{table*}

Overall, our evaluation results suggest that while the baseline method is successful in generating lexically diverse intents, \sysname{} produces intents that are semantically richer and more behaviorally grounded.

\sysname{} outperformed the baseline in terms of \emph{semantic diversity} ($\mu$ of average pairwise cosine similarity = 0.600 (\sysname{}) vs. 0.611 (Baseline)), showing significantly lower semantic similarity for 45 out of 60 queries. This indicates that our approach generated more semantically distinct intents encompassing diverse user tasks. In terms of lexical diversity, the baseline showed a higher n-gram diversity in 56 of 60 queries ($\mu$ of n-gram diversity = 0.59 (\sysname{}) vs. 0.70 (Baseline)). This suggests that the baseline's prompt produced more \emph{lexically diverse} intent statements.

Human evaluation focusing on different aspects of concreteness showed that \sysname{} generally outperformed the baseline in terms of evaluability and realism of the generated intents (Table~\ref{tab:techeval2-results}).

Among 20 intents per query, \sysname{} generated significantly more \emph{evaluable} intents ($M$ = $19.45$, $SD$ = 1.13), than the baseline ($M$ = $16.03$, $SD$ = $4.94$, $p$ < $0.01$, $W$ = $2.0$). 
Similarly, \sysname{} produced significantly more \emph{realistic} intents ($M$ = $18.91$, $SD$ = $2.27$), compared to the baseline ($M$ = $12.85$, $SD$ = $5.42$, $p$ < $0.01$, $W$ = $0.0$). 
The average number of novel intents between the two methods was not significantly different ($p$ > $0.05$, $W$ = $188.0$), but the significance varied between query categories. For Location queries, there were no significant differences. For Shopping queries, \sysname{} generated significantly more novel intents ($M$ = $5.00$, $SD$ = $2.61$, $p$ < $0.01$, $W$ = $1.0$) than the baseline ($M$ = $1.55$, $SD$ = $1.21$). Contrarily, for the Knowledge domain, baseline generated significantly more novel intents ($M$ = $4.09$, $SD$ = $3.18$, $p$ < $0.01$, $W$ = $0.0$) compared to \sysname{} ($M$ = $0.36$, $SD$ = $0.50$). 
The detailed human evaluation results are presented in Appendix~\ref{appendix:techeval2}.
These findings suggest that our pipeline not only produces semantically diverse intents but also ensures that those intents are realistic and evaluable---supporting high-quality intent-based evaluation. Depending on the query categories, the novelty of intents showed different trends, which should be considered when applying our framework across diverse domains.  


\subsection{Evaluating the Performance of Automated Evaluation Pipeline} 
\label{techeval-3}

To assess the effectiveness and reliability of our automated evaluation pipeline, we compared LLM-based judgments with human annotations across four evaluation dimensions, examining whether our pipeline could replicate nuanced human judgment across subjective, intent-grounded evaluation dimensions.

For each of the 180 queries (Section~\ref{sec:evaluation-dataset}), we selected up to 10 cluster centroids as representative intents (Section~\ref{sec:pipeline2-clustering}). This produced a total of 1,664 query-intent pairs (Shopping: 561, Location: 566, Knowledge: 537).

\subsubsection{Ground Truth Construction with Human Evaluation}
In collaboration with NAVER Corporation, a major commercial search engine company, we recruited a team of trained search engine evaluators who were working with the company and had substantial experience in query evaluation.
Evaluators were tasked with assessing whether the retrieved search page results (SERP) satisfy each query-intent pair across four evaluation criteria used by the pipeline (Section~\ref{sec:pipeline2-autoeval}): \textit{satisfaction}, \textit{relevance}, \textit{reliability}, and \textit{clarity}.

Annotators first determined whether each intent was \textit{valid} (1) or \textit{invalid} (0). Invalid intents---those that were incoherent, unrealistic, or incomprehensible---were excluded from the downstream evaluation. This ensured consistent application to meaningful query-intent pairs only. 

\subsubsection{LLM-Based Evaluation Procedure}
To evaluate the reliability of our automated evaluation, we computed the agreement of our automated evaluation compared to human judgments and also measured the alignment between human and LLM scores using Cohen's Kappa and Spearman's correlation.

\subsection{Result 3: The Performance of Automated Evaluation} 
\label{sec:techeval-result3}

\begin{table}[t]
\centering
\caption{Agreement between LLM-based and human evaluations across different metrics. We also report class-wise accuracy under each metric.}
\Description{
Table showing agreement between LLM-based and human evaluations across four metrics: satisfaction, relevance, reliability, and clarity. For overall accuracy, satisfaction is highest at 0.721, followed by reliability (0.612), relevance (0.572), and clarity (0.554). Cohen’s kappa values indicate moderate agreement for satisfaction (0.445) and lower for the other metrics (0.348 for relevance, 0.381 for reliability, and 0.310 for clarity). Class-wise accuracy is also reported: For Class 0, accuracy ranges from 0.613 (reliability) to 0.816 (relevance). For Class 1, satisfaction is highest at 0.812, while relevance is lowest at 0.275. Class 2 is not applicable for satisfaction but shows moderate accuracy for other metrics, with relevance at 0.627, reliability at 0.601, and clarity at 0.573.
}
\label{tab:techeval_3}
\small
\begin{tabular}{l|c|c|c|c}
\toprule
\textbf{Metric} & \textbf{Satisfaction} & \textbf{Relevance} & \textbf{Reliability} & \textbf{Clarity} \\
\midrule
Accuracy & 0.721 & 0.572 & 0.612 & 0.554 \\
$\kappa$ & 0.445 & 0.348 & 0.381 & 0.310 \\
\midrule
\multicolumn{5}{c}{\textbf{Class-wise Accuracy}} \\
\midrule
Class 0 & 0.638 & 0.816 & 0.613 & 0.775 \\
Class 1 & 0.812 & 0.275 & 0.617 & 0.291 \\
Class 2 & - & 0.627 & 0.601 & 0.573 \\
\bottomrule
\end{tabular}
\end{table}
Our results show that the automated pipeline achieves moderate alignment with human evaluation. Agreement rates between LLM and human ratings varied across dimensions: $72.1$\% for satisfaction ($\kappa$=$0.445$), $57.2$\% for relevance ($\kappa$=$0.348$), $61.2$\% for reliability ($\kappa$=$0.381$), and $55.4$\% for clarity ($\kappa$=$0.310$), as shown in Table~\ref{tab:techeval_3}. These kappa values indicate fair agreement, while accuracy levels suggest reasonable performance for large-scale query analysis. The detailed auto-evaluation statistics are reported in Appendix~\ref{appendix:techeval3}.

Further analysis revealed that the low human-LLM agreement resulted from inherent task ambiguity.
Our automated evaluation achieved high agreement with human judgments for clear-cut cases. For relevance, agreement was 81.6\% for clearly irrelevant results (Class 0) and 62.7\% for clearly relevant results (Class 2), but 27.5\% for ambiguous results (Class 1). For clarity, agreement was similarly high for Class 0 (77.5\%) and Class 2 (57.3\%), but dropped to 29.1\% for borderline Class 1 cases. 

While collecting ground truth data with human evaluation, the three expert raters also showed substantial disagreement: 44.3\% of relevance, 39.3\% of reliability, and 44.9\% of clarity judgments resulted in a 2-1 split rather than unanimous agreement (Appendix~\ref{appendix:techeval3} - Table~\ref{tab:agreement_ratio}).
Interestingly, when raters unanimously agreed, accuracy was higher (71.0\% for relevance, 67.9\% for reliability, and 65.3\% for clarity) compared to when ratings were split (40.1\% for relevance, 50.8\% for reliability, and 43.1\% for clarity) (Appendix~\ref{appendix:techeval3} - Table~\ref{tab:accuracy_analysis}).
These results suggest that while our system reliably handles clear successes or failures, it struggles with mid-spectrum judgments where human understanding is more subjective.



Although not designed to replace expert review, our automated evaluation enables scalable and diagnostic identification of low-performing or ambiguous cases, thereby helping to prioritize intents that warrant closer human examination, especially in contexts where intent-level granularity is critical.



\section{Case Study}

To investigate the real-world applicability of our framework, we conducted an observational case study with four search specialists working in the search domain and evaluations (N=4). We aimed to understand how they interpret intent-level evaluations and envision integrating our framework into their existing search evaluation workflows. To this end, we ask the following research question: 
\begin{itemize}
    \item \textbf{RQ3: How do search specialists perceive and use intent-based evaluation in the context of existing search evaluation workflows?}
\end{itemize}

In the study, we presented the same 60 query sets (20 per topic category) we used for the intent quality evaluation (Section~\ref{techeval-2}) to the participants. 
Each query had an average of 9.88 intent clusters (std = 20.33; min = 3, max = 13), comprising a total of 64.87 distinct intents on average (std = 2.26; min = 29, max = 125). 

\subsection{Participants}
We recruited four search domain specialists from a commercial search service company, each with moderate to extensive experience in designing and executing evaluation processes (minimum 1 year experience, maximum 16 years; on average 7 years). All participants were familiar with both human evaluation and behavioral metrics in large-scale search system evaluations.

\subsection{Procedure}
Each session was conducted remotely over Zoom and lasted approximately 90 minutes. The session followed a structured format consisting of five parts:

\textit{Pre-Interview (10 min).} We began with a semi-structured interview to understand the participants’ evaluation experience and context, decision-making strategies, and current challenges with search evaluation. The participants described their current practices around how they typically conduct the search quality evaluation and what kind of insights they usually derive from the results. 

\textit{System Tutorial (10 min).} We introduced \sysname{} interface and let participants freely explore its functions and features. To familiarize them with intent-level evaluation, we encouraged them to explore several queries and understand how the system works.

\textit{Task 1: Exploring Pre-Generated Queries (30 min).} From the set of 60 pre-selected queries, participants chose the queries of interest and reviewed the \sysname{}-generated intents and corresponding LLM-based evaluation scores. They were asked to think aloud as they interpreted the results, identified useful patterns, and considered how these outputs could inform actual evaluation decisions.

\textit{Task 2: Running a New Evaluation (20 min).} Participants selected a new, potentially ambiguous query of their own and used \sysname{} to generate intents and evaluation results in real-time. They then reflected on whether the result from \sysname{} helped them discover the search intents and how the results aligned with their expectations and goals.

\textit{Post-Interview (15 min).} 
We focused on gathering feedback on the strengths, limitations, and practical values of intent-based evaluation. Our interview questions included topics such as trust in system outputs, complementarity with traditional methods, and use cases where this framework would be particularly helpful.

Throughout the study, we used a think-aloud protocol and recorded the participants' screens and voices for qualitative analysis. Observations focused on how participants analyzed and used system outputs to guide hypothetical decisions, and how they compared these workflows to existing evaluation methods. We followed a standard inductive thematic analysis procedure to analyze the interview and observation transcripts.

\subsection{Findings}
Overall, participants saw a clear potential to integrate \sysname{} into existing evaluation workflows. Primarily, they viewed \sysname{} not as a replacement for human evaluation but as a complementary tool. They mentioned applications such as low-quality intent discovery, issue triage, and deep-dive analysis for each query.
We elaborate on participants' reflections and the potential application of \sysname{} for each key feature: generated intents, automated evaluation, and intent clusters. 

\subsubsection{\sysname{}-generated Intents Support Understanding of Fine-grained Search Task} 
From our pre-interviews, participants consistently reported difficulties in understanding the reasons for user dissatisfaction when using traditional methods such as behavioral metrics alone, as interaction history often lacked sufficient context for understanding users' underlying needs and motivations.
For example, P1 described how their proposed SERP updates based on click data sometimes paradoxically decreased satisfaction. 
On the other hand, with \sysname{}, participants highlighted that the generated intents helped understand users' fine-grained motivations and search tasks.

\paragraph{Fine-grained Intents Led to Actionable Insights}
Participants considered that our \emph{intent-level evaluations} offered significant advantages over traditional single-score assessments by ``pinpointing what is missing'' and ``justifying why a particular result was unsatisfactory'' --- especially for ambiguous or low-performing queries.
P2 said, ``\sysname{}'s intents reveal what type of content users are looking for a specific query, such as videos or detailed product specification. The system makes it easier to understand why satisfaction might be low in this SERP. However, human evaluators typically do not provide such specific comments.''
Similarly, P1, P3, and P4 thought that this fine-grained understanding helped them communicate problems and potential improvements in search results more effectively with other stakeholders (e.g., managers, coworkers), making intent-based results more actionable in practice.
For deeper diagnostic use, P2 and P3 proposed temporal intent tracking for low-performing queries--- e.g., examining seasonal differences for the same query like `microfiber blanket'. Participants envisioned that user needs may shift across time and such temporal patterns could help teams prioritize fixes and monitor emerging issues more proactively.

\paragraph{\sysname{}-generated Intents Beyond One's Imagination}
Participants also valued that \sysname{} enabled understanding the user's intents beyond human evaluators' imagination. 
P3 said, ``There are some common patterns in the search intents considered by human evaluators. For example, if the query is about shopping, they typically think of purchase links and reviews as the main intents. \sysname{} presents more diverse intents, which are different from current evaluations.''
P2 elaborated on the value of \sysname{}, ``It helps a lot when the query is in a domain that the evaluator may not know well. With \sysname{}-generated intents, human evaluators can potentially evaluate the unfamiliar query with a clearer understanding.'' 
This automation was seen as a major strength, allowing for scalable breadth and depth that would be challenging and costly for human evaluators to achieve.

\subsubsection{Automated Evaluation was Reasonable Enough to Guide Future Evaluation}

Participants considered the automated evaluation of \sysname{} to be cost-effective and reasonable, especially when paired with clear LLM-generated reasoning. 
While participants cautioned against trusting the automated judgments blindly, they thought the results would be useful for quickly identifying potential issues or low-performing queries for more thorough investigation. 
P4 mentioned, ``I don't trust it 100\%, but it definitely helps me spot where I should look deeper.''
P2 especially valued that \sysname{} could complement the human evaluation of trendy, time-sensitive queries by providing quick initial evaluation results.

P1 emphasized the need for evaluation explanations to be grounded in actual SERP content, which would help bridge the gap between micro-level judgments and system-level design strategies --- ``For each area in the search result, there is a responsible team for it. So, for example, if `comparing price and specification' intents are evaluated as satisfactory in general, I could assess how helpful the shopping area was and communicate with the shopping team.''

\subsubsection{Intent Clusters Enabled Discovery of Actionable Insights} 

Participants found that intent-level insights from \sysname{} helped them identify specific opportunities for improving search results. However, they noted that reviewing individual intents could be time-consuming. To reduce this burden, participants appreciated the \emph{clustering of similar intents}, which allowed them to think in terms of meaningful, actionable units.
For example, P4 tried the query ``carry-on luggage liquid'' and directly discovered an intent cluster with a low satisfaction score. They ideated ``This cluster is mainly about comparing regulations in different countries. If these intents are common, I will propose adding a link for the country-wise comparison of regulations in the search results page.''

However, participants also emphasized the challenge of prioritization. Since general web search engines serve the same SERP to all users, teams must decide which intents to address first. P4 emphasized this challenge: ``Overall balance is most important in general (web) search. We need to decide, from the most objective position possible, what information should be answered first.'' 
To better support prioritization, P1, P3, and P4 requested additional features to determine the queries or intents that require more attention (e.g., queries with high frequency), especially for large-scale evaluations. 
This led to a broader discussion about weighing different intents relative to their evaluation significance. Participants commonly suggested mapping intents to actual follow-up queries and using the occurrence frequencies as weights.

Participants raised concerns about the scalability of the system, noting that manually reviewing each intent cluster for every query would be impractical. As a solution, they requested higher-level summaries, heatmaps, and filtering mechanisms for faster decision-making. 
P3 explained, ``Beyond examining queries case-by-case, we need to identify intent patterns that current SERPs do not support well.'' 
P1 shared a potential workflow, ``If I were improving search for items from online duty-free shops, I would first run 10,000 shopping queries with \sysname{}, cluster the unsatisfied intents to find frequent but unmet needs, such as price comparison between duty-free and internet prices, and then propose feature updates.''

\section{Discussion}
In this section, we reflect on how \sysname{} contributes to a broader shift toward intent-based search evaluation. We discuss how this shift supports more user-centered evaluation, highlights the need for human-AI collaborative approaches, and informs future personalized and multi-turn search systems.

\subsection{Leveraging Fine-grained Search Intents for User-Centered Evaluation}

While human-centered AI emphasizes the evaluation of AI systems with user-defined goals~\cite{arawjo2024chainforge, kim2024evallm}, search evaluation poses unique challenges in articulating user goals from short and underspecified search queries. 
To address this gap, we simulate realistic intents underlying each query using LLMs to generate sentence-level intent statements grounded in user attributes and query contexts, enabling goal-grounded evaluation at scale.
Our approach provides practical benefits for search evaluation and improvement, from discovering actionable insights on improving the SERP to concrete discussion with different stakeholders grounded on user needs.

Our method complements existing evaluation approaches by addressing their limited analytic capability while preserving efficiency through strategic automation. While it may not replace all forms of large-scale metric-based evaluation, it provides a valuable option for surfacing actionable insights or understanding diverse user needs often obscured by aggregate signals.

We acknowledge that further considerations are needed for an effective intent-based evaluation, such as diversity of generated intents, understanding how intent types differ across domains and queries, and challenges associated with LLM usage. For instance, LLM-generated intents may reflect social biases embedded in language models, potentially amplifying or overlooking the needs of certain user groups. To mitigate these risks, future work should explore structural strategies such as post-generation auditing with automated bias detection tools~\cite{sheng-etal-2020-towards}.

Another key issue around intent quality involves the tradeoff between novelty and plausibility. While the baseline showed higher novelty scores for knowledge queries (Table~\ref{tab:techeval2-results}), many of those intents were implausible or incoherent, making them unsuitable for meaningful evaluation. In contrast, \sysname{} generates realistic and evaluable intents---even at the cost of reduced novelty.

These examples highlight that generating high-quality evaluation intents is multi-faceted---requiring attention to semantic validity, representational fairness, and practical evaluability. We expect that future work would investigate approaches to balance these dimensions.

\subsection{From LLM-based Evaluation to Human-AI Collaborative Evaluation}

\sysname{} aims to provide a scalable and cost-effective evaluation of SERPs using LLMs. However, our results (Section~\ref{sec:techeval-result3}) showed only moderate agreement with human raters, mainly due to nuanced or ambiguous cases.
This challenge reflects a common limitation in LLM-based evaluation: difficulty of distinguishing partially satisfactory from unsatisfactory results, especially near decision boundaries~\cite{faggioli2023perspectives, hashemi2024llm}. 

However, human raters also exhibited substantial disagreement, with around 30 to 45\% of judgments resulting in 2-1 splits (Section~\ref{sec:techeval-result3}, Appendix~\ref{app:human_eval_metrics}). The LLM-Human match accuracy drops considerably in these partially agreed cases, compared to those with full agreement. This suggests that the inconsistencies observed in automated evaluation stem not only from the LLM limitations, but also from the inherent ambiguity and subjectivity of the task itself~\cite{arxiv2406.00247}. 
Notably, while LLM-generated judgments are imperfect, their accompanying rationales were sometimes viewed as helpful by search specialists in understanding borderline cases. 

To incorporate these insights, one promising direction is human-in-the-loop calibration~\cite{wang2023large}, where human raters selectively intervene---especially for edge cases or high-stake decisions. This hybrid approach could improve both the accuracy and trustworthiness of the evaluations while maintaining scalability. Indeed, our case study participants envisioned such hybrid workflows, using \sysname{} to filter low-performing intents and identify critical patterns.
As such, our method offers a promising foundation for scalable and trustworthy evaluation practices grounded in both automation and expert oversight.

\subsection{Supporting Hyper-Personalized and Adaptive Search Evaluation}

Although \sysname{} does not directly target personalized search settings, its attribute-guided query expansion simulates realistic user personas, enabling evaluation from multiple user perspectives. 
To explore its potential for personalization, we discuss two feasible directions: (1) using intent diversity to diagnose personalized needs, and (2) extending evaluation capabilities to multi-turn or generative search where user intents evolve over time. 

\paragraph{Diagnosing Personalization Needs through Intent Diversity}

Case study participants noted that when a single query yields diverse intents with varying satisfaction levels, it signals the need for personalized SERPs, as a one-size-fits-all SERP may be insufficient. 
We suggest two potential applications of our method for supporting personalized search.
First, intent clusters with low satisfaction reveal unmet subgroup needs, guiding teams to identify where ranking algorithms, content modules, or UI elements might need tuning for specific subgroups. 
Second, the intent distribution serves as a proxy for user diversity. With access to user history or behavioral logs, systems could prioritize intents that align the most with individual users, enabling personalized ranking and fine-grained evaluation.
Similar to prior work on simulating user behaviors for UX evaluation~\cite{Zhang2024USimAgent, zhou2024cognitive}, \sysname{} can support auditing search systems for intent diversity. By 
integrating features such as customizable personas~\cite{proxona-chi25}, \sysname{} could enable user-in-the-loop testing and nuanced simulation for evaluating or auditing personalization~\cite{chen2025iseeingthisdemocratizing, Empathy-chi24}.

\paragraph{Extending to Multi-turn and Generative Search Evaluation}

We further envision that our approach could evolve beyond evaluation of single-turn, static queries to support session-level evaluation of search by modeling intent progression. For example, when evaluating generative, conversational search systems such as Perplexity.ai\footnote{https://www.perplexity.ai/}, it is critical to capture the evolving user intent across multi-turn interaction and model user satisfaction for each turn. By combining our intent generation pipeline with existing approaches for search user simulation~\cite{Zhang2024USimAgent, zhang2024llm, zhou2024cognitive}, we expect a more realistic search simulation that accurately captures user satisfaction considering diverse and evolving search goals. 

Our approach thus lays the foundation for more precise, adaptive, and user-aligned personalization strategies that reflect evolving search behavior.

\subsection{Limitations and Future Work}
Our work has several limitations. First, our automated evaluation pipeline showed moderate agreement with human judgments, particularly for borderline cases. This suggests that our approach is useful for identifying clear successes or failures but may require human oversight for ambiguous cases.
Second, we intentionally removed intents related to image or video content requests from our taxonomy, as our framework primarily focuses on textual information to be evaluated by LLMs. Furthermore, we parsed only SERP snippets, without the actual content of each result. Future web agents that can autonomously navigate and extract information from linked pages might enable more comprehensive evaluations reflecting how users engage with search results.
Third, our implementation used search logs and SERPs from a Korean commercial search engine on limited query categories, which may limit generalizability. Different search engines with varying result formats or other languages might yield different intent distributions and evaluation outcomes. Future work should address these limitations by exploring hybrid human-AI evaluation approaches, expanding to multimedia intents, and validating the approach across multiple search engines, categories, and languages.

\section{Conclusion}
In this paper, we introduce \sysname{}, a search evaluation method that centers on user intents as the core unit of analysis. By generating diverse, realistic intents and automatically evaluating how well search results fulfill each one, our approach enables a more user-centric assessment of search quality than traditional methods. Our technical evaluations demonstrate that \sysname{} produces high-quality, evaluable intents, and its automated assessments achieve reasonable alignment with expert judgments. Search specialists in case study found that intent-level evaluation helped them identify underserved user needs, diagnose specific issues, and prioritize improvements that traditional methods would overlook. By decomposing queries into fine-grained intents, \sysname{} offers an approach that could extend to diverse domains, including conversational and generative search. 
This work advocates for a shift in search evaluation toward a deeper understanding of how systems address the diverse goals that may underlie a single query.

\begin{acks}
This work was supported by NAVER Corporation. We thank the members of KIXLAB (KAIST Interaction Lab) and CSTL (Collaborative Social Technologies Lab) at KAIST for their encouragement, feedback, and countless cups of coffee. We are especially grateful to all our study participants for their time and valuable insights.

The first author sincerely thanks all team members for their unwavering support and dedication. The first author shares her gratitude to Yongju Lee, Chaneon Park, Jiwan Jeong, and HyungAe Park for accompanying the journey that led to the publication of this work. Last but not least, the first author wants to celebrate her best friends on signing the most romantic and meaningful lifelong contract. 

\end{acks}

\bibliographystyle{ACM-Reference-Format}
\bibliography{uist2025-searchgpt}


\begin{thebibliography}{83}


\ifx \showCODEN    \undefined \def \showCODEN     #1{\unskip}     \fi
\ifx \showISBNx    \undefined \def \showISBNx     #1{\unskip}     \fi
\ifx \showISBNxiii \undefined \def \showISBNxiii  #1{\unskip}     \fi
\ifx \showISSN     \undefined \def \showISSN      #1{\unskip}     \fi
\ifx \showLCCN     \undefined \def \showLCCN      #1{\unskip}     \fi
\ifx \shownote     \undefined \def \shownote      #1{#1}          \fi
\ifx \showarticletitle \undefined \def \showarticletitle #1{#1}   \fi
\ifx \showURL      \undefined \def \showURL       {\relax}        \fi
\providecommand\bibfield[2]{#2}
\providecommand\bibinfo[2]{#2}
\providecommand\natexlab[1]{#1}
\providecommand\showeprint[2][]{arXiv:#2}

\bibitem[Ai et~al\mbox{.}(2023)]%
        {ai2023information}
\bibfield{author}{\bibinfo{person}{Qingyao Ai}, \bibinfo{person}{Ting Bai}, \bibinfo{person}{Zhao Cao}, \bibinfo{person}{Yi Chang}, \bibinfo{person}{Jiawei Chen}, \bibinfo{person}{Zhumin Chen}, \bibinfo{person}{Zhiyong Cheng}, \bibinfo{person}{Shoubin Dong}, \bibinfo{person}{Zhicheng Dou}, \bibinfo{person}{Fuli Feng}, {et~al\mbox{.}}} \bibinfo{year}{2023}\natexlab{}.
\newblock \showarticletitle{Information retrieval meets large language models: a strategic report from chinese ir community}.
\newblock \bibinfo{journal}{\emph{AI Open}}  \bibinfo{volume}{4} (\bibinfo{year}{2023}), \bibinfo{pages}{80--90}.
\newblock


\bibitem[Alaofi et~al\mbox{.}(2023)]%
        {Alaofi2023Can}
\bibfield{author}{\bibinfo{person}{Marwah Alaofi}, \bibinfo{person}{Luke Gallagher}, \bibinfo{person}{Mark Sanderson}, \bibinfo{person}{Falk Scholer}, {and} \bibinfo{person}{Paul Thomas}.} \bibinfo{year}{2023}\natexlab{}.
\newblock \showarticletitle{Can Generative LLMs Create Query Variants for Test Collections? An Exploratory Study}. In \bibinfo{booktitle}{\emph{Proceedings of the 46th International ACM SIGIR Conference on Research and Development in Information Retrieval}} (Taipei, Taiwan) \emph{(\bibinfo{series}{SIGIR '23})}. \bibinfo{publisher}{Association for Computing Machinery}, \bibinfo{address}{New York, NY, USA}, \bibinfo{pages}{1869–1873}.
\newblock
\showISBNx{9781450394086}
\href{https://doi.org/10.1145/3539618.3591960}{doi:\nolinkurl{10.1145/3539618.3591960}}


\bibitem[Ali and Beg(2011)]%
        {ali2011overview}
\bibfield{author}{\bibinfo{person}{Rashid Ali} {and} \bibinfo{person}{MM~Sufyan Beg}.} \bibinfo{year}{2011}\natexlab{}.
\newblock \showarticletitle{An overview of Web search evaluation methods}.
\newblock \bibinfo{journal}{\emph{Computers \& Electrical Engineering}} \bibinfo{volume}{37}, \bibinfo{number}{6} (\bibinfo{year}{2011}), \bibinfo{pages}{835--848}.
\newblock


\bibitem[Anand et~al\mbox{.}(2024)]%
        {anand2024understanding}
\bibfield{author}{\bibinfo{person}{Abhijit Anand}, \bibinfo{person}{Jurek Leonhardt}, \bibinfo{person}{Venktesh V}, {and} \bibinfo{person}{Avishek Anand}.} \bibinfo{year}{2024}\natexlab{}.
\newblock \showarticletitle{Understanding the User: An Intent-Based Ranking Dataset}. In \bibinfo{booktitle}{\emph{Proceedings of the 33rd ACM International Conference on Information and Knowledge Management}} (Boise, ID, USA) \emph{(\bibinfo{series}{CIKM '24})}. \bibinfo{publisher}{Association for Computing Machinery}, \bibinfo{address}{New York, NY, USA}, \bibinfo{pages}{5323–5327}.
\newblock
\showISBNx{9798400704369}
\href{https://doi.org/10.1145/3627673.3679166}{doi:\nolinkurl{10.1145/3627673.3679166}}


\bibitem[Arawjo et~al\mbox{.}(2024)]%
        {arawjo2024chainforge}
\bibfield{author}{\bibinfo{person}{Ian Arawjo}, \bibinfo{person}{Chelse Swoopes}, \bibinfo{person}{Priyan Vaithilingam}, \bibinfo{person}{Martin Wattenberg}, {and} \bibinfo{person}{Elena~L Glassman}.} \bibinfo{year}{2024}\natexlab{}.
\newblock \showarticletitle{Chainforge: A visual toolkit for prompt engineering and llm hypothesis testing}. In \bibinfo{booktitle}{\emph{Proceedings of the 2024 CHI Conference on Human Factors in Computing Systems}}. \bibinfo{pages}{1--18}.
\newblock


\bibitem[Bai et~al\mbox{.}(2024)]%
        {bai2024intent}
\bibfield{author}{\bibinfo{person}{Yutong Bai}, \bibinfo{person}{Yujia Zhou}, \bibinfo{person}{Zhicheng Dou}, {and} \bibinfo{person}{Ji-Rong Wen}.} \bibinfo{year}{2024}\natexlab{}.
\newblock \showarticletitle{Intent-oriented dynamic interest modeling for personalized web search}.
\newblock \bibinfo{journal}{\emph{ACM Transactions on Information Systems}} \bibinfo{volume}{42}, \bibinfo{number}{4} (\bibinfo{year}{2024}), \bibinfo{pages}{1--30}.
\newblock


\bibitem[Bouchoucha et~al\mbox{.}(2013)]%
        {Bouchoucha2013Diversified}
\bibfield{author}{\bibinfo{person}{Arbi Bouchoucha}, \bibinfo{person}{Jing He}, {and} \bibinfo{person}{Jian-Yun Nie}.} \bibinfo{year}{2013}\natexlab{}.
\newblock \showarticletitle{Diversified query expansion using conceptnet}. In \bibinfo{booktitle}{\emph{Proceedings of the 22nd ACM International Conference on Information \& Knowledge Management}} (San Francisco, California, USA) \emph{(\bibinfo{series}{CIKM '13})}. \bibinfo{publisher}{Association for Computing Machinery}, \bibinfo{address}{New York, NY, USA}, \bibinfo{pages}{1861–1864}.
\newblock
\showISBNx{9781450322638}
\href{https://doi.org/10.1145/2505515.2507881}{doi:\nolinkurl{10.1145/2505515.2507881}}


\bibitem[Broder(2002)]%
        {broder2002taxonomy}
\bibfield{author}{\bibinfo{person}{Andrei Broder}.} \bibinfo{year}{2002}\natexlab{}.
\newblock \showarticletitle{A taxonomy of web search}.
\newblock \bibinfo{journal}{\emph{SIGIR Forum}} \bibinfo{volume}{36}, \bibinfo{number}{2}, \bibinfo{pages}{3–10}.
\newblock
\showISSN{0163-5840}
\href{https://doi.org/10.1145/792550.792552}{doi:\nolinkurl{10.1145/792550.792552}}


\bibitem[Cabrera et~al\mbox{.}(2023)]%
        {Cabrera2023ZenoAIA}
\bibfield{author}{\bibinfo{person}{{\'A}ngel~Alexander Cabrera}, \bibinfo{person}{Erica Fu}, \bibinfo{person}{Donald Bertucci}, \bibinfo{person}{Kenneth Holstein}, \bibinfo{person}{Ameet Talwalkar}, \bibinfo{person}{Jason~I. Hong}, {and} \bibinfo{person}{Adam Perer}.} \bibinfo{year}{2023}\natexlab{}.
\newblock \showarticletitle{Zeno: An Interactive Framework for Behavioral Evaluation of Machine Learning}.
\newblock \bibinfo{journal}{\emph{Proceedings of the 2023 CHI Conference on Human Factors in Computing Systems}} (\bibinfo{year}{2023}).
\newblock
\urldef\tempurl%
\url{http://dl.acm.org/citation.cfm?id=3581268}
\showURL{%
\tempurl}


\bibitem[Cambazoglu et~al\mbox{.}(2021)]%
        {cambazoglu2021intent}
\bibfield{author}{\bibinfo{person}{B.~Barla Cambazoglu}, \bibinfo{person}{Leila Tavakoli}, \bibinfo{person}{Falk Scholer}, \bibinfo{person}{Mark Sanderson}, {and} \bibinfo{person}{Bruce Croft}.} \bibinfo{year}{2021}\natexlab{}.
\newblock \showarticletitle{An Intent Taxonomy for Questions Asked in Web Search}. In \bibinfo{booktitle}{\emph{Proceedings of the 2021 Conference on Human Information Interaction and Retrieval}} (Canberra ACT, Australia) \emph{(\bibinfo{series}{CHIIR '21})}. \bibinfo{publisher}{Association for Computing Machinery}, \bibinfo{address}{New York, NY, USA}, \bibinfo{pages}{85–94}.
\newblock
\showISBNx{9781450380553}
\href{https://doi.org/10.1145/3406522.3446027}{doi:\nolinkurl{10.1145/3406522.3446027}}


\bibitem[Chapelle et~al\mbox{.}(2011)]%
        {Chapelle2011IntentbasedDOA}
\bibfield{author}{\bibinfo{person}{O. Chapelle}, \bibinfo{person}{Shihao Ji}, \bibinfo{person}{Ciya Liao}, \bibinfo{person}{Emre Velipasaoglu}, \bibinfo{person}{Larry Lai}, {and} \bibinfo{person}{Su-Lin Wu}.} \bibinfo{year}{2011}\natexlab{}.
\newblock \showarticletitle{Intent-based diversification of web search results: metrics and algorithms}.
\newblock \bibinfo{journal}{\emph{Information Retrieval}}  \bibinfo{volume}{14} (\bibinfo{year}{2011}), \bibinfo{pages}{572--592}.
\newblock
\urldef\tempurl%
\url{https://doi.org/10.1007/s10791-011-9167-7}
\showURL{%
\tempurl}


\bibitem[Chen et~al\mbox{.}(2025)]%
        {chen2025iseeingthisdemocratizing}
\bibfield{author}{\bibinfo{person}{Chaoran Chen}, \bibinfo{person}{Leyang Li}, \bibinfo{person}{Luke Cao}, \bibinfo{person}{Yanfang Ye}, \bibinfo{person}{Tianshi Li}, \bibinfo{person}{Yaxing Yao}, {and} \bibinfo{person}{Toby~Jia jun Li}.} \bibinfo{year}{2025}\natexlab{}.
\newblock \bibinfo{title}{Why am I seeing this: Democratizing End User Auditing for Online Content Recommendations}.
\newblock
\showeprint[arxiv]{2410.04917}~[cs.HC]
\urldef\tempurl%
\url{https://arxiv.org/abs/2410.04917}
\showURL{%
\tempurl}


\bibitem[Chen et~al\mbox{.}(2024)]%
        {Empathy-chi24}
\bibfield{author}{\bibinfo{person}{Chaoran Chen}, \bibinfo{person}{Weijun Li}, \bibinfo{person}{Wenxin Song}, \bibinfo{person}{Yanfang Ye}, \bibinfo{person}{Yaxing Yao}, {and} \bibinfo{person}{Toby Jia-Jun Li}.} \bibinfo{year}{2024}\natexlab{}.
\newblock \showarticletitle{An Empathy-Based Sandbox Approach to Bridge the Privacy Gap among Attitudes, Goals, Knowledge, and Behaviors}. In \bibinfo{booktitle}{\emph{Proceedings of the 2024 CHI Conference on Human Factors in Computing Systems}} (Honolulu, HI, USA) \emph{(\bibinfo{series}{CHI '24})}. \bibinfo{publisher}{Association for Computing Machinery}, \bibinfo{address}{New York, NY, USA}, Article \bibinfo{articleno}{234}, \bibinfo{numpages}{28}~pages.
\newblock
\showISBNx{9798400703300}
\href{https://doi.org/10.1145/3613904.3642363}{doi:\nolinkurl{10.1145/3613904.3642363}}


\bibitem[Chen et~al\mbox{.}(2021)]%
        {chen2021towards}
\bibfield{author}{\bibinfo{person}{Jia Chen}, \bibinfo{person}{Jiaxin Mao}, \bibinfo{person}{Yiqun Liu}, \bibinfo{person}{Fan Zhang}, \bibinfo{person}{Min Zhang}, {and} \bibinfo{person}{Shaoping Ma}.} \bibinfo{year}{2021}\natexlab{}.
\newblock \showarticletitle{Towards a Better Understanding of Query Reformulation Behavior in Web Search}. In \bibinfo{booktitle}{\emph{Proceedings of the Web Conference 2021}} (Ljubljana, Slovenia) \emph{(\bibinfo{series}{WWW '21})}. \bibinfo{publisher}{Association for Computing Machinery}, \bibinfo{address}{New York, NY, USA}, \bibinfo{pages}{743–755}.
\newblock
\showISBNx{9781450383127}
\href{https://doi.org/10.1145/3442381.3450127}{doi:\nolinkurl{10.1145/3442381.3450127}}


\bibitem[Chiang and Lee(2023)]%
        {chiang2023can}
\bibfield{author}{\bibinfo{person}{Cheng-Han Chiang} {and} \bibinfo{person}{Hung-yi Lee}.} \bibinfo{year}{2023}\natexlab{}.
\newblock \showarticletitle{Can large language models be an alternative to human evaluations?}
\newblock \bibinfo{journal}{\emph{arXiv preprint arXiv:2305.01937}} (\bibinfo{year}{2023}).
\newblock


\bibitem[Choi et~al\mbox{.}(2025)]%
        {proxona-chi25}
\bibfield{author}{\bibinfo{person}{Yoonseo Choi}, \bibinfo{person}{Eun~Jeong Kang}, \bibinfo{person}{Seulgi Choi}, \bibinfo{person}{Min~Kyung Lee}, {and} \bibinfo{person}{Juho Kim}.} \bibinfo{year}{2025}\natexlab{}.
\newblock \showarticletitle{Proxona: Supporting Creators' Sensemaking and Ideation with LLM-Powered Audience Personas}. In \bibinfo{booktitle}{\emph{Proceedings of the 2025 CHI Conference on Human Factors in Computing Systems}} \emph{(\bibinfo{series}{CHI '25})}. \bibinfo{publisher}{Association for Computing Machinery}, \bibinfo{address}{New York, NY, USA}, Article \bibinfo{articleno}{149}, \bibinfo{numpages}{32}~pages.
\newblock
\showISBNx{9798400713941}
\href{https://doi.org/10.1145/3706598.3714034}{doi:\nolinkurl{10.1145/3706598.3714034}}


\bibitem[Clarke et~al\mbox{.}(2008)]%
        {Clarke2008}
\bibfield{author}{\bibinfo{person}{Charles~L.A. Clarke}, \bibinfo{person}{Maheedhar Kolla}, \bibinfo{person}{Gordon~V. Cormack}, \bibinfo{person}{Olga Vechtomova}, \bibinfo{person}{Azin Ashkan}, \bibinfo{person}{Stefan B\"{u}ttcher}, {and} \bibinfo{person}{Ian MacKinnon}.} \bibinfo{year}{2008}\natexlab{}.
\newblock \showarticletitle{Novelty and diversity in information retrieval evaluation}. In \bibinfo{booktitle}{\emph{Proceedings of the 31st Annual International ACM SIGIR Conference on Research and Development in Information Retrieval}} (Singapore, Singapore) \emph{(\bibinfo{series}{SIGIR '08})}. \bibinfo{publisher}{Association for Computing Machinery}, \bibinfo{address}{New York, NY, USA}, \bibinfo{pages}{659–666}.
\newblock
\showISBNx{9781605581644}
\href{https://doi.org/10.1145/1390334.1390446}{doi:\nolinkurl{10.1145/1390334.1390446}}


\bibitem[Dou et~al\mbox{.}(2007)]%
        {dou2007large}
\bibfield{author}{\bibinfo{person}{Zhicheng Dou}, \bibinfo{person}{Ruihua Song}, {and} \bibinfo{person}{Ji-Rong Wen}.} \bibinfo{year}{2007}\natexlab{}.
\newblock \showarticletitle{A large-scale evaluation and analysis of personalized search strategies}. In \bibinfo{booktitle}{\emph{Proceedings of the 16th international conference on World Wide Web}}. \bibinfo{pages}{581--590}.
\newblock


\bibitem[Duan et~al\mbox{.}(2012)]%
        {duan2012click}
\bibfield{author}{\bibinfo{person}{Huizhong Duan}, \bibinfo{person}{Emre Kiciman}, {and} \bibinfo{person}{ChengXiang Zhai}.} \bibinfo{year}{2012}\natexlab{}.
\newblock \showarticletitle{Click patterns: An empirical representation of complex query intents}. In \bibinfo{booktitle}{\emph{Proceedings of the 21st ACM international conference on Information and knowledge management}}. \bibinfo{pages}{1035--1044}.
\newblock


\bibitem[Faggioli et~al\mbox{.}(2023)]%
        {faggioli2023perspectives}
\bibfield{author}{\bibinfo{person}{Guglielmo Faggioli}, \bibinfo{person}{Laura Dietz}, \bibinfo{person}{Charles~LA Clarke}, \bibinfo{person}{Gianluca Demartini}, \bibinfo{person}{Matthias Hagen}, \bibinfo{person}{Claudia Hauff}, \bibinfo{person}{Noriko Kando}, \bibinfo{person}{Evangelos Kanoulas}, \bibinfo{person}{Martin Potthast}, \bibinfo{person}{Benno Stein}, {et~al\mbox{.}}} \bibinfo{year}{2023}\natexlab{}.
\newblock \showarticletitle{Perspectives on large language models for relevance judgment}. In \bibinfo{booktitle}{\emph{Proceedings of the 2023 ACM SIGIR International Conference on Theory of Information Retrieval}}. \bibinfo{pages}{39--50}.
\newblock


\bibitem[Fox et~al\mbox{.}(2005)]%
        {fox2005Evaluating}
\bibfield{author}{\bibinfo{person}{Steve Fox}, \bibinfo{person}{Kuldeep Karnawat}, \bibinfo{person}{Mark Mydland}, \bibinfo{person}{Susan Dumais}, {and} \bibinfo{person}{Thomas White}.} \bibinfo{year}{2005}\natexlab{}.
\newblock \showarticletitle{Evaluating implicit measures to improve web search}.
\newblock \bibinfo{journal}{\emph{ACM Transactions on Information Systems}} \bibinfo{volume}{23}, \bibinfo{number}{2} (\bibinfo{date}{April} \bibinfo{year}{2005}), \bibinfo{pages}{147--168}.
\newblock
\showISSN{1046-8188}
\href{https://doi.org/10.1145/1059981.1059982}{doi:\nolinkurl{10.1145/1059981.1059982}}


\bibitem[Fu et~al\mbox{.}(2023)]%
        {fu2023gptscore}
\bibfield{author}{\bibinfo{person}{Jinlan Fu}, \bibinfo{person}{See-Kiong Ng}, \bibinfo{person}{Zhengbao Jiang}, {and} \bibinfo{person}{Pengfei Liu}.} \bibinfo{year}{2023}\natexlab{}.
\newblock \showarticletitle{Gptscore: Evaluate as you desire}.
\newblock \bibinfo{journal}{\emph{arXiv preprint arXiv:2302.04166}} (\bibinfo{year}{2023}).
\newblock


\bibitem[Gebreegziabher et~al\mbox{.}(2025)]%
        {gebreegziabher2025metricmate}
\bibfield{author}{\bibinfo{person}{Simret~Araya Gebreegziabher}, \bibinfo{person}{Charles Chiang}, \bibinfo{person}{Zichu Wang}, \bibinfo{person}{Zahra Ashktorab}, \bibinfo{person}{Michelle Brachman}, \bibinfo{person}{Werner Geyer}, \bibinfo{person}{Toby Jia-Jun Li}, {and} \bibinfo{person}{Diego G\'{o}mez-Zar\'{a}}.} \bibinfo{year}{2025}\natexlab{}.
\newblock \showarticletitle{MetricMate: An Interactive Tool for Generating Evaluation Criteria for LLM-as-a-Judge Workflow}.
\newblock , Article \bibinfo{articleno}{22} (\bibinfo{year}{2025}), \bibinfo{numpages}{18}~pages.
\newblock
\showISBNx{9798400713842}
\href{https://doi.org/10.1145/3729176.3729199}{doi:\nolinkurl{10.1145/3729176.3729199}}


\bibitem[Google(2025)]%
        {Google_2025}
\bibfield{author}{\bibinfo{person}{Google}.} \bibinfo{year}{2025}\natexlab{}.
\newblock \bibinfo{title}{Search Quality Rater Program Guidelines}.
\newblock
\urldef\tempurl%
\url{https://static.googleusercontent.com/media/guidelines.raterhub.com/en//searchqualityevaluatorguidelines.pdf}
\showURL{%
\tempurl}
\newblock
\shownote{Accessed: 2025-04-09}.


\bibitem[Hashemi et~al\mbox{.}(2024)]%
        {hashemi2024llm}
\bibfield{author}{\bibinfo{person}{Helia Hashemi}, \bibinfo{person}{Jason Eisner}, \bibinfo{person}{Corby Rosset}, \bibinfo{person}{Benjamin Van~Durme}, {and} \bibinfo{person}{Chris Kedzie}.} \bibinfo{year}{2024}\natexlab{}.
\newblock \showarticletitle{LLM-rubric: A multidimensional, calibrated approach to automated evaluation of natural language texts}.
\newblock \bibinfo{journal}{\emph{arXiv preprint arXiv:2501.00274}} (\bibinfo{year}{2024}).
\newblock


\bibitem[Hashemi et~al\mbox{.}(2021)]%
        {hashemi2021learning}
\bibfield{author}{\bibinfo{person}{Helia Hashemi}, \bibinfo{person}{Hamed Zamani}, {and} \bibinfo{person}{W~Bruce Croft}.} \bibinfo{year}{2021}\natexlab{}.
\newblock \showarticletitle{Learning multiple intent representations for search queries}. In \bibinfo{booktitle}{\emph{Proceedings of the 30th ACM International Conference on Information \& Knowledge Management}}. \bibinfo{pages}{669--679}.
\newblock


\bibitem[Hassan et~al\mbox{.}(2013)]%
        {hassan2013beyond}
\bibfield{author}{\bibinfo{person}{Ahmed Hassan}, \bibinfo{person}{Xiaolin Shi}, \bibinfo{person}{Nick Craswell}, {and} \bibinfo{person}{Bill Ramsey}.} \bibinfo{year}{2013}\natexlab{}.
\newblock \showarticletitle{Beyond clicks: query reformulation as a predictor of search satisfaction}. In \bibinfo{booktitle}{\emph{Proceedings of the 22nd ACM International Conference on Information \& Knowledge Management}} (San Francisco, California, USA) \emph{(\bibinfo{series}{CIKM '13})}. \bibinfo{publisher}{Association for Computing Machinery}, \bibinfo{address}{New York, NY, USA}, \bibinfo{pages}{2019–2028}.
\newblock
\showISBNx{9781450322638}
\href{https://doi.org/10.1145/2505515.2505682}{doi:\nolinkurl{10.1145/2505515.2505682}}


\bibitem[Hofmann et~al\mbox{.}(2016)]%
        {hofmann2016Online}
\bibfield{author}{\bibinfo{person}{Katja Hofmann}, \bibinfo{person}{Lihong Li}, {and} \bibinfo{person}{Filip Radlinski}.} \bibinfo{year}{2016}\natexlab{}.
\newblock \showarticletitle{Online {Evaluation} for {Information} {Retrieval}}.
\newblock \bibinfo{journal}{\emph{Foundations and Trends® in Information Retrieval}} \bibinfo{volume}{10}, \bibinfo{number}{1} (\bibinfo{year}{2016}), \bibinfo{pages}{1--117}.
\newblock
\showISSN{1554-0669, 1554-0677}
\href{https://doi.org/10.1561/1500000051}{doi:\nolinkurl{10.1561/1500000051}}


\bibitem[Huffman and Hochster(2007)]%
        {huffman2007well}
\bibfield{author}{\bibinfo{person}{Scott~B. Huffman} {and} \bibinfo{person}{Michael Hochster}.} \bibinfo{year}{2007}\natexlab{}.
\newblock \showarticletitle{How well does result relevance predict session satisfaction?}. In \bibinfo{booktitle}{\emph{Proceedings of the 30th Annual International ACM SIGIR Conference on Research and Development in Information Retrieval}} (Amsterdam, The Netherlands) \emph{(\bibinfo{series}{SIGIR '07})}. \bibinfo{publisher}{Association for Computing Machinery}, \bibinfo{address}{New York, NY, USA}, \bibinfo{pages}{567–574}.
\newblock
\showISBNx{9781595935977}
\href{https://doi.org/10.1145/1277741.1277839}{doi:\nolinkurl{10.1145/1277741.1277839}}


\bibitem[Jangwon~Park(2021)]%
        {park2021kobigbird}
\bibfield{author}{\bibinfo{person}{Donggyu~Kim Jangwon~Park}.} \bibinfo{year}{2021}\natexlab{}.
\newblock \bibinfo{title}{KoBigBird: Pretrained BigBird Model for Korean}.
\newblock \bibinfo{howpublished}{\url{https://github.com/monologg/KoBigBird}}.
\newblock


\bibitem[Jia et~al\mbox{.}(2024)]%
        {jia-etal-2024-mill}
\bibfield{author}{\bibinfo{person}{Pengyue Jia}, \bibinfo{person}{Yiding Liu}, \bibinfo{person}{Xiangyu Zhao}, \bibinfo{person}{Xiaopeng Li}, \bibinfo{person}{Changying Hao}, \bibinfo{person}{Shuaiqiang Wang}, {and} \bibinfo{person}{Dawei Yin}.} \bibinfo{year}{2024}\natexlab{}.
\newblock \showarticletitle{{MILL}: Mutual Verification with Large Language Models for Zero-Shot Query Expansion}. In \bibinfo{booktitle}{\emph{Proceedings of the 2024 Conference of the North American Chapter of the Association for Computational Linguistics: Human Language Technologies (Volume 1: Long Papers)}}. \bibinfo{publisher}{Association for Computational Linguistics}, \bibinfo{address}{Mexico City, Mexico}, \bibinfo{pages}{2498--2518}.
\newblock
\href{https://doi.org/10.18653/v1/2024.naacl-long.138}{doi:\nolinkurl{10.18653/v1/2024.naacl-long.138}}


\bibitem[Jiang et~al\mbox{.}(2017)]%
        {Jiang2017Generating}
\bibfield{author}{\bibinfo{person}{Zhengbao Jiang}, \bibinfo{person}{Zhicheng Dou}, {and} \bibinfo{person}{Ji-Rong Wen}.} \bibinfo{year}{2017}\natexlab{}.
\newblock \showarticletitle{Generating Query Facets Using Knowledge Bases}.
\newblock \bibinfo{journal}{\emph{IEEE Trans. on Knowl. and Data Eng.}} \bibinfo{volume}{29}, \bibinfo{number}{2} (\bibinfo{date}{Feb.} \bibinfo{year}{2017}), \bibinfo{pages}{315–329}.
\newblock
\showISSN{1041-4347}
\href{https://doi.org/10.1109/TKDE.2016.2623782}{doi:\nolinkurl{10.1109/TKDE.2016.2623782}}


\bibitem[Joachims(2002)]%
        {joachims2002optimizing}
\bibfield{author}{\bibinfo{person}{Thorsten Joachims}.} \bibinfo{year}{2002}\natexlab{}.
\newblock \showarticletitle{Optimizing search engines using clickthrough data}. In \bibinfo{booktitle}{\emph{Proceedings of the eighth ACM SIGKDD international conference on Knowledge discovery and data mining}}. \bibinfo{pages}{133--142}.
\newblock


\bibitem[Joshi et~al\mbox{.}(2025)]%
        {joshi2025coprompter}
\bibfield{author}{\bibinfo{person}{Ishika Joshi}, \bibinfo{person}{Simra Shahid}, \bibinfo{person}{Shreeya~Manasvi Venneti}, \bibinfo{person}{Manushree Vasu}, \bibinfo{person}{Yantao Zheng}, \bibinfo{person}{Yunyao Li}, \bibinfo{person}{Balaji Krishnamurthy}, {and} \bibinfo{person}{Gromit Yeuk-Yin Chan}.} \bibinfo{year}{2025}\natexlab{}.
\newblock \showarticletitle{CoPrompter: User-Centric Evaluation of LLM Instruction Alignment for Improved Prompt Engineering}. In \bibinfo{booktitle}{\emph{Proceedings of the 30th International Conference on Intelligent User Interfaces}}. \bibinfo{pages}{341--365}.
\newblock


\bibitem[Kim et~al\mbox{.}(2024)]%
        {kim2024evallm}
\bibfield{author}{\bibinfo{person}{Tae~Soo Kim}, \bibinfo{person}{Yoonjoo Lee}, \bibinfo{person}{Jamin Shin}, \bibinfo{person}{Young-Ho Kim}, {and} \bibinfo{person}{Juho Kim}.} \bibinfo{year}{2024}\natexlab{}.
\newblock \showarticletitle{Evallm: Interactive evaluation of large language model prompts on user-defined criteria}. In \bibinfo{booktitle}{\emph{Proceedings of the 2024 CHI Conference on Human Factors in Computing Systems}}. \bibinfo{pages}{1--21}.
\newblock


\bibitem[Kim et~al\mbox{.}(2014)]%
        {kim2014modeling}
\bibfield{author}{\bibinfo{person}{Youngho Kim}, \bibinfo{person}{Ahmed Hassan}, \bibinfo{person}{Ryen~W White}, {and} \bibinfo{person}{Imed Zitouni}.} \bibinfo{year}{2014}\natexlab{}.
\newblock \showarticletitle{Modeling dwell time to predict click-level satisfaction}. In \bibinfo{booktitle}{\emph{Proceedings of the 7th ACM international conference on Web search and data mining}}. \bibinfo{pages}{193--202}.
\newblock


\bibitem[Lee(2024)]%
        {Lee2024Kiwi}
\bibfield{author}{\bibinfo{person}{Min-chul Lee}.} \bibinfo{year}{2024}\natexlab{}.
\newblock \showarticletitle{Kiwi: Developing a Korean Morphological Analyzer Based on Statistical Language Models and Skip-Bigram}.
\newblock \bibinfo{journal}{\emph{Korean Journal of Digital Humanities}} \bibinfo{volume}{1}, \bibinfo{number}{1} (\bibinfo{year}{2024}).
\newblock
\showISSN{3058-311X}
\href{https://doi.org/10.23287/KJDH.2024.1.1.6}{doi:\nolinkurl{10.23287/KJDH.2024.1.1.6}}


\bibitem[Li et~al\mbox{.}(2024b)]%
        {li2024llms}
\bibfield{author}{\bibinfo{person}{Haitao Li}, \bibinfo{person}{Qian Dong}, \bibinfo{person}{Junjie Chen}, \bibinfo{person}{Huixue Su}, \bibinfo{person}{Yujia Zhou}, \bibinfo{person}{Qingyao Ai}, \bibinfo{person}{Ziyi Ye}, {and} \bibinfo{person}{Yiqun Liu}.} \bibinfo{year}{2024}\natexlab{b}.
\newblock \showarticletitle{Llms-as-judges: a comprehensive survey on llm-based evaluation methods}.
\newblock \bibinfo{journal}{\emph{arXiv preprint arXiv:2412.05579}} (\bibinfo{year}{2024}).
\newblock


\bibitem[Li et~al\mbox{.}(2024a)]%
        {li2024multi}
\bibfield{author}{\bibinfo{person}{Mingzhe Li}, \bibinfo{person}{Xiuying Chen}, \bibinfo{person}{Jing Xiang}, \bibinfo{person}{Qishen Zhang}, \bibinfo{person}{Changsheng Ma}, \bibinfo{person}{Chenchen Dai}, \bibinfo{person}{Jinxiong Chang}, \bibinfo{person}{Zhongyi Liu}, {and} \bibinfo{person}{Guannan Zhang}.} \bibinfo{year}{2024}\natexlab{a}.
\newblock \showarticletitle{Multi-Intent Attribute-Aware Text Matching in Searching}. In \bibinfo{booktitle}{\emph{Proceedings of the 17th ACM International Conference on Web Search and Data Mining}}. \bibinfo{pages}{360--368}.
\newblock


\bibitem[Li et~al\mbox{.}(2023b)]%
        {li2023agent4rankingsemanticrobustranking}
\bibfield{author}{\bibinfo{person}{Xiaopeng Li}, \bibinfo{person}{Lixin Su}, \bibinfo{person}{Pengyue Jia}, \bibinfo{person}{Xiangyu Zhao}, \bibinfo{person}{Suqi Cheng}, \bibinfo{person}{Junfeng Wang}, {and} \bibinfo{person}{Dawei Yin}.} \bibinfo{year}{2023}\natexlab{b}.
\newblock \bibinfo{title}{Agent4Ranking: Semantic Robust Ranking via Personalized Query Rewriting Using Multi-agent LLM}.
\newblock
\urldef\tempurl%
\url{https://arxiv.org/abs/2312.15450}
\showURL{%
\tempurl}


\bibitem[Li et~al\mbox{.}(2008)]%
        {Li2008learning}
\bibfield{author}{\bibinfo{person}{Xiao Li}, \bibinfo{person}{Ye-Yi Wang}, {and} \bibinfo{person}{Alex Acero}.} \bibinfo{year}{2008}\natexlab{}.
\newblock \showarticletitle{Learning query intent from regularized click graphs}. In \bibinfo{booktitle}{\emph{Proceedings of the 31st Annual International ACM SIGIR Conference on Research and Development in Information Retrieval}} (Singapore, Singapore) \emph{(\bibinfo{series}{SIGIR '08})}. \bibinfo{publisher}{Association for Computing Machinery}, \bibinfo{address}{New York, NY, USA}, \bibinfo{pages}{339–346}.
\newblock
\showISBNx{9781605581644}
\href{https://doi.org/10.1145/1390334.1390393}{doi:\nolinkurl{10.1145/1390334.1390393}}


\bibitem[Li et~al\mbox{.}(2010)]%
        {li2010learning}
\bibfield{author}{\bibinfo{person}{Xiao Li}, \bibinfo{person}{Ye-Yi Wang}, \bibinfo{person}{Dou Shen}, {and} \bibinfo{person}{Alex Acero}.} \bibinfo{year}{2010}\natexlab{}.
\newblock \showarticletitle{Learning with click graph for query intent classification}.
\newblock \bibinfo{journal}{\emph{ACM Transactions on Information Systems (TOIS)}} \bibinfo{volume}{28}, \bibinfo{number}{3} (\bibinfo{year}{2010}), \bibinfo{pages}{1--20}.
\newblock


\bibitem[Li et~al\mbox{.}(2023a)]%
        {li-etal-2023-contrastive}
\bibfield{author}{\bibinfo{person}{Xiang~Lisa Li}, \bibinfo{person}{Ari Holtzman}, \bibinfo{person}{Daniel Fried}, \bibinfo{person}{Percy Liang}, \bibinfo{person}{Jason Eisner}, \bibinfo{person}{Tatsunori Hashimoto}, \bibinfo{person}{Luke Zettlemoyer}, {and} \bibinfo{person}{Mike Lewis}.} \bibinfo{year}{2023}\natexlab{a}.
\newblock \showarticletitle{Contrastive Decoding: Open-ended Text Generation as Optimization}. In \bibinfo{booktitle}{\emph{Proceedings of the 61st Annual Meeting of the Association for Computational Linguistics (Volume 1: Long Papers)}}, \bibfield{editor}{\bibinfo{person}{Anna Rogers}, \bibinfo{person}{Jordan Boyd-Graber}, {and} \bibinfo{person}{Naoaki Okazaki}} (Eds.). \bibinfo{publisher}{Association for Computational Linguistics}, \bibinfo{address}{Toronto, Canada}, \bibinfo{pages}{12286--12312}.
\newblock
\href{https://doi.org/10.18653/v1/2023.acl-long.687}{doi:\nolinkurl{10.18653/v1/2023.acl-long.687}}


\bibitem[Liu et~al\mbox{.}(2018)]%
        {liu2018satisfaction}
\bibfield{author}{\bibinfo{person}{Mengyang Liu}, \bibinfo{person}{Yiqun Liu}, \bibinfo{person}{Jiaxin Mao}, \bibinfo{person}{Cheng Luo}, \bibinfo{person}{Min Zhang}, {and} \bibinfo{person}{Shaoping Ma}.} \bibinfo{year}{2018}\natexlab{}.
\newblock \showarticletitle{"Satisfaction with Failure" or "Unsatisfied Success": Investigating the Relationship between Search Success and User Satisfaction}. In \bibinfo{booktitle}{\emph{Proceedings of the 2018 World Wide Web Conference}} (Lyon, France) \emph{(\bibinfo{series}{WWW '18})}. \bibinfo{publisher}{International World Wide Web Conferences Steering Committee}, \bibinfo{address}{Republic and Canton of Geneva, CHE}, \bibinfo{pages}{1533–1542}.
\newblock
\showISBNx{9781450356398}
\href{https://doi.org/10.1145/3178876.3186065}{doi:\nolinkurl{10.1145/3178876.3186065}}


\bibitem[Liu et~al\mbox{.}(2023)]%
        {liu2023g}
\bibfield{author}{\bibinfo{person}{Yang Liu}, \bibinfo{person}{Dan Iter}, \bibinfo{person}{Yichong Xu}, \bibinfo{person}{Shuohang Wang}, \bibinfo{person}{Ruochen Xu}, {and} \bibinfo{person}{Chenguang Zhu}.} \bibinfo{year}{2023}\natexlab{}.
\newblock \showarticletitle{G-eval: NLG evaluation using gpt-4 with better human alignment}.
\newblock \bibinfo{journal}{\emph{arXiv preprint arXiv:2303.16634}} (\bibinfo{year}{2023}).
\newblock


\bibitem[MacAvaney et~al\mbox{.}(2021)]%
        {macavaney2021intent5}
\bibfield{author}{\bibinfo{person}{Sean MacAvaney}, \bibinfo{person}{Craig Macdonald}, \bibinfo{person}{Roderick Murray-Smith}, {and} \bibinfo{person}{Iadh Ounis}.} \bibinfo{year}{2021}\natexlab{}.
\newblock \showarticletitle{Intent5: Search result diversification using causal language models}.
\newblock \bibinfo{journal}{\emph{arXiv preprint arXiv:2108.04026}} (\bibinfo{year}{2021}).
\newblock


\bibitem[Malaviya et~al\mbox{.}(2024)]%
        {malaviya2024contextualizedevaluationstakingguesswork}
\bibfield{author}{\bibinfo{person}{Chaitanya Malaviya}, \bibinfo{person}{Joseph~Chee Chang}, \bibinfo{person}{Dan Roth}, \bibinfo{person}{Mohit Iyyer}, \bibinfo{person}{Mark Yatskar}, {and} \bibinfo{person}{Kyle Lo}.} \bibinfo{year}{2024}\natexlab{}.
\newblock \bibinfo{title}{Contextualized Evaluations: Taking the Guesswork Out of Language Model Evaluations}.
\newblock
\urldef\tempurl%
\url{https://arxiv.org/abs/2411.07237}
\showURL{%
\tempurl}


\bibitem[Mao et~al\mbox{.}(2016)]%
        {mao2016does}
\bibfield{author}{\bibinfo{person}{Jiaxin Mao}, \bibinfo{person}{Yiqun Liu}, \bibinfo{person}{Ke Zhou}, \bibinfo{person}{Jian-Yun Nie}, \bibinfo{person}{Jingtao Song}, \bibinfo{person}{Min Zhang}, \bibinfo{person}{Shaoping Ma}, \bibinfo{person}{Jiashen Sun}, {and} \bibinfo{person}{Hengliang Luo}.} \bibinfo{year}{2016}\natexlab{}.
\newblock \showarticletitle{When does relevance mean usefulness and user satisfaction in web search?}. In \bibinfo{booktitle}{\emph{Proceedings of the 39th International ACM SIGIR conference on Research and Development in Information Retrieval}}. \bibinfo{pages}{463--472}.
\newblock


\bibitem[Mehrdad et~al\mbox{.}(2024)]%
        {arxiv2406.00247}
\bibfield{author}{\bibinfo{person}{Navid Mehrdad}, \bibinfo{person}{Hrushikesh Mohapatra}, \bibinfo{person}{Mossaab Bagdouri}, \bibinfo{person}{Prijith Chandran}, \bibinfo{person}{Alessandro Magnani}, \bibinfo{person}{Xunfan Cai}, \bibinfo{person}{Ajit Puthenputhussery}, \bibinfo{person}{Sachin Yadav}, \bibinfo{person}{Tony Lee}, \bibinfo{person}{ChengXiang Zhai}, {and} \bibinfo{person}{Ciya Liao}.} \bibinfo{year}{2024}\natexlab{}.
\newblock \bibinfo{title}{Large Language Models for Relevance Judgment in Product Search}.
\newblock
\showeprint[arxiv]{2406.00247}~[cs.IR]
\urldef\tempurl%
\url{https://arxiv.org/abs/2406.00247}
\showURL{%
\tempurl}


\bibitem[Mitsui et~al\mbox{.}(2016)]%
        {mitsui16extracting}
\bibfield{author}{\bibinfo{person}{Matthew Mitsui}, \bibinfo{person}{Chirag Shah}, {and} \bibinfo{person}{Nicholas~J. Belkin}.} \bibinfo{year}{2016}\natexlab{}.
\newblock \showarticletitle{Extracting Information Seeking Intentions for Web Search Sessions}. In \bibinfo{booktitle}{\emph{Proceedings of the 39th International ACM SIGIR Conference on Research and Development in Information Retrieval}} (Pisa, Italy) \emph{(\bibinfo{series}{SIGIR '16})}. \bibinfo{publisher}{Association for Computing Machinery}, \bibinfo{address}{New York, NY, USA}, \bibinfo{pages}{841–844}.
\newblock
\showISBNx{9781450340694}
\href{https://doi.org/10.1145/2911451.2914746}{doi:\nolinkurl{10.1145/2911451.2914746}}


\bibitem[Niu et~al\mbox{.}(2024)]%
        {niu2024judgerank}
\bibfield{author}{\bibinfo{person}{Tong Niu}, \bibinfo{person}{Shafiq Joty}, \bibinfo{person}{Ye Liu}, \bibinfo{person}{Caiming Xiong}, \bibinfo{person}{Yingbo Zhou}, {and} \bibinfo{person}{Semih Yavuz}.} \bibinfo{year}{2024}\natexlab{}.
\newblock \showarticletitle{JudgeRank: Leveraging Large Language Models for Reasoning-Intensive Reranking}.
\newblock \bibinfo{journal}{\emph{arXiv preprint arXiv:2411.00142}} (\bibinfo{year}{2024}).
\newblock


\bibitem[Patel et~al\mbox{.}(2024)]%
        {patel2024aime}
\bibfield{author}{\bibinfo{person}{Bhrij Patel}, \bibinfo{person}{Souradip Chakraborty}, \bibinfo{person}{Wesley~A Suttle}, \bibinfo{person}{Mengdi Wang}, \bibinfo{person}{Amrit~Singh Bedi}, {and} \bibinfo{person}{Dinesh Manocha}.} \bibinfo{year}{2024}\natexlab{}.
\newblock \showarticletitle{AIME: AI System Optimization via Multiple LLM Evaluators}.
\newblock \bibinfo{journal}{\emph{arXiv preprint arXiv:2410.03131}} (\bibinfo{year}{2024}).
\newblock


\bibitem[Potey et~al\mbox{.}(2013)]%
        {6514421}
\bibfield{author}{\bibinfo{person}{Madhuri~A. Potey}, \bibinfo{person}{Dhanashri~A. Patel}, {and} \bibinfo{person}{P.~K. Sinha}.} \bibinfo{year}{2013}\natexlab{}.
\newblock \showarticletitle{A survey of query log processing techniques and evaluation of web query intent identification}. In \bibinfo{booktitle}{\emph{2013 3rd IEEE International Advance Computing Conference (IACC)}}. \bibinfo{pages}{1330--1335}.
\newblock
\href{https://doi.org/10.1109/IAdCC.2013.6514421}{doi:\nolinkurl{10.1109/IAdCC.2013.6514421}}


\bibitem[Radlinski et~al\mbox{.}(2010)]%
        {Radlinski2010Inferring}
\bibfield{author}{\bibinfo{person}{Filip Radlinski}, \bibinfo{person}{Martin Szummer}, {and} \bibinfo{person}{Nick Craswell}.} \bibinfo{year}{2010}\natexlab{}.
\newblock \showarticletitle{Inferring query intent from reformulations and clicks}. In \bibinfo{booktitle}{\emph{Proceedings of the 19th International Conference on World Wide Web}} (Raleigh, North Carolina, USA) \emph{(\bibinfo{series}{WWW '10})}. \bibinfo{publisher}{Association for Computing Machinery}, \bibinfo{address}{New York, NY, USA}, \bibinfo{pages}{1171–1172}.
\newblock
\showISBNx{9781605587998}
\href{https://doi.org/10.1145/1772690.1772859}{doi:\nolinkurl{10.1145/1772690.1772859}}


\bibitem[Rahimi et~al\mbox{.}(2021)]%
        {rahimi2021explaining}
\bibfield{author}{\bibinfo{person}{Razieh Rahimi}, \bibinfo{person}{Youngwoo Kim}, \bibinfo{person}{Hamed Zamani}, {and} \bibinfo{person}{James Allan}.} \bibinfo{year}{2021}\natexlab{}.
\newblock \showarticletitle{Explaining documents' relevance to search queries}.
\newblock \bibinfo{journal}{\emph{arXiv preprint arXiv:2111.01314}} (\bibinfo{year}{2021}).
\newblock


\bibitem[Rahmani et~al\mbox{.}(2024)]%
        {rahmani2024llmjudge}
\bibfield{author}{\bibinfo{person}{Hossein~A Rahmani}, \bibinfo{person}{Emine Yilmaz}, \bibinfo{person}{Nick Craswell}, \bibinfo{person}{Bhaskar Mitra}, \bibinfo{person}{Paul Thomas}, \bibinfo{person}{Charles~LA Clarke}, \bibinfo{person}{Mohammad Aliannejadi}, \bibinfo{person}{Clemencia Siro}, {and} \bibinfo{person}{Guglielmo Faggioli}.} \bibinfo{year}{2024}\natexlab{}.
\newblock \showarticletitle{Llmjudge: Llms for relevance judgments}.
\newblock \bibinfo{journal}{\emph{arXiv preprint arXiv:2408.08896}} (\bibinfo{year}{2024}).
\newblock


\bibitem[Rha et~al\mbox{.}(2016)]%
        {rha2016reformulations}
\bibfield{author}{\bibinfo{person}{Eun~Youp Rha}, \bibinfo{person}{Matthew Mitsui}, \bibinfo{person}{Nicholas~J. Belkin}, {and} \bibinfo{person}{Chirag Shah}.} \bibinfo{year}{2016}\natexlab{}.
\newblock \showarticletitle{Exploring the relationships between search intentions and query reformulations}.
\newblock \bibinfo{journal}{\emph{Proceedings of the Association for Information Science and Technology}} \bibinfo{volume}{53}, \bibinfo{number}{1} (\bibinfo{year}{2016}), \bibinfo{pages}{1--9}.
\newblock
\href{https://doi.org/10.1002/pra2.2016.14505301048}{doi:\nolinkurl{10.1002/pra2.2016.14505301048}}


\bibitem[Robertson et~al\mbox{.}(2023)]%
        {robertson2023angler}
\bibfield{author}{\bibinfo{person}{Samantha Robertson}, \bibinfo{person}{Zijie~J Wang}, \bibinfo{person}{Dominik Moritz}, \bibinfo{person}{Mary~Beth Kery}, {and} \bibinfo{person}{Fred Hohman}.} \bibinfo{year}{2023}\natexlab{}.
\newblock \showarticletitle{Angler: Helping machine translation practitioners prioritize model improvements}. In \bibinfo{booktitle}{\emph{Proceedings of the 2023 CHI Conference on Human Factors in Computing Systems}}. \bibinfo{pages}{1--20}.
\newblock


\bibitem[Rose and Levinson(2004)]%
        {rose2004understanding}
\bibfield{author}{\bibinfo{person}{Daniel~E Rose} {and} \bibinfo{person}{Danny Levinson}.} \bibinfo{year}{2004}\natexlab{}.
\newblock \showarticletitle{Understanding user goals in web search}. In \bibinfo{booktitle}{\emph{Proceedings of the 13th international conference on World Wide Web}}. \bibinfo{pages}{13--19}.
\newblock


\bibitem[Sadikov et~al\mbox{.}(2010)]%
        {sadikov2010clustering}
\bibfield{author}{\bibinfo{person}{Eldar Sadikov}, \bibinfo{person}{Jayant Madhavan}, \bibinfo{person}{Lu Wang}, {and} \bibinfo{person}{Alon Halevy}.} \bibinfo{year}{2010}\natexlab{}.
\newblock \showarticletitle{Clustering query refinements by user intent}. In \bibinfo{booktitle}{\emph{Proceedings of the 19th International Conference on World Wide Web}} (Raleigh, North Carolina, USA) \emph{(\bibinfo{series}{WWW '10})}. \bibinfo{publisher}{Association for Computing Machinery}, \bibinfo{address}{New York, NY, USA}, \bibinfo{pages}{841–850}.
\newblock
\showISBNx{9781605587998}
\href{https://doi.org/10.1145/1772690.1772776}{doi:\nolinkurl{10.1145/1772690.1772776}}


\bibitem[Sakai and Zeng(2021)]%
        {Sakai2021Retrieval}
\bibfield{author}{\bibinfo{person}{Tetsuya Sakai} {and} \bibinfo{person}{Zhaohao Zeng}.} \bibinfo{year}{2021}\natexlab{}.
\newblock \showarticletitle{Retrieval Evaluation Measures that Agree with Users’ SERP Preferences: Traditional, Preference-based, and Diversity Measures}.
\newblock \bibinfo{journal}{\emph{ACM Trans. Inf. Syst.}} \bibinfo{volume}{39}, \bibinfo{number}{2}, Article \bibinfo{articleno}{14} (\bibinfo{date}{Dec.} \bibinfo{year}{2021}), \bibinfo{numpages}{35}~pages.
\newblock
\showISSN{1046-8188}
\href{https://doi.org/10.1145/3431813}{doi:\nolinkurl{10.1145/3431813}}


\bibitem[Shah et~al\mbox{.}(2023)]%
        {shah2023using}
\bibfield{author}{\bibinfo{person}{Chirag Shah}, \bibinfo{person}{Ryen~W White}, \bibinfo{person}{Reid Andersen}, \bibinfo{person}{Georg Buscher}, \bibinfo{person}{Scott Counts}, \bibinfo{person}{Sarkar Snigdha~Sarathi Das}, \bibinfo{person}{Ali Montazer}, \bibinfo{person}{Sathish Manivannan}, \bibinfo{person}{Jennifer Neville}, \bibinfo{person}{Xiaochuan Ni}, {et~al\mbox{.}}} \bibinfo{year}{2023}\natexlab{}.
\newblock \showarticletitle{Using large language models to generate, validate, and apply user intent taxonomies}.
\newblock \bibinfo{journal}{\emph{arXiv preprint arXiv:2309.13063}} (\bibinfo{year}{2023}).
\newblock


\bibitem[Shankar et~al\mbox{.}(2024a)]%
        {shankar2024spade}
\bibfield{author}{\bibinfo{person}{Shreya Shankar}, \bibinfo{person}{Haotian Li}, \bibinfo{person}{Parth Asawa}, \bibinfo{person}{Madelon Hulsebos}, \bibinfo{person}{Yiming Lin}, \bibinfo{person}{J.~D. Zamfirescu-Pereira}, \bibinfo{person}{Harrison Chase}, \bibinfo{person}{Will Fu-Hinthorn}, \bibinfo{person}{Aditya~G. Parameswaran}, {and} \bibinfo{person}{Eugene Wu}.} \bibinfo{year}{2024}\natexlab{a}.
\newblock \bibinfo{title}{SPADE: Synthesizing Data Quality Assertions for Large Language Model Pipelines}.
\newblock
\urldef\tempurl%
\url{https://arxiv.org/abs/2401.03038}
\showURL{%
\tempurl}


\bibitem[Shankar et~al\mbox{.}(2024b)]%
        {shankar2024validates}
\bibfield{author}{\bibinfo{person}{Shreya Shankar}, \bibinfo{person}{JD Zamfirescu-Pereira}, \bibinfo{person}{Bj{\"o}rn Hartmann}, \bibinfo{person}{Aditya Parameswaran}, {and} \bibinfo{person}{Ian Arawjo}.} \bibinfo{year}{2024}\natexlab{b}.
\newblock \showarticletitle{Who validates the validators? aligning llm-assisted evaluation of llm outputs with human preferences}. In \bibinfo{booktitle}{\emph{Proceedings of the 37th Annual ACM Symposium on User Interface Software and Technology}}. \bibinfo{pages}{1--14}.
\newblock


\bibitem[Sheng et~al\mbox{.}(2020)]%
        {sheng-etal-2020-towards}
\bibfield{author}{\bibinfo{person}{Emily Sheng}, \bibinfo{person}{Kai-Wei Chang}, \bibinfo{person}{Prem Natarajan}, {and} \bibinfo{person}{Nanyun Peng}.} \bibinfo{year}{2020}\natexlab{}.
\newblock \showarticletitle{Towards {C}ontrollable {B}iases in {L}anguage {G}eneration}. In \bibinfo{booktitle}{\emph{Findings of the Association for Computational Linguistics: EMNLP 2020}}, \bibfield{editor}{\bibinfo{person}{Trevor Cohn}, \bibinfo{person}{Yulan He}, {and} \bibinfo{person}{Yang Liu}} (Eds.). \bibinfo{publisher}{Association for Computational Linguistics}, \bibinfo{address}{Online}, \bibinfo{pages}{3239--3254}.
\newblock
\href{https://doi.org/10.18653/v1/2020.findings-emnlp.291}{doi:\nolinkurl{10.18653/v1/2020.findings-emnlp.291}}


\bibitem[Sivaraman et~al\mbox{.}(2025)]%
        {sivaraman2025divisi}
\bibfield{author}{\bibinfo{person}{Venkatesh Sivaraman}, \bibinfo{person}{Zexuan Li}, {and} \bibinfo{person}{Adam Perer}.} \bibinfo{year}{2025}\natexlab{}.
\newblock \showarticletitle{Divisi: Interactive Search and Visualization for Scalable Exploratory Subgroup Analysis}.
\newblock \bibinfo{journal}{\emph{arXiv preprint arXiv:2502.10537}} (\bibinfo{year}{2025}).
\newblock


\bibitem[Sung et~al\mbox{.}(2025)]%
        {sung2025verila}
\bibfield{author}{\bibinfo{person}{Yoo~Yeon Sung}, \bibinfo{person}{Hannah Kim}, {and} \bibinfo{person}{Dan Zhang}.} \bibinfo{year}{2025}\natexlab{}.
\newblock \bibinfo{title}{VeriLA: A Human-Centered Evaluation Framework for Interpretable Verification of LLM Agent Failures}.
\newblock
\urldef\tempurl%
\url{https://arxiv.org/abs/2503.12651}
\showURL{%
\tempurl}


\bibitem[Szymanski et~al\mbox{.}(2024)]%
        {szymanski2024comparing}
\bibfield{author}{\bibinfo{person}{Annalisa Szymanski}, \bibinfo{person}{Simret~Araya Gebreegziabher}, \bibinfo{person}{Oghenemaro Anuyah}, \bibinfo{person}{Ronald~A Metoyer}, {and} \bibinfo{person}{Toby Jia-Jun Li}.} \bibinfo{year}{2024}\natexlab{}.
\newblock \showarticletitle{Comparing Criteria Development Across Domain Experts, Lay Users, and Models in Large Language Model Evaluation}.
\newblock \bibinfo{journal}{\emph{arXiv preprint arXiv:2410.02054}} (\bibinfo{year}{2024}).
\newblock


\bibitem[Thomas et~al\mbox{.}(2024)]%
        {thomas2024large}
\bibfield{author}{\bibinfo{person}{Paul Thomas}, \bibinfo{person}{Seth Spielman}, \bibinfo{person}{Nick Craswell}, {and} \bibinfo{person}{Bhaskar Mitra}.} \bibinfo{year}{2024}\natexlab{}.
\newblock \showarticletitle{Large language models can accurately predict searcher preferences}. In \bibinfo{booktitle}{\emph{Proceedings of the 47th International ACM SIGIR Conference on Research and Development in Information Retrieval}}. \bibinfo{pages}{1930--1940}.
\newblock


\bibitem[Vaughan(2004)]%
        {vaughan2004new}
\bibfield{author}{\bibinfo{person}{Liwen Vaughan}.} \bibinfo{year}{2004}\natexlab{}.
\newblock \showarticletitle{New measurements for search engine evaluation proposed and tested}.
\newblock \bibinfo{journal}{\emph{Information Processing \& Management}} \bibinfo{volume}{40}, \bibinfo{number}{4} (\bibinfo{year}{2004}), \bibinfo{pages}{677--691}.
\newblock


\bibitem[Voorhees(2002)]%
        {voorhees2002Philosophy}
\bibfield{author}{\bibinfo{person}{Ellen~M. Voorhees}.} \bibinfo{year}{2002}\natexlab{}.
\newblock \showarticletitle{The {Philosophy} of {Information} {Retrieval} {Evaluation}}. In \bibinfo{booktitle}{\emph{Evaluation of {Cross}-{Language} {Information} {Retrieval} {Systems}}}, \bibfield{editor}{\bibinfo{person}{Carol Peters}, \bibinfo{person}{Martin Braschler}, \bibinfo{person}{Julio Gonzalo}, {and} \bibinfo{person}{Michael Kluck}} (Eds.). \bibinfo{publisher}{Springer}, \bibinfo{address}{Berlin, Heidelberg}, \bibinfo{pages}{355--370}.
\newblock
\showISBNx{978-3-540-45691-9}
\href{https://doi.org/10.1007/3-540-45691-0_34}{doi:\nolinkurl{10.1007/3-540-45691-0_34}}


\bibitem[Wang et~al\mbox{.}(2023b)]%
        {wang-etal-2023-query2doc}
\bibfield{author}{\bibinfo{person}{Liang Wang}, \bibinfo{person}{Nan Yang}, {and} \bibinfo{person}{Furu Wei}.} \bibinfo{year}{2023}\natexlab{b}.
\newblock \showarticletitle{Query2doc: Query Expansion with Large Language Models}. In \bibinfo{booktitle}{\emph{Proceedings of the 2023 Conference on Empirical Methods in Natural Language Processing}}, \bibfield{editor}{\bibinfo{person}{Houda Bouamor}, \bibinfo{person}{Juan Pino}, {and} \bibinfo{person}{Kalika Bali}} (Eds.). \bibinfo{publisher}{Association for Computational Linguistics}, \bibinfo{address}{Singapore}, \bibinfo{pages}{9414--9423}.
\newblock
\href{https://doi.org/10.18653/v1/2023.emnlp-main.585}{doi:\nolinkurl{10.18653/v1/2023.emnlp-main.585}}


\bibitem[Wang et~al\mbox{.}(2023a)]%
        {wang2023large}
\bibfield{author}{\bibinfo{person}{Peiyi Wang}, \bibinfo{person}{Lei Li}, \bibinfo{person}{Liang Chen}, \bibinfo{person}{Zefan Cai}, \bibinfo{person}{Dawei Zhu}, \bibinfo{person}{Binghuai Lin}, \bibinfo{person}{Yunbo Cao}, \bibinfo{person}{Qi Liu}, \bibinfo{person}{Tianyu Liu}, {and} \bibinfo{person}{Zhifang Sui}.} \bibinfo{year}{2023}\natexlab{a}.
\newblock \showarticletitle{Large language models are not fair evaluators}.
\newblock \bibinfo{journal}{\emph{arXiv preprint arXiv:2305.17926}} (\bibinfo{year}{2023}).
\newblock


\bibitem[Wilcoxon et~al\mbox{.}(1970)]%
        {wilcoxon1970critical}
\bibfield{author}{\bibinfo{person}{Frank Wilcoxon}, \bibinfo{person}{S Katti}, \bibinfo{person}{Roberta~A Wilcox}, {et~al\mbox{.}}} \bibinfo{year}{1970}\natexlab{}.
\newblock \showarticletitle{Critical values and probability levels for the Wilcoxon rank sum test and the Wilcoxon signed rank test}.
\newblock \bibinfo{journal}{\emph{Selected tables in mathematical statistics}}  \bibinfo{volume}{1} (\bibinfo{year}{1970}), \bibinfo{pages}{171--259}.
\newblock


\bibitem[Wu et~al\mbox{.}(2023)]%
        {wu2023scattershot}
\bibfield{author}{\bibinfo{person}{Sherry Wu}, \bibinfo{person}{Hua Shen}, \bibinfo{person}{Daniel~S Weld}, \bibinfo{person}{Jeffrey Heer}, {and} \bibinfo{person}{Marco~Tulio Ribeiro}.} \bibinfo{year}{2023}\natexlab{}.
\newblock \showarticletitle{Scattershot: Interactive in-context example curation for text transformation}. In \bibinfo{booktitle}{\emph{Proceedings of the 28th International Conference on Intelligent User Interfaces}}. \bibinfo{pages}{353--367}.
\newblock


\bibitem[Wu et~al\mbox{.}(2019)]%
        {wu2019errudite}
\bibfield{author}{\bibinfo{person}{Tongshuang Wu}, \bibinfo{person}{Marco~Tulio Ribeiro}, \bibinfo{person}{Jeffrey Heer}, {and} \bibinfo{person}{Daniel~S Weld}.} \bibinfo{year}{2019}\natexlab{}.
\newblock \showarticletitle{Errudite: Scalable, reproducible, and testable error analysis}. In \bibinfo{booktitle}{\emph{Proceedings of the 57th Annual Meeting of the Association for Computational Linguistics}}. \bibinfo{pages}{747--763}.
\newblock


\bibitem[Xiong et~al\mbox{.}(2024)]%
        {xiong2024search}
\bibfield{author}{\bibinfo{person}{Haoyi Xiong}, \bibinfo{person}{Jiang Bian}, \bibinfo{person}{Yuchen Li}, \bibinfo{person}{Xuhong Li}, \bibinfo{person}{Mengnan Du}, \bibinfo{person}{Shuaiqiang Wang}, \bibinfo{person}{Dawei Yin}, {and} \bibinfo{person}{Sumi Helal}.} \bibinfo{year}{2024}\natexlab{}.
\newblock \showarticletitle{When search engine services meet large language models: visions and challenges}.
\newblock \bibinfo{journal}{\emph{IEEE Transactions on Services Computing}} (\bibinfo{year}{2024}).
\newblock


\bibitem[Yu et~al\mbox{.}(2022)]%
        {yu2022towards}
\bibfield{author}{\bibinfo{person}{Puxuan Yu}, \bibinfo{person}{Razieh Rahimi}, {and} \bibinfo{person}{James Allan}.} \bibinfo{year}{2022}\natexlab{}.
\newblock \showarticletitle{Towards explainable search results: a listwise explanation generator}. In \bibinfo{booktitle}{\emph{Proceedings of the 45th International ACM SIGIR Conference on Research and Development in Information Retrieval}}. \bibinfo{pages}{669--680}.
\newblock


\bibitem[Zhang et~al\mbox{.}(2024c)]%
        {Zhang2024USimAgent}
\bibfield{author}{\bibinfo{person}{Erhan Zhang}, \bibinfo{person}{Xingzhu Wang}, \bibinfo{person}{Peiyuan Gong}, \bibinfo{person}{Yankai Lin}, {and} \bibinfo{person}{Jiaxin Mao}.} \bibinfo{year}{2024}\natexlab{c}.
\newblock \showarticletitle{USimAgent: Large Language Models for Simulating Search Users}. In \bibinfo{booktitle}{\emph{Proceedings of the 47th International ACM SIGIR Conference on Research and Development in Information Retrieval}} (Washington DC, USA) \emph{(\bibinfo{series}{SIGIR '24})}. \bibinfo{publisher}{Association for Computing Machinery}, \bibinfo{address}{New York, NY, USA}, \bibinfo{pages}{2687–2692}.
\newblock
\showISBNx{9798400704314}
\href{https://doi.org/10.1145/3626772.3657963}{doi:\nolinkurl{10.1145/3626772.3657963}}


\bibitem[Zhang* et~al\mbox{.}(2020)]%
        {bert-score}
\bibfield{author}{\bibinfo{person}{Tianyi Zhang*}, \bibinfo{person}{Varsha Kishore*}, \bibinfo{person}{Felix Wu*}, \bibinfo{person}{Kilian~Q. Weinberger}, {and} \bibinfo{person}{Yoav Artzi}.} \bibinfo{year}{2020}\natexlab{}.
\newblock \showarticletitle{BERTScore: Evaluating Text Generation with BERT}. In \bibinfo{booktitle}{\emph{International Conference on Learning Representations}}.
\newblock
\urldef\tempurl%
\url{https://openreview.net/forum?id=SkeHuCVFDr}
\showURL{%
\tempurl}


\bibitem[Zhang et~al\mbox{.}(2024a)]%
        {zhang2024agentic}
\bibfield{author}{\bibinfo{person}{Weinan Zhang}, \bibinfo{person}{Junwei Liao}, \bibinfo{person}{Ning Li}, \bibinfo{person}{Kounianhua Du}, {and} \bibinfo{person}{Jianghao Lin}.} \bibinfo{year}{2024}\natexlab{a}.
\newblock \showarticletitle{Agentic information retrieval}.
\newblock \bibinfo{journal}{\emph{arXiv preprint arXiv:2410.09713}} (\bibinfo{year}{2024}).
\newblock


\bibitem[Zhang et~al\mbox{.}(2024b)]%
        {zhang2024llm}
\bibfield{author}{\bibinfo{person}{Zijian Zhang}, \bibinfo{person}{Shuchang Liu}, \bibinfo{person}{Ziru Liu}, \bibinfo{person}{Rui Zhong}, \bibinfo{person}{Qingpeng Cai}, \bibinfo{person}{Xiangyu Zhao}, \bibinfo{person}{Chunxu Zhang}, \bibinfo{person}{Qidong Liu}, {and} \bibinfo{person}{Peng Jiang}.} \bibinfo{year}{2024}\natexlab{b}.
\newblock \showarticletitle{LLM-Powered User Simulator for Recommender System}.
\newblock \bibinfo{journal}{\emph{arXiv preprint arXiv:2412.16984}} (\bibinfo{year}{2024}).
\newblock


\bibitem[Zhou et~al\mbox{.}(2024)]%
        {zhou2024cognitive}
\bibfield{author}{\bibinfo{person}{Yujia Zhou}, \bibinfo{person}{Qiannan Zhu}, \bibinfo{person}{Jiajie Jin}, {and} \bibinfo{person}{Zhicheng Dou}.} \bibinfo{year}{2024}\natexlab{}.
\newblock \showarticletitle{Cognitive personalized search integrating large language models with an efficient memory mechanism}. In \bibinfo{booktitle}{\emph{Proceedings of the ACM Web Conference 2024}}. \bibinfo{pages}{1464--1473}.
\newblock


\end{thebibliography}

\appendix
\captionsetup{width=.8\textwidth}
\clearpage

\onecolumn

\section*{Appendices}

\renewcommand{\thesection}{A.\arabic{section}}
\renewcommand{\thesubsection}{\thesection.\arabic{subsection}}
\renewcommand{\thefigure}{A.\arabic{figure}}
\renewcommand{\thetable}{A.\arabic{table}}

\setcounter{section}{0}
\setcounter{figure}{0}
\setcounter{table}{0}
\setcounter{page}{1}
\appendix
\section{Intent Generation Pipeline} \label{appendix:intent_generation}

\subsection{User Attribute Generation} \label{appendix:user_attribute_generation}

\subsubsection{Attribute Generation Prompt} \label{appendix:attribute_gen_prompt}

\def\lstlistingname{Prompt}
\lstset{
  basicstyle=\ttfamily\footnotesize,
  columns=fullflexible,
  frame=single,
  breaklines=true,
  postbreak=\mbox{\textcolor{red}{$\hookrightarrow$}\space},
  breakatwhitespace=true,
  captionpos=b,
}

\leavevmode\begin{lstlisting}
You are an AI assistant specializing in analyzing user behaviors to identify user segments.
From the given **large-scale search logs (query sequences, dwell times, clicked URLs, etc.)**, infer attributes of users and search contexts that hint the intents behind search queries. You must define attributes that explain **how the same query may yield different expected answers depending on the user's context.** The generated attributes will be used to infer how different users would issue a follow-up query for a given search query. The inferred attributes could be either directly based on the search queries and behaviors (e.g., users' interest, search purpose) or describing more general user attributes in general (e.g., expertise, sensitivity towards trends). Be very creative, and do not limit yourself into only superficial features from the logs.

---
### ** TASK **
0. **Infer the Search Domain**
   - Analyze the search logs to understand the domain of the search logs provided.
   - While identifying the attributes, consider the domain of the search logs.
   - Example Domains: Shopping, Local, Knowledge

1. **Define Search Context Dimensions**
  - Identify **all mutually exclusive dimensions** that capture key factors influencing user expectations for search results. There should be at least six dimensions.
  - These dimensions should describe either a) "user differences/categories" such as **who they are, why they search, what they are interested in, or what constraints affect their decisions** or b) "search contexts" such as **final purpose of the search**.
  - Example dimensions include (but are not limited to):
   	- **Temporal relevance**: How up-to-date information the user requires.
   	- **Overall Task**: The overall task the user is trying to accomplish with the information retrieved from the search.

2. **Define Values within Each Dimension**
  - Identify **at least five distinct and mutually exclusive values** per dimension.
  - Each value must include the following components:
    - **Value Name**: A label representing the search context
    - **Definition**: A concise description of the circumstances in which a user fits this context
    - **Expected Answer Difference**: How a user with this context would expect different results from the same query
\end{lstlisting}

\clearpage
\subsubsection{Shopping Category Attributes} \label{appendix:shopping_attributes}
Here is the set of user attributes we used for Shopping domain queries to generate expanded queries.
\renewcommand{\arraystretch}{1.5} 

\begin{table}[htb]\centering
\caption{User attribute set for shopping domain queries}
\Description{Table titled "User attribute set for shopping domain queries." It defines six user attribute dimensions with five values each. The dimensions are:
1. Price Sensitivity, with values: Discount Seeker, Quality Focused, Value Hunter, Luxury Buyer, and Price Neutral.
2. Purchase Intent, with values: Exploratory Browsing, Detailed Comparison, Targeted Purchase, After-Sales Inquiry, and Reorder/Repeat Purchase.
3. Interaction Complexity, with values: Single Query Simplicity, Iterative Refinement, Ambiguous Query Issuer, Multi-Intent Exploration, and Batch/High-Interaction Shopper.
4. Query Specificity, with values: Product Code Lookup, Model Name Search, Brand/Category Browsing, Promotional Search, and Technical Specification Inquiry.
5. Search Goals, with values: Transactional, Informational, Navigational, Exploratory Browsing, and Store/Brand Specific.
6. Temporal Urgency, with values: Immediate Requirement, Seasonal Shopping, Long-Term Research, Impulsive Trend, and Casual, Non-Urgent.
7. User Expertise, with values: Novice Shopper, Informed Consumer, Expert Reviewer, Brand Loyalist, and Trend Enthusiast.
Each value is accompanied by a definition describing the user's shopping behavior or search characteristics.}
\scriptsize
\begin{tabular}{l r >{\raggedright\arraybackslash}p{0.5\textwidth}}
\toprule
\textbf{Dimension} & \textbf{Value} & \textbf{Definition} \\
\midrule
\multirow{5}{*}{Price Sensitivity} 
    & Discount Seeker        
    & The user prioritizes finding products with the best deals, discounts, or coupon offers. \\
\cline{2-3}
    & Quality Focused        
    & The shopper values high quality and is willing to pay a premium for reliable and reputable products. \\
\cline{2-3}
    & Value Hunter           
    & The customer seeks a balance between affordable pricing and acceptable quality, optimizing cost-effectiveness. \\
\cline{2-3}
    & Luxury Buyer           
    & The search is aimed at high-end products, with less concern about cost but a focus on exclusivity or advanced features. \\
\cline{2-3}
    & Price Neutral          
    & The buyer does not emphasize price in queries; instead, they focus on features or compatibility, treating cost as a secondary factor. \\
\cline{1-3}
\multirow{5}{*}{Purchase Intent}
    & Exploratory Browsing   
    & The user is casually navigating product categories without a clear intent to buy immediately, simply gathering broad information. \\
\cline{2-3}
    & Detailed Comparison    
    & The user is comparing multiple products or variants, scrutinizing specifications, reviews, and price differences before deciding on a purchase. \\
\cline{2-3}
    & Targeted Purchase      
    & The user knows exactly what he/she wants, often using specific product codes or model numbers to locate a particular item quickly. \\
\cline{2-3}
    & After-Sales Inquiry    
    & The user has already made a purchase or is post-purchase and seeks additional information such as manuals, installation tips or support details. \\
\cline{2-3}
    & Reorder/Repeat Purchase
    & This user is revisiting a previously purchased product or a trusted brand, indicating loyalty or routine reordering behavior. \\
\cline{1-3}
\multirow{5}{*}{Interaction Complexity} 
    & Single Query Simplicity
    & The user issues a one-off, straightforward query and makes a quick decision based on first impressions. \\
\cline{2-3}
    & Iterative Refinement   
    & The user actively reformulates queries and explores multiple related queries to narrow down choices. \\
\cline{2-3}
    & Ambiguous Query Issuer 
    & Queries in this category are vague or have multiple interpretations (e.g., misspellings or generic terms). \\
\cline{2-3}
    & Multi-Intent Exploration
    & The user's session covers several different facets of a product search, possibly mixing product specifications, reviews, and how-to guides in one session. \\
\cline{2-3}
    & Batch/High-Interaction Shopper
    & This user generates many clicks and longer sessions, showing high engagement with product pages and frequent query reformulations. \\
\cline{1-3}
\multirow{5}{*}{Query Specificity}
    & Product Code Lookup    
    & The query is composed of alphanumeric codes or specific model identifiers indicating that the user knows exactly which product to find. \\
\cline{2-3}
    & Model Name Search      
    & The user uses the product model name or series, indicating targeted but slightly broader search intent. \\
\cline{2-3}
    & Brand/Category Browsing
    & The query uses broad brand or category terms, showing a general interest in a product type without a specific model in mind. \\
\cline{2-3}
    & Promotional Search     
    & Queries that hint at deals, discounts, or seasonal promotions, sometimes embedded with sale-related keywords. \\
\cline{2-3}
    & Technical Specification Inquiry
    & The query reflects a need for detailed technical or performance information, often through model numbers with technical labels. \\
\cline{1-3}
\multirow{5}{*}{Search Goals}
    & Transactional          
    & User's primary goal is to complete a purchase; the query implies a direct buying intent. \\
\cline{2-3}
    & Informational          
    & User is researching product details, reading reviews, or seeking expert opinions without immediate purchase intent. \\
\cline{2-3}
    & Navigational           
    & User intends to navigate to a specific website, brand store, or product page and uses search terms to get there quickly. \\
\cline{2-3}
    & Exploratory Browsing   
    & User is casually browsing or window-shopping without a specific product in mind. \\
\cline{2-3}
    & Store/Brand Specific   
    & User searches with a focus on a particular retailer or brand experience. \\
\cline{1-3}
\multirow{5}{*}{Temporal Urgency}
    & Immediate Requirement  
    & The search is time-sensitive, indicating the need to purchase or find information immediately. \\
\cline{2-3}
    & Seasonal Shopping      
    & The query is influenced by seasonal factors (e.g., holiday or event-related), and the user is looking for timely, season-specific items. \\
\cline{2-3}
    & Long-Term Research     
    & The user is not in a rush but is investing time in gathering comprehensive information for future decisions. \\
\cline{2-3}
    & Impulsive Trend        
    & The query reflects spontaneous interest driven by trending or viral products with low commitment. \\
\cline{2-3}
    & Casual, Non-Urgent     
    & The search lacks any urgency, and the user is casually browsing without immediate plans for purchase. \\
\cline{1-3}
\multirow{5}{*}{User Expertise}
    & Novice Shopper         
    & The user is relatively new to the product category and uses generic or less refined queries. \\
\cline{2-3}
    & Informed Consumer      
    & The user demonstrates familiarity with the product, using specific codes and terminology. \\
\cline{2-3}
    & Expert Reviewer        
    & The user is experienced, often cross-checking multiple sources, and looking into minute details of product performance. \\
\cline{2-3}
    & Brand Loyalist         
    & The shopper consistently searches for a particular brand or supplier, indicating reliance on known quality. \\
\cline{2-3}
    & Trend Enthusiast       
    & The user is driven by current trends and often seeks the latest or most popular products. \\
\cline{1-3}
\bottomrule
\end{tabular}
\end{table}

\renewcommand{\arraystretch}{1} 

\clearpage
\subsubsection{Location Category Attributes} \label{appendix:location_attributes}
Here is the set of user attributes we used for Location domain queries to generate expanded queries.

\renewcommand{\arraystretch}{1.5} 

\begin{table}[H]\centering
\caption{User attribute set for location domain queries}
\Description{
This table presents six dimensions of user attributes relevant to location-based queries, with each dimension listing five specific values and their definitions. The dimensions are: Content Format Preference, Geographic Relevance, Search Purpose, Social Influence & Sentiment, Temporal Urgency, and User Expertise.
1. Content Format Preference, values include map-centric, list or directory, visual media, textual details, and aggregated data. 
2. Geographic Relevance ranges from hyperlocal to global. 
3. Search Purpose includes navigational, informational, transactional, social and review-oriented, and exploratory intents. 
4. Social Influence & Sentiment covers trend-driven, community recommendation, personalized preference, safety and authority seeking, and price sensitivity. 
5. Temporal Urgency spans immediate need to historical inquiry.
6. User Expertise ranges from novice to critical evaluator.
Each row specifies a dimension, its value, and a brief definition describing user expectations or behavior patterns in location domain searches.
}
\scriptsize
\begin{tabular}{l r >{\raggedright\arraybackslash}p{0.5\textwidth}}
\toprule
\textbf{Dimension} & \textbf{Value} & \textbf{Definition} \\
\midrule
\multirow{5}{*}{Content Format Preference} 
    & Map-centric 
    & Users prefer visual maps for navigation and location-based search results. \\
\cline{2-3}
    & List/Directory 
    & The user expects a listing or directory format, often in a ranked order or categorized view. \\
\cline{2-3}
    & Visual Media 
    & Users seek rich imagery or video content over textual descriptions. \\
\cline{2-3}
    & Textual Details 
    & A preference for comprehensive text-based content that provides detailed descriptions and narratives. \\
\cline{2-3}
    & Aggregated Data 
    & The user favors summaries of user reviews, star ratings, and aggregated scores. \\
\cline{1-3}
\multirow{5}{*}{Geographic Relevance}
    & Hyperlocal 
    & The search is restricted to an immediate, neighborhood-scale area. \\
\cline{2-3}
    & City-level 
    & Results are intended to cover the entire city, balancing local details with city wide data. \\
\cline{2-3}
    & Regional 
    & Users are interested in information spanning a larger area, such as a province or region. \\
\cline{2-3}
    & National 
    & The query is general and not restricted to a specific locale, expecting nationwide coverage. \\
\cline{2-3}
    & Global 
    & Query indicates interest in international options or comparisons beyond one's own country. \\
\cline{1-3}
\multirow{5}{*}{Search Purpose}
    & Navigational 
    & The user intends to find a specific website, location, or entity already known to them. \\
\cline{2-3}
    & Informational 
    & The user is looking to gain knowledge or understand details about a subject, event, or entity. \\
\cline{2-3}
    & Transactional 
    & The user wants to perform an action like placing an order, making a booking, or initiating a call. \\
\cline{2-3}
    & Social \& Review-Oriented 
    & The user's intent is to gauge community sentiment, opinions, or reviews on a service or location. \\
\cline{2-3}
    & Exploratory 
    & The user is casually browsing and open to discovering new options without a fixed goal. \\
\cline{1-3}
\multirow{5}{*}{Social Influence \& Sentiment}
    & Trend-Driven 
    & The user is influenced by current trends, expecting results to reflect what's hot or popular right now. \\
\cline{2-3}
    & Community Recommendation 
    & The user relies on social proof and reviews from peers and community sources. \\
\cline{2-3}
    & Personalized Preference 
    & The user expects results tailored to their past behaviors, preferences, and interaction history. \\
\cline{2-3}
    & Safety \& Authority Seeking 
    & The user's intent is underpinned by a desire for credible, authoritative, or certified information. \\
\cline{2-3}
    & Price Sensitive/Deal-Oriented 
    & The query is motivated by budget constraints, special offers or discount opportunities. \\
\cline{1-3}
\multirow{5}{*}{Temporal Urgency}
    & Immediate Need 
    & Queries that reflect a need for results and actions right away (e.g. phone calls, maps). \\
\cline{2-3}
    & Short-term Planning 
    & Users planning to act in the near future; results should support decision-making for the upcoming hours or day. \\
\cline{2-3}
    & Event-Based 
    & Searches linked to a specific event, occasion, or special offer tied to time. \\
\cline{2-3}
    & Research-Oriented 
    & Queries where the user is gathering background information with less pressure to act immediately. \\
\cline{2-3}
    & Historical Inquiry 
    & Users search with an interest in past trends, legacy information, or comparative historical data. \\
\cline{1-3}
\multirow{5}{*}{User Expertise}
    & Novice 
    & Users with minimal prior knowledge seeking straightforward, fundamental information. \\
\cline{2-3}
    & Intermediate 
    & Users who have a fair level of familiarity but seek additional context or comparisons. \\
\cline{2-3}
    & Expert 
    & Users with a deep understanding who require granular and technical details. \\
\cline{2-3}
    & Trend-sensitive 
    & Users who strongly value current popularity, trending options, or hot spots. \\
\cline{2-3}
    & Critical Evaluator 
    & Users who scrutinize offerings deeply and need comprehensive, unbiased comparisons. \\
\cline{1-3}
\bottomrule
\end{tabular}
\end{table}

\clearpage
\subsubsection{Knowledge Category Attributes} \label{appendix:knowledge_attributes}
Here is the set of user attributes we used for Knowledge domain queries to generate expanded queries.

\renewcommand{\arraystretch}{1.5} 

\begin{table}[!htp]\centering
\caption{User attribute set for knowledge domain queries}
\Description{Table titled "User attribute set for knowledge domain queries." It lists six dimensions of user attributes, each with five possible values and their definitions:
1. Content Domain Interest – Shopping/Product, News \& Current Affairs, Entertainment \& Social Media, Academic/Professional, Local Services. Defines what type of content users seek.
2. Content Format Preference – Text Articles, Video Content, Image Galleries, Interactive Tools, Aggregated Summaries. Defines preferred content presentation.
3. Task Complexity – Single-Step, Multi-Step, Comparative, Exploratory, Problem-Solving. Defines complexity and structure of the user's information task.
4. Regional/Cultural Context – Local (Korean Focus), Global Perspective, Traditional Culture, Pop Culture/Trendy, Multilingual/Hybrid. Defines cultural or regional relevance.
5. Search Goals – Navigational, Informational, Transactional, Entertainment, Social/Community. Defines the user's search intent.
6. Temporal Relevance – Real-Time, Recent News, Scheduled/Planned, Historical Archives, Evergreen Information. Defines time sensitivity of the query.
7. User Expertise – Novice, Intermediate, Advanced, Expert, Enthusiast. Defines user's familiarity with the topic.
Each value is accompanied by a short definition explaining the user's interest, preference, or behavior.
}
\scriptsize
\begin{tabular}{l r >{\raggedright\arraybackslash}p{0.5\textwidth}}
\toprule
\textbf{Dimension} & \textbf{Value} & \textbf{Definition} \\
\midrule
\multirow{5}{*}{Content Domain Interest} 
    & Shopping/Product 
    & Users are primarily interested in product details, reviews, pricing and purchasing directives. \\
\cline{2-3}
    & News \& Current Affairs
    & Queries in this category target the latest updates on local, national or international news and current events. \\
\cline{2-3}
    & Entertainment \& Social Media 
    & The focus here is on trending entertainment, celebrity news, social interactions, and cultural content. \\
\cline{2-3}
    & Academic/Professional 
    & Users are looking for scholarly articles, research studies, job information, college details or professional guidelines. \\
\cline{2-3}
    & Local Services 
    & The interest is in nearby businesses, local news, community events and practical services available in the region. \\
\cline{1-3}
\multirow{5}{*}{Content Format Preference}
    & Text Articles 
    & The user prefers detailed written content such as long-form articles, blog posts and in-depth written analyses. \\
\cline{2-3}
    & Video Content 
    & The user is inclined towards audiovisual material, seeking video explanations or product reviews. \\
\cline{2-3}
    & Image Galleries 
    & The search features a strong visual component, and the user expects image-heavy results such as product photos or infographics. \\
\cline{2-3}
    & Interactive Tools 
    & The user expects dynamic content such as calculators, map-based tools, or interactive widgets to explore information. \\
\cline{2-3}
    & Aggregated Summaries 
    & The user prefers concise summaries or aggregated content that condenses multiple sources into one overview. \\
\cline{1-3}
\multirow{5}{*}{Task Complexity}
    & Single-Step 
    & A straightforward query where the answer is expected with minimal follow‐up; the task completes with one search. \\
\cline{2-3}
    & Multi-Step 
    & The query is part of a larger research process, requiring several refined questions and detailed follow-up. \\
\cline{2-3}
    & Comparative 
    & The user wants to compare multiple items, products or opinions side by side. \\
\cline{2-3}
    & Exploratory 
    & The search is broad and open-ended, often to discover ideas or directions rather than a specific answer. \\
\cline{2-3}
    & Problem-Solving 
    & The query is aimed at troubleshooting or resolving a specific issue with a detailed stepwise explanation. \\
\cline{1-3}
\multirow{5}{*}{Regional/Cultural Context}
    & Local (Korean Focus) 
    & The user is primarily interested in content that is specific to the Korean market, language, and cultural context. \\
\cline{2-3}
    & Global Perspective 
    & The user looks for diverse international insights, incorporating sources from different countries. \\
\cline{2-3}
    & Traditional Culture 
    & The search emphasizes authoritative or classical sources rooted in long-standing cultural traditions. \\
\cline{2-3}
    & Pop Culture/Trendy 
    & The user is tuned into contemporary and trending cultural phenomena, including celebrity news and viral content. \\
\cline{2-3}
    & Multilingual/Hybrid 
    & The user is comfortable with multiple languages and expects answers that may blend local and international elements. \\
\cline{1-3}
\multirow{5}{*}{Search Goals}
    & Navigational 
    & The user intends to reach a specific website or page rather than explore information broadly. \\
\cline{2-3}
    & Informational 
    & The user seeks to learn or understand a topic and needs detailed background data, definitions or explanations. \\
\cline{2-3}
    & Transactional 
    & The user is aiming to perform a commercial action such as shopping, booking, or monitoring product listings. \\
\cline{2-3}
    & Entertainment 
    & The query is driven by a desire to consume leisure or fun content such as music, dramas, celebrity news, or blogs. \\
\cline{2-3}
    & Social/Community 
    & The user is looking for discussions, community opinions, forum posts or social interactions. \\
\cline{1-3}
\multirow{5}{*}{Temporal Relevance}
    & Real-Time 
    & The user needs immediate, live updates and information reflecting the current moment. \\
\cline{2-3}
    & Recent News 
    & The user is interested in information updated over the past few days, such as recent events or announcements. \\
\cline{2-3}
    & Scheduled/Planned 
    & The search is focused on events or releases planned for the near future (e.g., event scheduling, future product launches). \\
\cline{2-3}
    & Historical Archives 
    & The user is seeking background information, historical context, or archived data from the past. \\
\cline{2-3}
    & Evergreen Information 
    & The query's relevance is timeless; the answer is expected to be stable and factual regardless of the date. \\
\cline{1-3}
\multirow{5}{*}{User Expertise}
    & Novice 
    & A user with little background knowledge about the topic who often uses simple or ambiguous queries. \\
\cline{2-3}
    & Intermediate 
    & A user with moderate familiarity who sometimes uses industry terms but still benefits from clear definitions. \\
\cline{2-3}
    & Advanced 
    & A user who has substantial knowledge on the subject and expects more in‐depth discussion and nuanced information. \\
\cline{2-3}
    & Expert 
    & Highly knowledgeable users, often professionals or specialists, who use precise, technical queries. \\
\cline{2-3}
    & Enthusiast 
    & Users who are passionate about a niche topic, often exploring multiple facets and seeking insider perspectives. \\
\cline{1-3}
\bottomrule
\end{tabular}
\end{table}

\clearpage
\subsection{Query Background Knowledge Retrieval}
\label{appendix:query_background_knowledge_retrieval}

\def\lstlistingname{Prompt}
\lstset{
  basicstyle=\ttfamily\footnotesize,
  columns=fullflexible,
  frame=single,
  breaklines=true,
  postbreak=\mbox{\textcolor{red}{$\hookrightarrow$}\space},
  breakatwhitespace=true,
  captionpos=b,
}

\begin{lstlisting}
You are an expert in summarizing search results. Given a set of search results from Google and Naver, provide a brief explanation of the context around the search query in Korean. 
The explanation should be concise and informative, and should not exceed 200 words.
\end{lstlisting}

\subsection{Expanded Query Generation} \label{appendix:expanded_query_prompt}

\subsubsection{Expanded Query Generation with Attributes}

\def\lstlistingname{Prompt}
\lstset{
  basicstyle=\ttfamily\footnotesize,
  columns=fullflexible,
  frame=single,
  breaklines=true,
  postbreak=\mbox{\textcolor{red}{$\hookrightarrow$}\space},
  breakatwhitespace=true,
  captionpos=b,
}

\leavevmode\begin{lstlisting}
```
You are an expert in analyzing people's search behaviors. You will be given a search query, a brief explanation of the context around the query, a query category, and searcher attributes from a Korean search engine. Your task is to generate a list of follow-up queries that reflect what a person with the given attributes might actually want to search for.

**Limit the expansion to NO MORE THAN TWO additional words.**
The follow-up query **must not contain more than two additional words compared to the original query**.

# Step 1: Consider All Searcher Attributes and Query Context Equally
- Do NOT over-prioritize user attributes over the query context.
- Carefully analyze both:
   - **Searcher Attributes**: How the given attributes might shape the way a person refines their search.
   - **Query Category:** The provided category will help determine the most relevant follow-up queries. 
   - **Query Context**: What current trends, controversies, or relevant information about the query topic that may influence the search.

- Some attributes may naturally have more influence than others, but ensure that the query context remains a central guiding factor.

# Step 2: Identify Key Subtopics or Representative Entities (if applicable)
- You will be provided with the category of the query, so that you can better create follow-up queries.
- If the query refers to a well-known entity (e.g., a brand, location, product, person, or concept), identify common subtopics or associated terms that are frequently searched in relation to it.
- Examples:
   - For a shopping/brand-related query: Key products, best-selling items, reviews, pricing information, controversies, history, etc.
   - For a location/place-related query: Popular attractions, travel itineraries, cultural context, historical significance, local news, etc.
   - For a person-related query: Biography, achievements, controversies, related figures, etc.
   - For a general concept/knowledge-based query: Related terms, frequently associated discussions, subtopics, current debates, etc.
- Ensure that the follow-up queries naturally reflect these common expansions in a way that aligns with the search intent.

# Step 3: Create Realistic Combinations of Searcher Attributes
- Generate MEANINGFUL,PLAUSIBLE, REALISTIC combinations of searcher attributes that represent different types of users searching for the given query.
- It is NOT necessary to select one attribute from each category.
   - Some user groups may be defined by only one or two key attributes, while others may involve multiple attributes.
   - Ensure that each combination is natural and realistic.
- Ensure that each combination considers both the searcher attributes and the query context (including key products, if relevant).
- Make AT LEAST 10 different and mutually-exclusive combinations of searcher attributes. 
- Briefly justify each combination, explaining why a person with those attributes would search in that particular way.

-------------------------------------------------------------

# Step 4: Generate Follow-up Queries
For each attribute combination:
- Provide at least five follow-up queries that a user with those attributes would likely search for after their initial query did not return sufficient or relevant results.
- Ensure the follow-up queries:
   - Use natural and concise keyword combinations that resemble real-world search behavior.
   - Maintain the original search intent, refining or re-expressing it while keeping the core meaning intact.
   - Balance both searcher attributes and query context in a natural way.
   - If the query is a brand name, include both brand-related searches and product-specific searches.

# Step 5: Enforce Logical Consistency in Follow-up Queries
To ensure that the follow-up queries are realistic, relevant, and logically consistent, apply the following constraints:

1. Follow-up queries must align with the original query's context.
   - Incorrect Example: "Patagonia smartphone launch" (Patagonia is an outdoor apparel brand, so this query is not logically valid)
   - Correct Example: "Patagonia Synchilla vs Retro-X comparison"

2. Queries must be based on actual search behavior and natural language patterns.
   - Incorrect Example: "Patagonia best thing" (Unnatural phrasing)
   - Correct Example: "Patagonia top-rated products"

3. Avoid ambiguous or overly broad queries that lack a clear intent.
   - Incorrect Example: "Patagonia popular" (It is unclear what is popular: products? destinations?)
   - Correct Example: "Popular Patagonia fleece recommendations"

4. Follow-up queries should not contradict the known facts or nature of the entity.
   - Incorrect Example: "Patagonia home appliance price comparison" (Patagonia does not sell electronics)
   - Correct Example: "Patagonia fleece price comparison"

5. Ensure that the follow-up queries add meaningful refinements rather than repeating the same concept.
   - Bad Example:
      - "Patagonia clothes"
      - "Patagonia apparel"
      - "Patagonia fashion" (These queries repeat the same general idea)
   - Good Example:
      - "Patagonia Synchilla reviews"
      - "Patagonia fleece vs The North Face comparison"
      - "Patagonia Retro-X where to buy"

6. Exclude follow-up queries that attempt to ask multiple unrelated things at once.
   - Incorrect Example:
      - "Patagonia Synchilla price, material, discount, and store location all in one" (Too many elements in one query)
      - "Patagonia fleece features and how to wash" (Overly broad request)
   - Correct Example:
      - "Patagonia Synchilla price comparison"
      - "Patagonia Synchilla material features"
      - "Patagonia Synchilla discount info"

# Step 6: Enforce Query Format Constraints
- Follow-up queries should be structured for optimal searchability.
- Constraints:
   - **Follow-up queries must be noun-based** (avoid full sentences or complex phrasing).
   - **A follow-up query should only add up to two additional words** to the original search query.
   - **Avoid question-style queries** (e.g., "Is Patagonia Synchilla good?" => "Patagonia Synchilla reviews" preferred).

   - Incorrect Example:
      - "I'm thinking about buying Patagonia Synchilla, any reviews?" (Too complex)
      - "Where is the cheapest place to buy Patagonia Synchilla?" (Question-style)
      - "Patagonia Synchilla discount code 2024 and size comparison" (More than two additional words)
   - Correct Example:
      - "Patagonia Synchilla reviews"
      - "Patagonia Synchilla discount"
      - "Patagonia Synchilla size"
```
\end{lstlisting}

\subsubsection{Expanded Query Generation without Attributes}

\def\lstlistingname{Prompt}
\lstset{
  basicstyle=\ttfamily\footnotesize,
  columns=fullflexible,
  frame=single,
  breaklines=true,
  postbreak=\mbox{\textcolor{red}{$\hookrightarrow$}\space},
  breakatwhitespace=true,
  captionpos=b,
}

\leavevmode
\begin{lstlisting}
```
You are an expert in analyzing people's search behaviors. You will be given a search query, a brief explanation of the context around the query, and a query category from a Korean search engine. Your task is to generate a list of follow-up queries that reflect what a person might actually want to search for.

**Limit the expansion to NO MORE THAN TWO additional words.**
The follow-up query **must not contain more than two additional words compared to the original query**.     

# Step 1: Consider Query Context Carefully
- Carefully analyze the given **query context** to refine and expand the search intent.
- Consider:
   - **Query Context:** What current trends, controversies, or relevant information about the query topic might influence the search?
   - **Query Category:** The provided category will help determine the most relevant follow-up queries.

# Step 2: Identify Key Subtopics or Representative Entities (if applicable)
- You will be provided with the **category of the query**, which will help guide the generation of follow-up queries.
- If the query refers to a **well-known entity** (e.g., a brand, location, product, person, or concept), identify **commonly associated subtopics or related terms** that users frequently search in connection with that entity.
- Examples:
   - **Shopping/Brand-related query:** Popular products, bestsellers, reviews, pricing, controversies, brand history, etc.
   - **Location/Place-related query:** Attractions, itineraries, culture, history, local events or news, etc.
   - **Person-related query:** Biography, achievements, controversies, related individuals, etc.
   - **Concept/Knowledge-based query:** Related terms, discussions, common subtopics, ongoing debates, etc.
- Ensure the follow-up queries **naturally reflect these expansions** while aligning with the original intent.

# Step 3: Generate Follow-up Queries
- **Provide at least 50 follow-up queries** that a user might search if the original query didn't yield enough relevant results.
- Ensure that the follow-up queries:
   - Use **natural and concise keyword combinations** that reflect real-world search behavior.
   - **Preserve the original intent**, refining or narrowing it while maintaining the core goal.
   - **Balance the query context and category** in a realistic manner.
   - If the query is about a brand, include both **brand-level** and **product-level** refinements.

# Step 4: Enforce Logical Consistency in Follow-up Queries
To ensure that the follow-up queries are **realistic, relevant, and logically sound**, apply the following constraints:

1. **Follow-up queries must align with the original query's context.**  
   - Incorrect Example: "Patagonia smartphone release" (Patagonia is an outdoor apparel brand, not a smartphone company)
   - Correct Example: "Patagonia Synchilla vs Retro-X comparison"

2. **Queries must reflect actual search behavior and natural phrasing.**  
   - Incorrect Example: "The best thing from Patagonia" (unnatural phrasing)
   - Correct Example: "Top Patagonia products ranked"

3. **Avoid overly vague or broad queries with unclear intent.**  
   - Incorrect Example: "Patagonia popular" (what is popular? products? stores?)
   - Correct Example: "Popular Patagonia fleece recommendations"

4. **Do not contradict facts about the entity.**  
   - Incorrect Example: "Patagonia appliance price comparison" (Patagonia does not sell appliances)
   - Correct Example: "Patagonia fleece price comparison"

5. **Follow-up queries should meaningfully refine the original query instead of repeating it.**
   - Bad Example:
      - "Patagonia clothes"
      - "Patagonia apparel"
      - "Patagonia fashion" (all too similar)
   - Good Example:
      - "Patagonia Synchilla reviews"
      - "Patagonia fleece vs The North Face comparison"
      - "Where to buy Patagonia Retro-X"

6. **Avoid follow-up queries that combine multiple unrelated elements.**  
   - Incorrect Example:
      - "Patagonia Synchilla price, material, discount, and store locations" (too much in one query)
      - "Patagonia fleece features and washing instructions" (overly broad)
   - Correct Example:
      - "Patagonia Synchilla price comparison"
      - "Patagonia Synchilla material features"
      - "Patagonia Synchilla discount info"

# Step 5: Enforce Query Format Constraints
- Follow-up queries must be structured for easy searchability.
- Constraints:
   - **Follow-up queries must be noun-based**, not full sentences.
   - **They should add no more than two extra words** to the original query.
   - **Avoid question-style phrasing.**  
     (e.g., "Is Patagonia Synchilla good?" => use "Patagonia Synchilla reviews" instead)

   - Incorrect Example:
      - "Thinking about buying Patagonia Synchilla, any reviews?" (too long, informal)
      - "Where's the cheapest place to buy Patagonia Synchilla?" (question-style)
      - "Patagonia Synchilla discount code 2024 and sizing guide" (too many added words)
   - Correct Example:
      - "Patagonia Synchilla reviews"
      - "Patagonia Synchilla discount"
      - "Patagonia Synchilla size"
```
\end{lstlisting}

\clearpage
\subsection{Intent Types Selection} 
\label{appendix:intent_types}

\subsubsection{List of Final Intent Types}

Table~\ref{tab:intent-types} shows the final set of 11 intent types used for generating search intents out of the original 20 types~\cite{mitsui16extracting}. 
We excluded IS (Identify something to get started), KR (Keep record of a link), AS (Access a specific item), AC (Access items with common characteristics), AP (Access a web site/home page or similar), ED (Evaluate duplication of an item), OS (Obtain specific information), OP (Obtain part of the item), and OW (Obtain a whole item(s)). 

\begin{table}[!htp]\centering
\caption{Final set of 11 intent types used for generating search intents out of original 20 types~\cite{mitsui16extracting}. We excluded IS (Identify something to get started), KR (Keep record of a link), AS (Access a specific item), AC (Access items with common characteristics), AP (Access a web site/home page or similar), ED (Evaluate duplication of an item), OS (Obtain specific information), OP (Obtain part of the item), and OW (Obtain a whole item(s)). }
\Description{
Table summarizing the final set of 11 intent types used for generating search intents, adapted from Mitsui et al. (2016). The table has two columns: Intent Type (with abbreviations) and Definition. It lists types such as IM for exploring a topic more broadly, LK for learning domain knowledge, LD for learning database content, FK for finding a known item, FS for finding specific information, FC for finding items with a named characteristic, FP for finding useful items without predefined criteria, EC for evaluating correctness, EU for evaluating usefulness, EB for picking the best item, and ES for evaluating specificity. Nine original types were excluded, including intents like accessing specific items and obtaining specific or whole items.}
\label{tab:intent-types}
\scriptsize
\begin{tabular}{lll}\toprule
Intent Type &Definition \\\midrule
IM (Identify something more to search) &Explore a topic more broadly \\
LK (Learn domain knowledge) &Learn about the topic of a search \\
LD (Learn database content) &Learn the type of information/resources available at a particular website \\
FK (Find a known item) &Searching for an item that you were familiar with in advance \\
FS (Find specific information) &Finding a predetermined piece of information \\
FC (Find items sharing a named characteristic) &Finding items with something in common \\
FP (Find items without predefined criteria) &Finding items that will be useful for a task, but which haven't been specified in advance \\
EC (Evaluate correctness of an item) &Determine whether an item is factually correct \\
EU (Evaluate usefulness of an item) &Determine whether an item is useful \\
EB (Pick best item(s) from all the useful ones) &Determine the best item among a set of items \\
ES (Evaluate specificity of an item) &Determine whether an item is specific or general enough \\
\bottomrule
\end{tabular}
\end{table}

\subsubsection{Intent Type Selection Prompt}

\def\lstlistingname{Prompt}
\lstset{
  basicstyle=\ttfamily\footnotesize,
  columns=fullflexible,
  frame=single,
  breaklines=true,
  postbreak=\mbox{\textcolor{red}{$\hookrightarrow$}\space},
  breakatwhitespace=true,
  captionpos=b,
}


\leavevmode
\begin{lstlisting}
You are an agent that will specify the implicit intent from an expanded query based on predefined intent types.

## **[Understanding the Task]**
- We provide **Expanded Queries**, which represent **realistic search refinements** that users would naturally input.
- Your goal is to **map each Expanded Query to the most appropriate intent types** that reflect the **user's true search goal**.
- The intent types you assign **will later be used to construct a clear Final Intent statement**.

---

## **[Intent Selection Guidelines]**
- **Select only relevant intent types.** Do not assign an intent type unless it **naturally** aligns with the query. It can be connected to **any number of intent types**, as long as you are confident in the reasoning.
- **Each intent should reflect a realistic user motivation** and be **distinct from others** in the same set.
- **Limit each intent reasoning to ~20 words** for clarity.
- Ensure that **selected intents are mutually distinguishable** and map to **different types of documents** in search results.

---

## **[Available Intent Types]**

Below are predefined **Information Seeking Intentions**, along with their definitions and examples:

- **IM (Identify something more to search)** => Expand or explore a topic further.  
  - **Expanded Query**: `Happycall blender models`  
  - **Intent**: Wants to explore the differences between old and new models.

- **LK (Learn domain knowledge)** => Gain general knowledge about a subject.  
  - **Expanded Query**: `Origin of Jeonju Yi clan`  
  - **Intent**: Wants to learn about the origin and history of Jeonju Yi clan.

- **LD (Learn database content)** => Understand what information is available on a particular website or database.  
  - **Expanded Query**: `Iseongdang TripAdvisor reviews`  
  - **Intent**: Wants to check reviews of Iseongdang on TripAdvisor.

- **FK (Find a known item)** => Search for something already familiar to the user.  
  - **Expanded Query**: `Happycall blender product code`  
  - **Intent**: Wants to find the product code of the latest model.  
  - **Expanded Query**: `Iseongdang main store location`  
  - **Intent**: Wants to locate the main branch of Iseongdang in Gunsan.

- **FS (Find specific information)** => Locate a predetermined piece of information.  
  - **Expanded Query**: `Happycall blender noise level`  
  - **Intent**: Wants to know the exact noise level when using the blender.

- **FC (Find items sharing a named characteristic)** => Find a set of items with a common feature (e.g., "best affordable laptops").  
  - **Expanded Query**: `Best budget wireless earphones`  
  - **Intent**: Looking for cost-effective wireless earphones under $150.  
  - **Expanded Query**: `Vegan restaurant reviews Seoul`  
  - **Intent**: Wants to read reviews of vegan restaurants in Seoul.

- **FP (Find items without predefined criteria)** => Look for useful but undefined items (e.g., "unique birthday gift ideas").  
  - **Expanded Query**: `Jeju travel spots recommendation`  
  - **Intent**: Looking for good summer vacation destinations in Jeju Island.

- **EC (Evaluate correctness of an item)** => Verify whether an item is factually accurate.  
  - **Expanded Query**: `Brandy Melville one-size controversy`  
  - **Intent**: Wants to check whether Brandy Melville's one-size policy is actually controversial.

- **EU (Evaluate usefulness of an item)** => Assess whether an item is valuable or helpful.  
  - **Expanded Query**: `Happycall blender user reviews`  
  - **Intent**: Wants to understand pros and cons based on user reviews.

- **EB (Pick best item(s) from all the useful ones)** => Compare and determine the best option.  
  - **Expanded Query**: `Best hotels in Paris`  
  - **Intent**: Wants to choose the best value-for-money hotel in Paris.  
  - **Expanded Query**: `Compare laptops`  
  - **Intent**: Wants to compare popular laptop models to pick the most suitable one.

- **ES (Evaluate specificity of an item)** => Check if an item is too broad or too specific.  
  - **Expanded Query**: `Laptop recommendations`  
  - **Intent**: Wants more refined suggestions like gaming, work, or student laptops due to overly broad results.

---

## **[Output Format]**
Return the query analysis in **structured JSON format** as follows:

```json
{
  "query_analyses": [
    {
      "query": "query text",
      "intents": [
        {
          "intent_type": "intent code (e.g., IM, FK, EB)",
          "reasoning": "Clear explanation of why this intent type applies"
        },
        {
          "intent_type": "another intent code",
          "reasoning": "Clear explanation of why this intent type applies"
        }
      ]
    }
  ]
}
\end{lstlisting}

\clearpage

\subsection{Intent Generation} \label{appendix:intent_gen_prompt}

\def\lstlistingname{Prompt}
\lstset{
  basicstyle=\ttfamily\footnotesize,
  columns=fullflexible,
  frame=single,
  breaklines=true,
  postbreak=\mbox{\textcolor{red}{$\hookrightarrow$}\space},
  breakatwhitespace=true,
  captionpos=b,
}

\begin{lstlisting}
You are an expert in search behavior analysis and user intent specification. Your task is to generate the single most relevant, realistic **user intent** for a given **expanded query**, ensuring that each intent aligns with a **predefined intent type**.
Strictly limit to one intent per each expanded query-intent type pair.
  
---

### **[Guidelines for Generating Intents]**  

1. **Ensure Direct Alignment with the Expanded Query**  
   - The user intent MUST precisely correspond to the given expanded query.  
   - If the query is about **price**, the intent should be about price.  
   - If the query is about **store location**, the intent should be about location, NOT unrelated details.  

2. **Strictly Follow the Given Intent Type**  
   - The intent must fit within the constraints of the provided intent type.  
   - **DO NOT** introduce intents that fall outside the given intent category.  
   - Example: If intent type = "FS" (Finding Specific Information), **do not generate direct navigational purchase  

3. **Ensure Distinct and Non-Redundant Intents**  
   - Each intent should reflect a **unique user need** that can be supported by distinct search results.  

4. **Maintain Clear and Concise Intent Statements**  
   - Keep each intent **within 15 words** to make it **precise and scannable**.  
   - The intent should be actionable and describe **what the user wants to accomplish**.  

5. **Ensure Realism and Logical Consistency**  
   - The generated intent must reflect how real users would actually refine their searches.  
   - Avoid hypothetical or exaggerated intents that do not match real-world search behavior.  

---

### **[Example]**  

#### **Input:**  
- **Expanded Query:** Brandy Melville Seongsu store sale  
- **Search Context:** Brandy Melville has opened a store in Seongsu and is currently running a sale event for a limited time.  
- **Intent Type:** FS - Finding Specific Information  

#### **Output (Valid User Intent for This Query):**  
**Wants to know which products are currently on sale at the Brandy Melville Seongsu store.**  
 

---
#### **Expanded Query:** {expanded_query}  
#### **Intent Type:** {intent_type}  
#### **Search Context:** {serp_summary}  

---
#### **Response Format:**  
Return a maximum of 1 intents. DO NOT generate more than one intent. 
Again, generate only the single most relevant, realistic intent. 
Ensure the response is in **Korean**.
Return the query analysis in **structured JSON format** as follows: 

           {{
               "intents": [
                           "Intent 1",
                   ]
}}
---


\end{lstlisting}

\clearpage
\section{Evaluation Pipeline} 

\renewcommand{\thefigure}{B.\arabic{figure}}
\renewcommand{\thetable}{B.\arabic{table}}

\setcounter{figure}{0}
\setcounter{table}{0}

\label{appendix:evaluation_pipeline}

\subsection{Evaluation Prompts} 
\label{appendix:evaluation_prompts}

\subsubsection{Satisfaction Evaluation Prompt} \label{appendix:satisfaction_prompt}

\def\lstlistingname{Prompt}
\lstset{
  basicstyle=\ttfamily\footnotesize,
  columns=fullflexible,
  frame=single,
  breaklines=true,
  postbreak=\mbox{\textcolor{red}{$\hookrightarrow$}\space},
  breakatwhitespace=true,
  captionpos=b,
}


\leavevmode
\begin{lstlisting}
You are a **search engine evaluator** responsible for assessing user satisfaction with search results.  
Your task is to classify the search engine result page (SERP) as **Satisfactory (1) or Dissatisfactory (0)** based on whether the user's information need is reasonably met.

### **Evaluation Criteria**

**Class 1 (Satisfactory - "1")**
- The content provides **sufficient, relevant, and actionable information** to reasonably satisfy the search intent.
- The user is likely to find **clear and useful paths** to fulfill their information need, either through the snippet content or the linked sources.
- Even if the results don't answer the query directly, they may still be considered satisfactory if the information is **clearly related and useful**.
- User experiences that illustrate outcomes---positive or negative---may be considered relevant to intents involving reviews or success stories.
- Information related to reviews or user experiences may be expressed through ratings, summaries, or visible user-generated content previews.
- Personal stories or scattered user experiences may still contribute meaningfully to satisfaction if they align with the core user intent.


**Class 0 (Dissatisfactory - "0")**
- The content is **insufficient**, irrelevant, outdated, or overly generic for the given intent.
- The user is unlikely to gain meaningful information without further effort or reformulating the query.
- The snippets may be vague, misleading, repetitive, or require excessive inference.

----
 ### **Expected Output Format**
    Return the classification in **JSON format**, including a short explanation for the decision.
```json
{
    "Classification": Class 0/1
    "Reasoning": The reason for this classification, which should be written in Korean (Maximum two sentences).
}```
\end{lstlisting}

\subsubsection{Relevance Evaluation Prompt} \label{appendix:relevance_prompt}

\def\lstlistingname{Prompt}
\lstset{
  basicstyle=\ttfamily\footnotesize,
  columns=fullflexible,
  frame=single,
  breaklines=true,
  postbreak=\mbox{\textcolor{red}{$\hookrightarrow$}\space},
  breakatwhitespace=true,
  captionpos=b,
}


\leavevmode
\begin{lstlisting}
You are a **search engine evaluator** responsible for assessing the **relevance of search results**. 

A relevant result must provide information that matches the type of information the user is seeking (e.g., explanations, comparisons, recommendations, or usage guides).

If the user asks for **a specific type of informational goal** --- such as alternative suggestions, product comparisons, or scientific mechanisms --- simply listing related keywords or products without satisfying that goal is not sufficient.

- For example, if the user asks for alternatives to a product, results should mention other usable options and their context of use.
- If the intent is to understand a chemical reaction, results should contain a clear explanation of the process, not just a mention of the compounds.
- When the user requests comparisons or rankings, results should go beyond product listings: they must include structured comparison, ranking rationale, or expert/user-driven synthesis.
- A product list without structured comparison, recommendation rationale, or summary insights from reviews should NOT be considered sufficient for fulfilling a comparison or recommendation intent.
- If the query explicitly asks for a list (e.g., discount items, substitute products), the result must include that list or closely resemble it (e.g., table, summary, enumerated mentions).
- Mentions of related locations or products without fulfilling the specific format or content (e.g., discount item list) should be considered irrelevant (Class 0).

Relevance must be judged by whether the content enables the user to fulfill their original goal, not merely whether it is topically similar.
Diverse content formats are acceptable --- bullet points, narratives, or reviews --- as long as they help meet the user's intent.

If none of the search results meet the specified constraint, the result should be considered irrelevant (Class 0), regardless of general topical similarity.

Your task is to classify the search engine result page (SERP) into three classes: Directly Relevant (class 2), Indirectly Relevant (class 1), and Irrelevant (class 0). 

---
### **Evaluation Criteria**
**Class 2 (Directly Relevant - "2")**: Fully satisfies the user's intent.
- There is at least one search result that suffices the search intent
- The result provides a clear and direct answer or solution to the user's search intent

**Class 1 (Indirectly Relevant - "1")**: Topically related but does not fully meet the intent.
- There is at least one search result that is related to the search intent, but does not fully satisfy it
- The relevance may be subjective and the degree of satisfaction can vary among users

**Class 0 (Irrelevant - "0")**: Not meaningfully related or does not address the user's need.
- No search results are meaningfully related to or satisfy the search intent 

---
### **Expected Output Format**
Return the classification in **JSON format**, including a short explanation for the decision.
```json
{
  "Classification": The class for the metric. ("Class 0", "Class 1", "Class 2")
  "Reasoning": The reason for this classification, which should be written in Korean (Maximum two sentences).
}```
\end{lstlisting}

\subsubsection{Clarity Evaluation Prompt} 
\label{appendix:clarity_prompt}

\def\lstlistingname{Prompt}
\lstset{
  basicstyle=\ttfamily\footnotesize,
  columns=fullflexible,
  frame=single,
  breaklines=true,
  postbreak=\mbox{\textcolor{red}{$\hookrightarrow$}\space},
  breakatwhitespace=true,
  captionpos=b,
}


\leavevmode
\begin{lstlisting}
### Pre-processing SERP
### Use the context up till the third appearance of "Section"
	splitted_context = serp_context.split("Section")
	if len(splitted_context) > 3:
    	serp_context = "Section".join(splitted_context[:3])

### System Prompt
You are a **search engine evaluator** responsible for assessing the clarity of search results.  
Your task is to classify the excerpt of the top 2 sections from a search engine result page (SERP) into three classes: **Highly clear (Class 2)**, **Moderately Clear (Class 1)**, **Not Clear (Class 0)**.

Clarity is defined as whether the provided search result page provides information that satisfies the user's search intent at the top two sections. You should consider:
- Whether the ranking of the documents are appropriate
- Whether users would need scrolling and exploration beyond the top two sections to satisfy their search intent

Clarity should be assessed based on ranking appropriateness:
- Do the results on the top two sections contain key information that resolves the search intent?
- Is the information that primarily satisfies the search intent placed at top ranks?

---
### **Evaluation Criteria **
**Class 2**
- At least one search result (snippet) within the top 2 sections fully satisfies the search intent
- With intents on the objective information of the products (e.g. price), shopping, price comparison and other shopping platforms are considered as fully satisfying intents.

**Class 1**
- The information provided by the top two sections are only partially satisfying the search intent

**Class 0**
- None of the search results in the top 2 collections satisfy the user's search intent
- For intents on subjective information (e.g. reviews, recommendations), shopping, price comparison and other shopping platforms are not considered as satisfying the intent.

---
### **Expected Output Format**
Return the classification in **JSON format**, including a short explanation for the decision.
```json
{
  "Classification": Class 0/1/2
  "Reasoning": The reason for this classification, which should be written in Korean (Maximum two sentences).
}```
\end{lstlisting}

\subsubsection{Reliability Evaluation Prompt} \label{appendix:reliability_prompt}

\def\lstlistingname{Prompt}
\lstset{
  basicstyle=\ttfamily\footnotesize,
  columns=fullflexible,
  frame=single,
  breaklines=true,
  postbreak=\mbox{\textcolor{red}{$\hookrightarrow$}\space},
  breakatwhitespace=true,
  captionpos=b,
}

\leavevmode
\begin{lstlisting}

### System Prompt
You are a **search engine evaluator** responsible for assessing the reliability of search results. You will be given a search query, the search intent, and the search engine result page. The search engine result page consists of sections, which may include snippets of documents.
Your task is to classify the search engine result page (SERP) into three classes: Highly reliable (Class 2), Moderately Reliable (Class 1), Ureliabile (Class 0). First, locate the documents that are most relevant to the search intent. Then, evaluate the reliability of the documents based on the following criteria:
---
### **Evaluation Criteria **
**Class 2**
- There exists at least one search result that fully satisfies the search intent and from a fully reliable source
- Reliable sources include official websites, professionally verified sources, news articles, and knowledge snippets on movies, places, people, etc.
- With intents on the objective information of the products (e.g. price), shopping, price comparison and other shopping platforms are considered fully reliable sources satisfying intents. However, for intents on subjective information (e.g. reviews, recommendations), they are not considered as fully reliable sources satisfying the intent.
- With intents on obejctive information about places and locations, map & place are considered reliable sources.

**Class 1**
- There exists at least one search result that satisfies the search intent and from a moderately reliable source
- Moderately reliable sources include Wikis and user-created contents with substantiated information such as blogs, cafe, user-generated content, etc.

**Class 0**
- There does not exist any search result that satisfies the search intent or all the results satisfying the serach intent are from unreliable sources.
- Unreliable sources include heavy advertising content and user-created contents without factual support or credible references.
---
### **Expected Output Format**
Return the classification in **JSON format**, including a short explanation for the decision.
```json
{
  "Classification": "0/1/2"
  "Reasoning": The reason for this classification, which should be written in Korean (Maximum two sentences).
}```
\end{lstlisting}

\clearpage
\section{Technical Evaluation Details}

\renewcommand{\thefigure}{C.\arabic{figure}}
\renewcommand{\thetable}{C.\arabic{table}}

\setcounter{figure}{0}
\setcounter{table}{0}

\subsection{Technical Evaluation 1}
\label{appendix:techeval1}

\begin{figure}[ht!]
\centering
\includegraphics[width=0.4\textwidth]{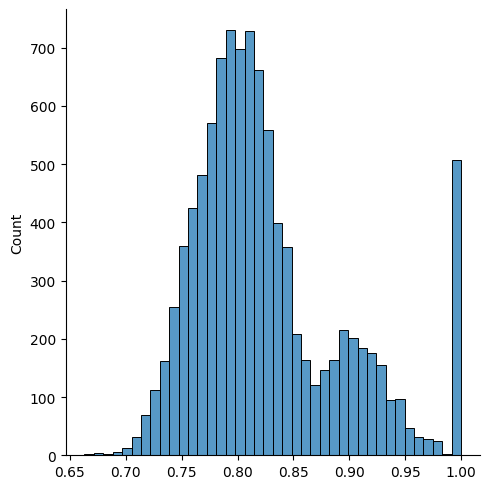}
\caption{Distribution of recall scores for generated expanded queries. Mean = 0.83, Std = 0.067.}
\Description{Distribution of recall scores for generated expanded queries. Mean = 0.83, Std = 0.067.}
\label{fig:techeval1}
\end{figure}


\clearpage
\subsection{Technical Evaluation 2}
\label{appendix:techeval2}

\begin{table}[!htp]\centering
\caption{For 11 queries in the Shopping domain, we report the frequency of the majority voted intents out of 20 in total.}
\Description{Table showing the frequency of majority-voted intents for 11 shopping queries, comparing Baseline and BloomIntent methods across three intent categories: Evaluativeness, Realism, and Novelty. For each query (S1 to S11), frequencies out of 20 are reported for each category and method. The Baseline method's Evaluativeness scores range from 7 to 20 with a mean of 14.64 and standard deviation 4.99; Realism ranges from 5 to 16 with mean 9.82 and SD 3.54; Novelty ranges from 0 to 3 with mean 1.55 and SD 1.21. The BloomIntent method shows consistently higher frequencies, with Evaluativeness scores close to or at 20 (mean 19.45, SD 1.04), Realism nearly always 20 (mean 19.91, SD 0.3), and Novelty ranging from 2 to 10 (mean 5, SD 2.61).
}
\scriptsize
\scalebox{1.2}{
\begin{tabular}{lrrrrrrr}\toprule
\textbf{} &\multicolumn{3}{c}{\textbf{Baseline}} &\multicolumn{3}{c}{\textbf{\sysname{}}} \\\midrule
Shopping &1. Evaluativeness &2. Realism &3. Novelty &1. Evaluativeness &2. Realism &3. Novelty \\
\midrule
Query\_S1 &8 &9 &3 &19 &20 &6 \\
Query\_S2 &9 &9 &0 &17 &20 &2 \\
Query\_S3 &7 &8 &1 &20 &20 &8 \\
Query\_S4 &13 &5 &2 &20 &20 &7 \\
Query\_S5 &20 &16 &3 &20 &20 &6 \\
Query\_S6 &18 &10 &2 &20 &20 &10 \\
Query\_S7 &17 &9 &0 &20 &20 &3 \\
Query\_S8 &19 &7 &0 &18 &20 &4 \\
Query\_S9 &18 &7 &2 &20 &19 &4 \\
Query\_S10 &20 &16 &1 &20 &20 &3 \\
Query\_S11 &12 &12 &3 &20 &20 &2 \\
\midrule
MEAN &14.64 &9.82 &1.55 &19.45 &19.91 &5 \\
STDEV &4.99 &3.54 &1.21 &1.04 &0.3 &2.61 \\
\bottomrule
\end{tabular}
}
\end{table}

\begin{table}[!htp]\centering
\caption{For 11 queries in Location domain, we report the frequency of the majority voted intents out of 20 in total.}
\Description{Table reporting frequencies of majority-voted intents for 11 location queries under Baseline and BloomIntent methods, across three categories: Evaluativeness, Realism, and Novelty, with scores out of 20. For each query (L1 to L11), frequencies are shown per category and method. Baseline Evaluativeness ranges from 4 to 19 (mean 13.45, SD 5.07), Realism from 1 to 20 (mean 11.91, SD 6.09), and Novelty mostly 0 or 1 (mean 0.55, SD 0.69). BloomIntent shows higher scores in Evaluativeness (mean 18.91, SD 1.51) and Realism (mean 17.27, SD 3.17), with Novelty remaining low (mean 0.36, SD 0.5). Some queries have zero Novelty frequency in both methods.
}
\scriptsize
\scalebox{1.2}{
\begin{tabular}{lrrrrrrr}\toprule
&\multicolumn{3}{c}{\textbf{Baseline}} &\multicolumn{3}{c}{\textbf{\sysname{}}} \\\midrule
Location &1. Evaluativeness &2. Realism &3. Novelty &1. Evaluativeness &2. Realism &3. Novelty \\
\midrule
Query\_L1 &14 &11 &0 &20 &20 &0 \\
Query\_L2 &6 &5 &0 &16 &17 &0 \\
Query\_L3 &17 &17 &1 &20 &18 &0 \\
Query\_L4 &12 &7 &1 &20 &10 &0 \\
Query\_L5 &4 &1 &0 &17 &15 &1 \\
Query\_L6 &16 &17 &0 &19 &20 &0 \\
Query\_L7 &18 &18 &1 &20 &20 &1 \\
Query\_L8 &19 &20 &1 &20 &20 &0 \\
Query\_L9 &16 &15 &0 &17 &18 &1 \\
Query\_L10 &17 &12 &2 &19 &18 &1 \\
Query\_L11 &9 &8 &0 &20 &14 &0 \\
\midrule
MEAN &13.45 &11.91 &0.55 &18.91 &17.27 &0.36 \\
STDEV &5.07 &6.09 &0.69 &1.51 &3.17 &0.5 \\
\bottomrule
\end{tabular}
}
\end{table}

\begin{table}[!htp]\centering
\caption{For 11 queries in the Knowledge domain, we report the frequency of the majority voted intents out of 20 in total.}
\Description{Table showing frequencies of majority-voted intents for 11 knowledge queries, comparing Baseline and BloomIntent methods across Evaluativeness, Realism, and Novelty categories, out of 20 votes. For all queries, Evaluativeness scores are consistently 20 for both methods. Baseline Realism ranges from 8 to 20 (mean 16.82, SD 3.63), while BloomIntent's Realism is higher, ranging from 16 to 20 (mean 19.55, SD 1.21). Novelty scores are higher in Baseline (mean 4.09, SD 3.18), with values between 0 and 9, whereas BloomIntent scores are mostly 0 or 1 (mean 0.36, SD 0.5).

}
\label{tab: }
\scriptsize
\scalebox{1.2}{
\begin{tabular}{lrrrrrrr}\toprule
&\multicolumn{3}{c}{\textbf{Baseline}} &\multicolumn{3}{c}{\textbf{\sysname{}}} \\\midrule
Knowledge &1. Evaluativeness &2. Realism &3. Novelty &1. Evaluativeness &2. Realism &3. Novelty \\
\midrule
Query\_K1 &20 &13 &6 &20 &16 &0 \\
Query\_K2 &20 &15 &2 &20 &20 &1 \\
Query\_K3 &20 &20 &4 &20 &20 &0 \\
Query\_K4 &20 &19 &5 &20 &20 &0 \\
Query\_K5 &20 &19 &0 &20 &20 &0 \\
Query\_K6 &20 &18 &6 &20 &20 &1 \\
Query\_K7 &20 &19 &9 &20 &20 &1 \\
Query\_K8 &20 &19 &9 &20 &20 &0 \\
Query\_K9 &20 &19 &1 &20 &20 &0 \\
Query\_K10 &20 &16 &2 &20 &19 &1 \\
Query\_K11 &20 &8 &1 &20 &20 &0 \\
\midrule
MEAN &20 &16.82 &4.09 &20 &19.55 &0.36 \\
STDEV &0 &3.63 &3.18 &0 &1.21 &0.5 \\
\bottomrule
\end{tabular}
}
\end{table}

\clearpage
\subsection{Technical Evaluation 3}
\label{appendix:techeval3}

\subsubsection{LLM-based Evaluation Results across Four Metrics}
\label{app:eval_metrics}

Performance comparison of LLM-based evaluation against human judgments across all four dimensions: Satisfaction, Relevance, Reliability, and Clarity. In the main text, we report the average accuracy across all four dimensions (0.6145) and average Cohen's Kappa (0.3709).

\begin{table*}[ht!]
\centering
\caption{The results demonstrate that our LLM-based evaluation approach achieves moderate to substantial agreement with human evaluators, with particularly strong performance in Satisfaction assessment (0.7206 accuracy). The confusion matrices reveal that the LLM evaluation tends to be most accurate for extreme judgments (Class 0 and Class 2), with somewhat lower performance on borderline cases (Class 1). This pattern is consistent with findings in prior work on automated evaluation~\cite{li2024llms, chiang2023can}, suggesting that LLMs, like human evaluators, have higher confidence in clear-cut cases.
}
\Description{Table presenting overall accuracy, Cohen's kappa, confusion matrices, and class-wise accuracy for LLM-based evaluation across four metrics: Satisfaction, Relevance, Reliability, and Clarity. Overall accuracy ranges from 0.5536 (Clarity) to 0.7206 (Satisfaction). Cohen's kappa ranges from 0.3098 to 0.4453. Confusion matrices show LLM predictions versus human ratings across three classes (0, 1, 2). For Satisfaction, LLM predicts class 0 correctly 540 times and class 1 correctly 623 times, with fewer errors in extreme classes. Class-wise accuracy indicates higher accuracy for extreme classes (Class 0 and Class 2) than for borderline Class 1 across all metrics, with Class 1 accuracy as low as 0.2753 for Relevance. This pattern aligns with prior work showing LLMs perform better on clear-cut cases.
}
\label{tab:full_eval_metrics}
\small
\begin{tabular}{l|cc|ccc|ccc|ccc}
\toprule
& \multicolumn{2}{c|}{\textbf{Satisfaction}} & \multicolumn{3}{c|}{\textbf{Relevance}} & \multicolumn{3}{c|}{\textbf{Reliability}} & \multicolumn{3}{c}{\textbf{Clarity}} \\
\midrule
Overall Accuracy & \multicolumn{2}{c|}{0.7206} & \multicolumn{3}{c|}{0.5719} & \multicolumn{3}{c|}{0.6119} & \multicolumn{3}{c}{0.5536} \\
Cohen's Kappa & \multicolumn{2}{c|}{0.4453} & \multicolumn{3}{c|}{0.3479} & \multicolumn{3}{c|}{0.3808} & \multicolumn{3}{c}{0.3098} \\
\midrule
\multicolumn{12}{c}{\textbf{Confusion Matrix}} \\
\midrule
& \multicolumn{2}{c|}{\textbf{LLM Predictions}} & \multicolumn{3}{c|}{\textbf{LLM Predictions}} & \multicolumn{3}{c|}{\textbf{LLM Predictions}} & \multicolumn{3}{c}{\textbf{LLM Predictions}} \\
\textbf{Human} & \textbf{0} & \textbf{1} & \textbf{0} & \textbf{1} & \textbf{2} & \textbf{0} & \textbf{1} & \textbf{2} & \textbf{0} & \textbf{1} & \textbf{2} \\
\midrule
0 & 540 & 307 & 311 & 47 & 23 & 233 & 108 & 39 & 498 & 116 & 29 \\
1 & 144 & 623 & 189 & 125 & 140 & 127 & 516 & 194 & 288 & 165 & 115 \\
2 & - & - & 109 & 171 & 471 & 20 & 133 & 230 & 47 & 121 & 225 \\
\midrule
\multicolumn{12}{c}{\textbf{Class-wise Accuracy}} \\
\midrule
Class 0 & \multicolumn{2}{c|}{0.6375} & \multicolumn{3}{c|}{0.8163} & \multicolumn{3}{c|}{0.6132} & \multicolumn{3}{c}{0.7745} \\
Class 1 & \multicolumn{2}{c|}{0.8123} & \multicolumn{3}{c|}{0.2753} & \multicolumn{3}{c|}{0.6165} & \multicolumn{3}{c}{0.2905} \\
Class 2 & \multicolumn{2}{c|}{-} & \multicolumn{3}{c|}{0.6272} & \multicolumn{3}{c|}{0.6005} & \multicolumn{3}{c}{0.5725} \\
\bottomrule
\end{tabular}
\end{table*}

\clearpage
\subsubsection{Human Expert Evaluation Results across Four Metrics}
\label{app:human_eval_metrics}
We present human expert evaluation results collected as ground truth. We first analyzed internal agreement among expert raters to characterize the level of ambiguity in human evaluation, as shown in Table~\ref{tab:agreement_ratio}. Table~\ref{tab:accuracy_analysis} further shows that LLM accuracy was significantly higher for cases with unanimous human rating (full agreement) than for those with split decisions (partial agreement), highlighting how task ambiguity limits automated evaluation performance.

\begin{table*}[ht!]
\centering
\caption{Descriptive analysis of full vs. partial agreement among expert raters across four evaluation metrics. A substantial portion of cases exhibit partial agreement, indicating the inherent ambiguity of the evaluation task.}
\label{tab:agreement_ratio}
\Description{This table presents a descriptive analysis of human expert agreement across four evaluation metrics: Satisfaction, Relevance, Reliability, and Clarity. For each metric, the number and percentage of cases with full agreement and partial agreement among raters are shown, along with the total number of evaluated cases. Satisfaction shows the highest full agreement rate at 68.96\% (1113 cases), while Clarity and Relevance have lower full agreement rates of approximately 55\%. Each metric also includes a significant proportion of partially agreed-upon cases, highlighting the inherent ambiguity of the evaluation task.}
\begin{tabularx}{\textwidth}{@{}l|>{\centering\arraybackslash}X|>{\centering\arraybackslash}X|>{\centering\arraybackslash}X@{}}
\toprule
\textbf{Metric} & \textbf{Full Agreement Count (Ratio)} & \textbf{Partial Agreement Count (Ratio)} & \textbf{Total Cases} \\
\midrule
Satisfaction & 1113 (68.96\%) & 501 (31.04\%) & 1614 \\
Relevance    & 876 (55.23\%)  & 710 (44.77\%) & 1586 \\
Reliability  & 974 (60.88\%)  & 626 (39.13\%) & 1600 \\
Clarity      & 883 (55.05\%)  & 721 (44.95\%) & 1604 \\
\bottomrule
\end{tabularx}
\end{table*}

\begin{table*}[ht!]
\centering
\caption{Evaluation accuracy varies significantly between unanimous and split human ratings, demonstrating the impact of task ambiguity on automated assessment. When human raters unanimously agree, LLM accuracy ranges from 65.35\% to 78.98\% with substantial inter-rater agreement (Cohen's $\kappa$ = 0.42-0.58). However, accuracy drops considerably for cases with split ratings (40.14\% to 56.49\%), with much lower agreement ($\kappa$ = 0.1-0.17). This pattern confirms that LLM evaluation performance is fundamentally limited by inherent task ambiguity, consistent with prior findings~\cite{arxiv2406.00247}.}
\Description{Table presenting LLM evaluation accuracy and inter-rater agreement (Cohen's kappa) by human rating agreement type: unanimous, split, and overall, across four metrics---Satisfaction, Relevance, Reliability, and Clarity. For unanimous ratings, accuracy ranges from 65.35\% to 78.98\%, with kappa values between 0.42 and 0.58. For split ratings, accuracy drops to 40.14\%–56.49\%, with much lower kappa (0.1–0.17). Overall accuracy lies between 55.36\% and 72.0\%, with kappa between 0.31 and 0.44. This illustrates that LLM performance is higher when human raters agree and decreases significantly with task ambiguity reflected by split ratings.}
\label{tab:accuracy_analysis}
\footnotesize\begin{tabularx}{\textwidth}{@{}l|XXll|XXll|XXll@{}}
\toprule
             & \multicolumn{4}{l|}{Full Agreement Ratings (All Raters Agree)} & \multicolumn{4}{l|}{Partial Agreement Ratings (Mixed Ratings)} & \multicolumn{4}{l}{Overall} \\ \midrule
Metric &
  LLM-Human Match Count &
  Total Human Ratings &
  Accuracy (\%) &
  Kappa &
  LLM-Human Match Count &
  Total Human Ratings &
  Accuracy (\%) &
  Kappa &
  LLM-Human Match Count &
  Total Human Ratings &
  Accuracy (\%) &
  Kappa \\ \midrule
Satisfaction & 879     & 1113    & 78.98    & 0.58    & 283    & 501    & 56.49   & 0.1    & 1162  & 1614 & 72.0  & 0.44 \\
Relevance    & 622     & 876     & 71.0     & 0.52    & 285    & 710    & 40.14   & 0.12   & 907   & 1586 & 57.19 & 0.35 \\
Reliability  & 661     & 974     & 67.86    & 0.5     & 318    & 626    & 50.8    & 0.17   & 979   & 1600 & 61.19 & 0.38 \\
Clarity      & 577     & 883     & 65.35    & 0.42    & 311    & 721    & 43.13   & 0.16   & 888   & 1604 & 55.36 & 0.31 \\ \bottomrule
\end{tabularx}
\end{table*}

\end{document}